\documentclass[desactivate]{aa}  

\usepackage{graphicx}
\usepackage{float}
\usepackage{natbib}
\bibpunct{(}{)}{;}{a}{}{,}
\usepackage{txfonts}

\usepackage{color}
\usepackage{multirow}
\usepackage{caption} 
\usepackage{lscape}
\usepackage{float}
\usepackage{placeins}

\usepackage[colorlinks=true, linkcolor=blue, citecolor=blue, urlcolor=blue]{hyperref}

\begin{document}

   \title{Cosmography via stellar archaeology of \\low-redshift early-type galaxies from SDSS}

   \author{Carlos A. Álvarez \inst{1}
          \and
          Marcos M. Cueli \inst{1,2}
          \and
          Alessandro Bressan \inst{1}
          \and
          Lumen Boco \inst{3}
          \and
          Balakrishna S. Haridasu \inst{1, 2}
          \and
          Michele Bosi \inst{1, 4}
          \and
          Luigi Danese \inst{1}
          \and
          Andrea Lapi \inst{1,2,5,6}
          }

   \institute{Scuola Internazionale Superiore di Studi Avanzati (SISSA), Via Bonomea 265, 34136 Trieste (TS), Italy\\
              \email{calonsoa@sissa.it}
         \and
    Institute for Fundamental Physics of the Universe (IFPU), Via Beirut 2, 34014 Trieste, Italy
        \and
Universitat Heidelberg, Zentrum fur Astronomie, Institut fur theoretische Astrophysik, Albert-Ueberle-Str. 3, 69120 Heidelberg, Germany
        \and
Department of Physics, University of Trento, Via Sommarive 14, 38123 Povo (TN), Italy
        \and
IRA-INAF, Via Gobetti 101, 40129 Bologna, Italy
        \and
INFN-Sezione di Trieste, via Valerio 2, 34127 Trieste, Italy}

  \abstract 
   {Cosmic chronometers offer a model-independent way to trace the expansion history of the Universe via the dating of passively evolving objects. This enables testing the validity of cosmological models without concrete assumptions of their energy content.}
   {The main goal of this work is to derive model-independent constraints on the Hubble parameter up to $z \sim 0.4$ using stellar ages from the fitting of Lick index absorption lines in passively evolving galaxies. Contrary to recent related works that rely on finite differences to obtain a discrete measurement of the expansion of the Universe at an average redshift, our goal is to perform a cosmographic fit of $H(z)$ in terms of the Hubble constant ($H_0$) and the deceleration ($q_0$) and jerk ($j_0$) parameters.}
   {We carefully select spectra of massive and passively evolving galaxies from the SDSS Legacy Survey. After applying a stacking procedure to ensure a high signal-to-noise ratio, the strength of Lick indices is fit using two stellar population models (TMJ and Knowles) to derive stellar population parameters. A cosmographic fit to the stellar ages is performed, which in turn enables the sampling of the Hubble parameter within the considered redshift range.}
   {The baseline result comes from using the TMJ-modelled ages, and it yields a value of $H_0 = 70.0^{+4.1}_{-7.6} \text{ km s}^{-1} \text{ Mpc}^{-1}$ for the Hubble constant, where uncertainties refer only to the statistical treatment of the data. The sampling of the Hubble parameter at $0.05 < z < 0.35$ is competitive with discreet model-independent measurements from the literature. As a by-product of the Lick index fitting procedure, we provide scaling and dispersion relations of stellar population parameters with respect to velocity dispersion using the low-redshift end of our sample. We finally draw attention to an unexpected oscillating pattern in a number of critical indices with respect to redshift, which translates into a similar behaviour in the $t-z$ relations. These features have never been discussed before, although they are present in previous measurements. We show that they do not originate from our methodology, suggesting a possible origin in the data reduction process.}
   {}


   \maketitle

\section{Introduction}

The $\Lambda$CDM model is regarded as the simplest theoretical framework to provide a remarkable fit to the bulk of cosmological observables, namely the cosmic microwave background (CMB) temperature and polarization spectra \citep{planck18,aiola20,DUTCHER21}, baryon acoustic oscillation (BAO) measurements \citep{EISENSTEIN05,BEUTLER11,ROSS15,ZHAO22,ADAME25}, the distance-redshift relation with type Ia supernovae \citep{PERLMUTTER99,SCOLNIC18,BROUT22} or weak gravitational lensing \citep{ABBOTT20,ABBOTT22,DVORNIK23,DES+KIDS}, among others. However, the era of precision cosmology has revealed a number of observational inconsistencies that could point to a fundamental defect in the model, the most famous of which is the Hubble tension \citep[see][for a review]{DIVALENTINO21}. Indeed, ever since the first release of the \emph{Planck} cosmological analysis of the CMB \citep{PLANCK14}, a pervasive discrepancy between early- and late-time measurements of the cosmological parameter $H_0$ has plagued the scientific community. The former, represented by CMB, BAO and Big Bang Nucleosynthesis (BBN) analyses, assume a $\Lambda$CDM model and agree upon low values of the Hubble constant \citep[\emph{e.g.} $H_0=67.27\pm0.60$ km$\,\text{s}^{-1}\text{Mpc}^{-1}$ as reported by][]{planck18}. The latter, characterized by "direct" model-independent approaches, are represented (although not limited to) the cosmic distance ladder approach and systematically predict high values, such as the last determination of $H_0=73.04\pm1.05$ km$\,\text{s}^{-1}\text{Mpc}^{-1}$ by \cite{RIESS22}.

In this troubled arena, the importance of emergent, independent and complementary cosmological probes should not be overlooked \citep{moresco22}, especially if they require no assumptions on the underlying cosmological model. This is the case of cosmic chronometers, which have come forth as a promising alternative to probe the expansion history of the Universe. In general terms, cosmic chronometers are astrophysical objects that can be statistically observed and dated at different redshifts, thus allowing us to reconstruct the time evolution of a large-scale homogeneous and isotropic Universe without assuming a specific cosmological model. This idea was first put forward by \cite{jimenez02}, who suggested the use of passively evolving galaxies as cosmic chronometers, arguing that identifying the shift between the age distribution of two such populations at different epochs would yield an estimate of the Hubble parameter at an effective redshift between them. Although this exact approach was not used, the subsequent work by \cite{simon05} put the underlying idea to the test and provided measurements of the Hubble parameter at different redshifts using data from the Gemini Deep Deep Survey \citep[GDDS;][]{abraham04} via the single age differences of galaxies determined to be in passive evolution after a very short burst of star formation. Indeed, within the downsizing scenario \citep{COWIE96}, the most massive elliptical galaxies are older, formed earlier and in a shorter burst, after which they have been in passive evolution \citep{KAUFFMANN03,GALLAZZI05,thomas05,thomas10,CONROY14}. In other words, at a given redshift, the stellar populations of passive elliptical galaxies with the same stellar mass should be increasingly coeval as one examines progressively more massive or $\alpha$-enhanced galaxies. As a consequence, the cosmic chronometer approach relies on an accurate identification of massive and passively evolving elliptical galaxies at any given redshift \citep{borghi22a}.

The other essential aspect is the determination of the stellar age of galaxies. Although the cosmic chronometer approach is differential in nature (in the sense that the relevant quantity for the analysis is not the absolute ages but their redshift evolution), a robust dating of the stellar populations is still needed. Two main methods have been used to derive stellar ages from passive galaxy spectra, the first of which is full spectral fitting (FSF). FSF techniques \citep{HEAVENS00,CID05,TOJEIRO07,CHEVALLARD16,WILKINSON17,CARNALL18} involve deriving the physical properties of a galaxy by fitting the entire observed spectrum with synthetic stellar population models, thus considering all available spectral features as well as the continuum. This approach has recently been applied in the cosmic chronometer literature \citep{simon05,zhang14,ratsimbazafy17,jiao22,tomasetti23}, producing 16 distinct data points in the Hubble diagram. The second method involves the modelling of specific spectral features that can correlate with age, metallicity and star formation history. One example of this is the characteristic D$4000_\text{n}$ break in passive galaxy spectra, which has been primarily employed by \cite{moresco11,moresco12,moresco16a} owing to its simple tight correlation to stellar age at fixed metallicity that can be calibrated using stellar population synthesis models. A total of 15 measurements of $H(z)$ have so far been obtained for $z<2$ using this method, with overall tighter constraints with respect to FSF approaches. However, several other spectral features have been traditionally employed to determine the age of passive elliptical galaxies given their sensitivity to stellar population properties.

Lick indices \citep{worthey94,worthey97} are a set of 25 well-studied absorption features in the spectra of elliptical galaxies. They are defined by relatively wide bandpasses and two additional windows characterizing pseudocontinua blue- and redward of the central region. Their measurement in early-type galaxy (ETG) spectra signalled the presence of supersolar Mg/Fe ratios \citep{WORTHEY92,DAVIES93,carollo94,MEHLERT8,jorgensen99,LONGHETTI00}, which strongly suggested short star formation timescales, in agreement with the downsizing scenario. This led to the development of stellar population models that predict the strength of Lick absorption features as a function of single-burst stellar age, metallicity and varying element abundances \citep{thomas03,thomas04,tantalo04,schiavon07,thomas11,chung13}. In addition, the alternative avenue of using models that predict the full spectral energy distribution (SED) at a moderate-high resolution \citep{schiavon02,BRUZUAL03,vazdekis10,conroy12,vazdekis15,conroy18,knowles23} can also be exploited to predict Lick index strengths and derive stellar population parameters. However, the Lick index method has only been employed by \cite{borghi22b} in the cosmic chronometer context, providing a single measurement of the Hubble parameter at $z\sim 0.7$.

The aim of this work is to provide independent measurements of the Hubble parameter at low redshifts $(z\lesssim 0.4)$ via the cosmic chronometer technique. These measurements are alternative to those based on FSF \citep{simon05,zhang14} and the D$4000_\text{n}$ break \citep{moresco12}, not only due to the method of estimating the stellar ages, but also because of the nature of the approach. Indeed, instead of using finite differences to estimate the redshift derivative of cosmic time, we trace the continuous redshift evolution of the ages via a cosmographic model-independent approach. The cosmographic approach \citep{capozziello12} consists in directly fitting $H(z)$ (or derivative/integral functions) when $H(z)$ is expressed as a Taylor series on kinematical parameters (deceleration $q_0$, jerk $j_0$, snap $s_0$). The expansion on the $y-$redshift holds well up to $y\sim 0.4$ \citep{capozziello12}, which translates to $z \sim 0.7$.

We provide a thorough description of the careful selection of the cosmic chronometer sample and employ two different stellar population (SPS) models, namely \cite{thomas11} and \cite{knowles23}, to determine the age of the stellar populations from Lick index strength predictions. It should be noted that some models allow for a variety of internal configurations, and different models can significantly alter the final readout of $H(z)$. For example, one can find in literature the work by \cite{moresco20}, which analyses the impact of the chosen IMF, stellar library and overall SPS model on the final $H(z)$ estimation. The results from that study highlighted that the global SPS model and the stellar library are the main drivers of disagreement on the final $H$ readout, while the IMF played only a minor role. A comprehensive analysis of the various contributions to the systematic uncertainty budget \citep{moresco22} is necessary in order to predict the current and future power of cosmic chronometers in providing a reliable estimation of the Hubble parameter. However, this work aims to produce a robust statistical measurement of $H(z)$, through the ages derived with the aforementioned models, accounting only for the systematic uncertainty contribution driven by data management.

The paper is structured as follows. Section \ref{sec:CC} introduces the cosmic chronometer framework, outlines the cosmographic approach to the Hubble parameter and provides a detailed description of the selection criteria for the cosmic chronometer sample. Section \ref{sec:DATATREATMENT} details the methodology involved in the process of measuring and fitting Lick absorption indices to the models, including a careful description of different aspects, namely spectrum stacking, instrumental resolution and velocity dispersion corrections. The fit of the stellar ages to sample the Hubble parameter within the cosmographic framework is also explained. Section \ref{sec:RESULTS} presents the outcome of the cosmographic fit for both the TMJ and Knowles modelled ages. With the former, we draw the estimation of the Hubble parameter as our central result, accompanied by a brief discussion about the sources of systematic uncertainty. Section \ref{sec:otrosRESULTS} is included to discuss other implications of our results beyond the cosmographic analysis. We have leveraged the low-redshift data to obtain an estimate of the scaling relations of the ETGs using both SPS models. The final part of this section covers the analysis of the oscillations in the $t-z$ relation and the role that index strengths play. Finally, Section \ref{sec:conclusions} summarises the main conclusions.

\section{Cosmic chronometers\label{sec:CC}}

In this section, we briefly introduce the theoretical framework of cosmic chronometers, focusing on the cosmographic approach that we will follow throughout the paper. Then, the cosmic chronometer sample will be presented, with a detailed description of the selection criteria and its physical properties.

\subsection{Theoretical framework \label{sec:cosmoframe}}

The cosmic chronometer approach typically relies on approximating the differential expression of the Hubble parameter, given by
\begin{equation}
    H\equiv \frac{1}{a}\frac{\text{d}a}{\text{d}t_{\text{\scriptsize{U}}}}=-\frac{1}{1+z}\frac{\text{d}z}{\text{d}t_{\text{\scriptsize{U}}}} ,
    \label{eq:Hubble}
\end{equation}
by the ratio of the intervals in redshift ($\Delta z$) and age\footnote{In this work, we denote the age of the Universe at redshift z by $t_{\text{\scriptsize{U}}}(z)$, while $t(z)$ refers to the age of a galaxy at redshift $z$.} ($\Delta t$) between two well-defined samples of cosmic chronometers. These two groups statistically trace the same population at different redshifts \citep{jimenez02}, i.e. $(t_1,z_1)$ and $(t_2, z_2)$.  Indeed, since the average formation time of cosmic chronometers observed at different redshifts is defined to be be equal, it factors out when computing the difference in cosmic time between two redshifts, that is,  $\Delta t_{\text{\scriptsize{U}}} = t_{\text{\scriptsize{U}}}(z_2) - t_{\text{\scriptsize{U}}}(z_1) = t_2 + t_{\text{\scriptsize{U}}}(z_{\text{form}}) - (t_1 + t_{\text{\scriptsize{U}}}(z_{\text{form}}) )= \Delta t $. In other words, the time elapsed between two redshifts and the difference in the age of the stellar populations of cosmic chronometers observed at those redshifts are equivalent. This gives a local measurement of $H$ at the mean redshift $\langle z \rangle \equiv \left(z_1 +z_2\right)/2$ of the ensemble. Therefore,
\begin{equation}
    H\left(\langle z \rangle\right) \approx -\frac{1}{1+\langle z \rangle} \frac{\Delta z}{\Delta t} 
    \label{eq:Hubble_approx}
\end{equation}

However, we decide to take a different path to that of equation (\ref{eq:Hubble_approx}) and follow a cosmographic approach, valid for relatively low redshifts \citep{capozziello12}. As will be observed when the full results are presented in Section \ref{sec:RESULTS}, the overall distribution of ages for each coeval population can feature fluctuations that would critically affect (i.e. reading $H < 0$) the point estimation of $H$ obtained following (\ref{eq:Hubble_approx}). This issue is not new; indeed \cite{tomasetti23} found an anomaly at higher redshift by which they would measure a negative value for $H$ if used (Appendix A: The $z<1.07$ anomaly). Since we aim at drawing an estimation for $H$ based on a wealth of data from a rich catalogue, it could be argued that these anomalies could be addressed statistically, with $H<0$ measurements being outliers. Despite this, there are two more caveats that make us prefer the cosmographic analysis based on a functional fit. The first is the manual pairing of groups of galaxies (or stacks in our case), which implies a further arbitrary choice that shall be accompanied with a sensible analysis of its effect as a systematic uncertainty \citep{borghi22b}. The other is the fact that a continuous fit of $t(z)$ allows us not to reduce the redshift window for which our estimation is solid. In other words, a point measurement of $H$, based on a pair of points, takes as fiducial redshift the average redshift of the pair. Instead, following our methodology, each point individually contributes with its own spectroscopic redshift to the estimation and so our original range of redshift is kept.

Working directly with $t(z)$ requires integrating the inverse of the Hubble function. By doing so, one gets the evolution of cosmic time as a function of redshift,
\begin{equation}
    t_{\text{\scriptsize{U}}}(z) = t_{\text{\scriptsize{U}}}(0) - \int_0 ^z \frac{1}{1+z'}\frac{\text{d}z'}{H(z')}
    \label{eq:tUdezdeHz},
\end{equation}
where $t_{\text{\scriptsize{U}}}(0)$ is the current age of the Universe. In terms of the stellar ages of cosmic chronometers, we can rewrite (\ref{eq:tUdezdeHz}) as
\begin{equation}
    t(z)= t_0-\int_0 ^z \frac{1}{1+z'}\frac{\text{d}z'}{H(z')},
    \label{eq:tdezdeHz}
\end{equation}
where $t_0\equiv t_{\text{\scriptsize{U}}}(0)- t_{\text{\scriptsize{U}}}(z_{\text{form}})$ is the age of the cosmic chronometer population at $z=0$. Therefore, the continuous redshift evolution of the ages can be exploited to derive the expansion history of the Universe via the parametrisation of the Hubble parameter. Following \cite{capozziello12}, one can perform a series expansion that overcomes convergence issues at high redshift, with the advantage of being independent of the underlying cosmological model. Under the introduction of the $y$-redshift, defined as $y\equiv z/(1+z)$, the expansion to order $k$ of $H(y)$ reads
\begin{equation}
    H(y) = H_0 \cdot \sum_{i = 0}^k\left(\mathcal{H}^i_0 \cdot y^i\right) + \mathcal{O}(y^{k+1})
    \label{eq:Hcosmographic},
\end{equation}
where the terms written as $\mathcal{H}^i_0$ are dimensionless as the Hubble constant has been factored out so that the sum reflects just the kinematics of the expansion of the Universe. Given the redshifts of interest in this work, it is enough to use the expansion up to second order ($k = 2$). The relevant coefficients are $\mathcal{H}^0_0\equiv 1$, $\mathcal{H}^1_0 \equiv \mathcal{H}^1_0(q_0)$ and $\mathcal{H}^2_0 \equiv \mathcal{H}^2_0(q_0, j_0)$, where $q_0$ and $j_0$ are the so-called deceleration and jerk parameters. With all this, we can rewrite the cosmographic expansion of $H(z)$ as
\begin{equation}
   \frac{H(z)}{H_0} \approx1 + \left(1+q_0\right)\frac{z}{1+z} + \left(1+\frac{j_0}{2} + q_0 - \frac{q_0^2}{2}\right)\left(\frac{z}{1+z}\right)^2
    \label{eq:Hcosmographic_ord2}.
\end{equation}

Introducing the above equation into (\ref{eq:tdezdeHz}) yields the cosmographic model for the redshift evolution of the stellar ages of cosmic chronometers. In particular, the $\Lambda$CDM model is recovered by setting $q_0 = \Omega_m/2 - \Omega_\Lambda$ and $j_0 = 1$.

\subsection{Cosmic chronometer sample\label{sec:Sample}}

As described in the introduction, we aim to select the most massive and passively evolving sample of ETGs. This is the group of red and metal-rich elliptical galaxies that have not undergone any major process of star formation since the end of the formation of the bulk of their stellar mass \citep{carretero07,sargent15}. This happened at $z\gtrsim1$ \citep{yildirim17} and in a short burst \citep{cimatti06}, implying $\alpha$-enhanced stellar populations. It is well established in literature that through a cross-selection in morphology \citep{clemens06, shimasaku01}, colour \citep{moresco16a}, emission lines \citep{moresco11} and the ratio between the CaIIK and CaIIH absorption lines \citep{borghi22a}, a clean sample of these objects can be obtained.

Since this work aims at low $(z \lesssim 0.4)$ redshift, the sources will be extracted from the Sloan Digital Sky Survey (SDSS), in particular from the SDSS Legacy Survey corresponding to SDSS-I/II \citep{YORK00,ABAZAJIAN09}. In a first step, we submit a query to the SDSS catalogue selecting all galaxies satisfying a series of criteria including morphology, velocity dispersion (a proxy for stellar mass) and other quality flags for their spectra. This process is performed via an SQL query and is described as follows:

\begin{figure}
	\includegraphics[width=\linewidth]{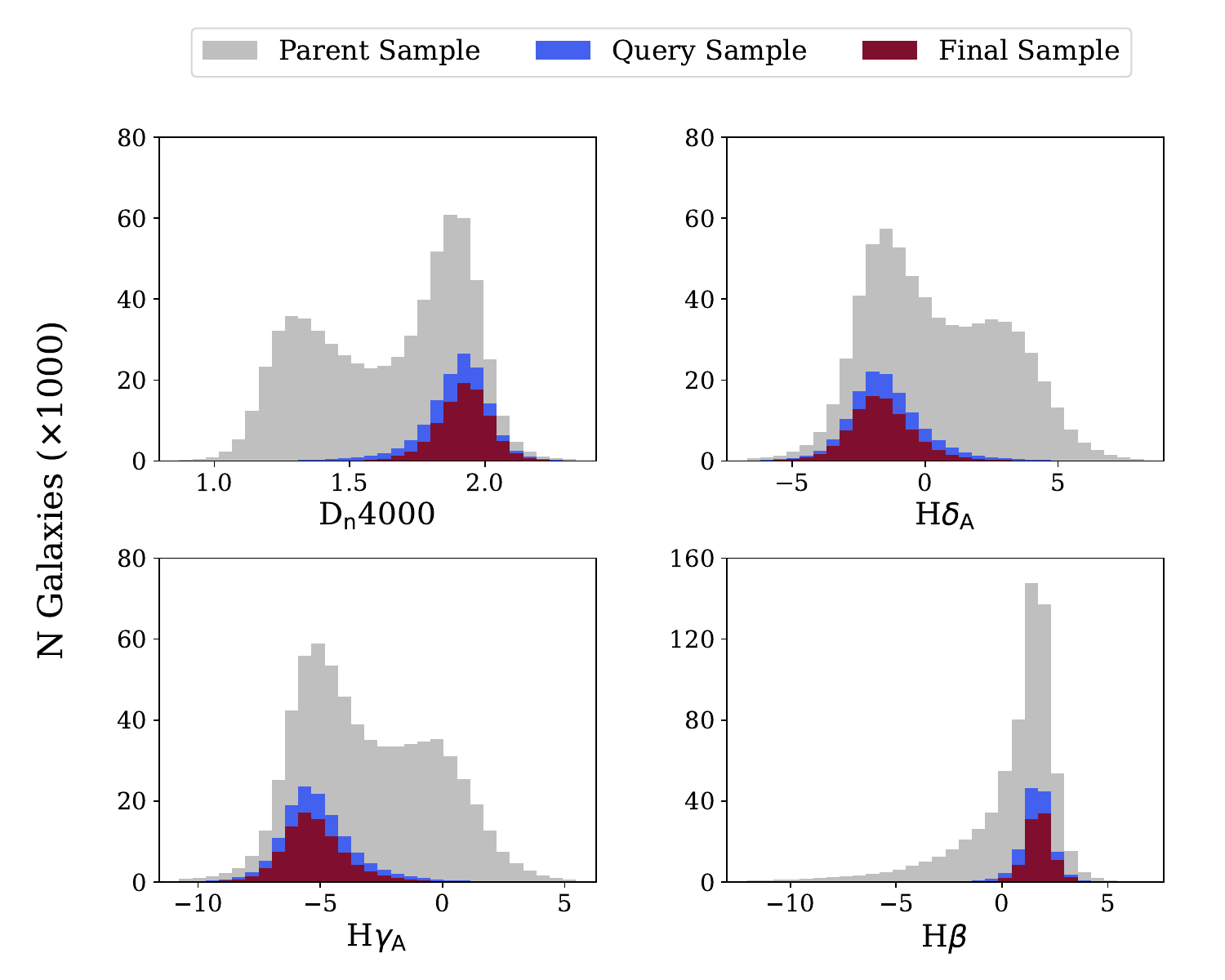}
    \caption{Distribution of D$4000_\text{n}$ and three Balmer-line indices over the parent (grey), query (blue) and final (maroon) galaxy samples.}
    \label{fig:distribslines}
\end{figure}

\begin{enumerate}
\item Objects identified as a galaxy with no star formation activity, using the columns \texttt{Class} and \texttt{subClass} from the \texttt{specObj} view.
\item Velocity dispersion restricted to lie between $100$ and $400 \text{ km s}^{-1}$, in order to select massive galaxies with stellar masses greater than $\sim 10^{10.5} \text{M}_\odot$, preventing contamination from rotation-dominated objects \citep{veale17} that are typically star-forming \citep{diteodoro16}.
\item Elliptical morphology, imposed through the restriction of the concentration/compactness parameter as indicated in \cite{shimasaku01}. In other words, the ratio between the radii containing $50\%$ and $90\%$ of the Petrosian flux has to be less than $0.33$.
\item Rest-frame wavelength window to measure spectral features. Indeed, requiring that all 25 Lick indices (and the CaII H and CaII K lines for reasons that will be explained below) be observed would imply that the spectra should be available in rest-frame wavelengths between $3600$ and $6500$ \AA. Although this would be a conservative cut, as the reddest indices (NaD, TiO$_1$ and TiO$_2$) will not be used in the analysis, the window is actually extended to $6700$ \AA \ to be able to later exclude objects with emission in Balmer H$\alpha$ ($6526$ \AA). Since the SDSS observed window reaches up to $\approx 10000$ \AA, this imposes the restriction that the redshift be below $z \sim 0.5$. This is not a problem for the scope of this work, but it is important to keep in mind should one wish to extend the analysis to higher redshifts using SDSS data. In that case, one could drop the upper limit on the rest-frame window down to some value around $5100$ \AA \  (effectively including $\text{[OIII]}\lambda5007$), or slightly higher if one wishes to measure the Fe$5270$ and Fe$5335$ indices, used to build the metallicity-sensitive [MgFe] and [MgFe']. For reference, $z \sim 0.9$ can be achieved by only dropping the requirement on the absence of emission in $\text{H}\alpha$.

\end{enumerate}

We start from a parent sample, constructed by requiring that the objects be galaxies and satisfy the wavelength coverage condition, which amounts to a total of 669475 galaxies. This number is reduced to 506475 once the velocity dispersion constraint is imposed; then, both the star-forming flag and the concentration parameter conditions further diminish the number to 135554 galaxies, which constitute what we call the SQL query sample. It should be noted that the most restrictive criterion is the concentration parameter, which reduces the number of sources by two thirds.

Once the SQL query sample is defined, we can proceed with the download of the spectra to perform a further spectroscopic selection and remove objects that could show some remaining star-formation. To this end, we inspected emission lines associated with young stellar systems and ongoing star formation, namely $\text{[OII]}\lambda 3727$, $\text{[OIII]}\lambda 5007$, $\text{H}\beta$ and $\text{H}\alpha$ \citep{sobral15,suzuki16,khostovan20}. The SDSS provides the measurements of the equivalent width (EW) of emission lines as performed by \cite{tremonti04}. For completeness, we decided to also measure the EW ourselves to cross-check with the SDSS values. We found good agreement and for the purposes of this work we decided to make a combined (stricter) cut requiring galaxies to meet the selection criteria for both measurements. In particular, we imposed that any of the EWs be larger than $-2$ \AA, although values between $-5$ and $-2$ \AA \ were allowed if the $\text{S}/\text{N}$ of the line was smaller than $3$, meaning that some noise can be mistakenly identified as an emission. For reference, previous works with cosmic chronometers agree to exclude any galaxy with some of these lines showing emissions stronger than $-5$ \AA \ \citep{moresco11,moresco12,tomasetti23,borghi22a}. The criterion on the $\text{S}/\text{N}$ varies slightly instead; for example, \cite{moresco16a} chose a maximum threshold of $2$ in $\text{S}/\text{N}$ while \cite{tomasetti23} relaxed it to a value of $3$. \cite{borghi22a} did not impose any restrictions, although they found that the final sample rarely showed a $\text{S}/\text{N}$ higher than $3$.

\begin{figure}
	\includegraphics[width=\columnwidth]{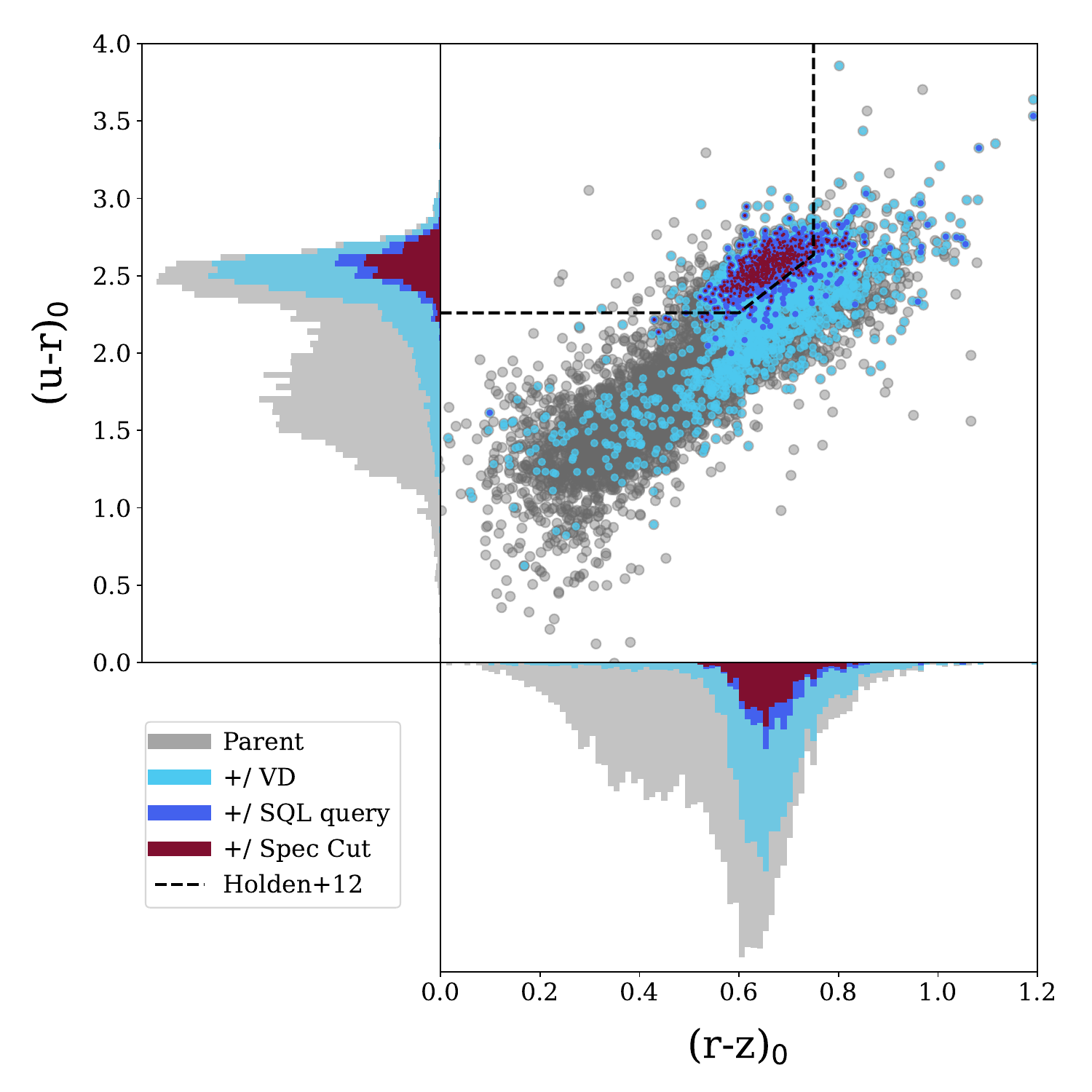}
    \caption{Distribution of sources in the $(\text{u} - \text{r})_0 \text{ vs } (\text{r}-\text{z})_0$ colour-colour plane. The parent sample is shown in grey, while the increasingly restrictive subsamples (adding velocity dispersion, query and spectroscopic information) are shown in light blue, dark blue and maroon, respectively. The dashed black line defines the passive galaxy region \citep{holden12}}
    \label{fig:colour_ur0_rz0}
\end{figure}

To complement this criterion, we decided to put in practice one of the results from \cite{borghi22a} by making a cut in $\text{H}/\text{K}$, the ratio between the $\text{CaII H}$ and the $\text{CaII K}$ pseudo-Lick indices. Larger values of this quantity indicate the presence of star formation \citep{cincunegui07,moresco18,pimbblet19,werle20} as $\text{CaII H}$ deepens due to its blending with the Balmer H$\varepsilon$ from young hot stars, while $\text{CaII K}$ remains unaffected. Following \cite{borghi22a}, we select sources that satisfy $\text{H}/\text{K} < 1.2$. 

The final sample is made up of $90396$ sources and extends up to $z \sim 0.38$. Fig. \ref{fig:distribslines} shows how the distribution of some absorption features ($\text{D}4000_\text{n}$ and three Balmer indices) changes from the parent to the final sample. The typical bimodality separating early- from late-type galaxies is clearly observed in the $\text{D}4000_{\text{n}}$, $\text{H}\delta_\text{A}$ and $\text{H}\gamma_\text{A}$ distributions of the parent sample (in grey), which is already broken with the query sample (in blue), clearly defining a much purer passive ETG sample, which is culminated in the final step after adding spectrometry (maroon). Note how the distributions for D$4000_\text{n}$ and H$\delta_\text{A}$ are equivalent to the ones observed by \cite{borghi22a}.

Lastly, studies with ETGs typically rely on a photometric selection, including cuts in galaxy colour that separate between early and late types. These criteria have been known for a long time, as \cite{strateva01} already showed that the distribution of SDSS galaxies on the $\text{u}-\text{g}/\text{g}-\text{r}$ plane presented a strong bimodality and proposed $\text{u}-\text{r}= 2.22$ as an optimal separator. More recently, cuts in colour have been applied for the selection of cosmic chronometer candidates. For instance, \cite{moresco16a} selected early-type galaxies from BOSS from the $\text{g}-\text{r}/\text{r}-\text{i}$ plane following \cite{eisenstein11}, \cite{masters11}, \cite{white11}, and \cite{maraston13}, while \cite{borghi22a} used $(\text{NUV}-\text{r})_0/(\text{r}-\text{J})_0$ \citep{ilbert13} and \cite{tomasetti23} used $(\text{U}-\text{V})_0/(\text{V}-\text{J})_0$ colour\footnote{The subscript $_0$ denotes rest-frame colours.} planes \citep{mclure18} to identify passive galaxies in the LEGA-C \citep{vanderwel16} and VANDALS \citep{mclure18} surveys, respectively.

Since criteria based on colours typically have non-negligible percentages of contamination or incompleteness, we do not intend to apply a further cut based on colours. However, we do want to make sure that the final sample is physically reasonable in terms of its colour distribution. For the SDSS photometric bands at our disposal, \cite{holden12} defined a passive galaxy region on the $(\text{u}-\text{r})_0/(\text{r}-\text{z})_0$ plane by the polygon
\begin{equation}
    \left\{
    \begin{aligned}
        (\text{u} - \text{r})_0 &> 2.26 \\
        (\text{r} - \text{z})_0 &< 0.75 \\
        (\text{u} - \text{r})_0 &> 0.76 + 2.5 (\text{r} - \text{z})_0
    \end{aligned}
    \right. \ \ \ ,
    \label{eq:polygon_Holden}
\end{equation}
which is plotted with a black dashed line in Fig. \ref{fig:colour_ur0_rz0}. We also depict, for the $0.05<z<0.06$ redshift bin, the rest-frame colours of all galaxies in the parent sample (grey) and all the subsequent subsamples, namely adding a velocity dispersion cut (cyan), the star-forming flag and concentration parameter (blue) and the spectroscopic selection (maroon) defining the final sample. K-corrections have been applied to the observed magnitudes following \cite{chilingarian10} and \cite{omill11}. The behaviour is qualitatively similar for all redshifts in the sample, although with different incompleteness fractions. In all cases, these numbers never exceed the expected 18\% of missing passive galaxies \citep{holden12}. We can safely conclude that the selection procedure is physically solid.

\section{Data treatment and methodology\label{sec:DATATREATMENT}}

The cosmic chronometer framework involves deriving stellar ages from the spectra of the final galaxy sample detailed in the previous section. To this end, Lick absorption features are measured and fit to the predictions of stellar population models for Lick index strengths. However, the process is not straightforward and needs to be carefully explained. In particular, high signal-to-noise spectra are needed for the index measurements to be reliable. This is accomplished by stacking the flux densities of single galaxies of similar nature \citep{ferreras19,sanchez20,parikh21}. Once these "raw" measurements are performed, corrections are applied to account for the astrophysical and observational effects that are tangled in the spectra and affect the real value of the indices, namely instrumental resolution and velocity dispersion. Indices will then be fit with two flux-calibrated models, the Lick-index model of \cite{thomas11} (referred to as TMJ for the rest of the paper) and the SSP model based on the sMILES library \cite{knowles23} (referred to as Knowles) via a Markov chain Monte Carlo (MCMC) algorithm. An estimation of the luminosity-weighted ages, metallicities and $\alpha$-enhancement of the populations is obtained. Finally, the age-redshift relations are fit within a cosmographic approach to derive a model-independent sampling of the Hubble parameter up to $z \sim 0.4$.

\subsection{Stacking\label{sec:stacking}}

\begin{figure*}
	\includegraphics[width=\textwidth]{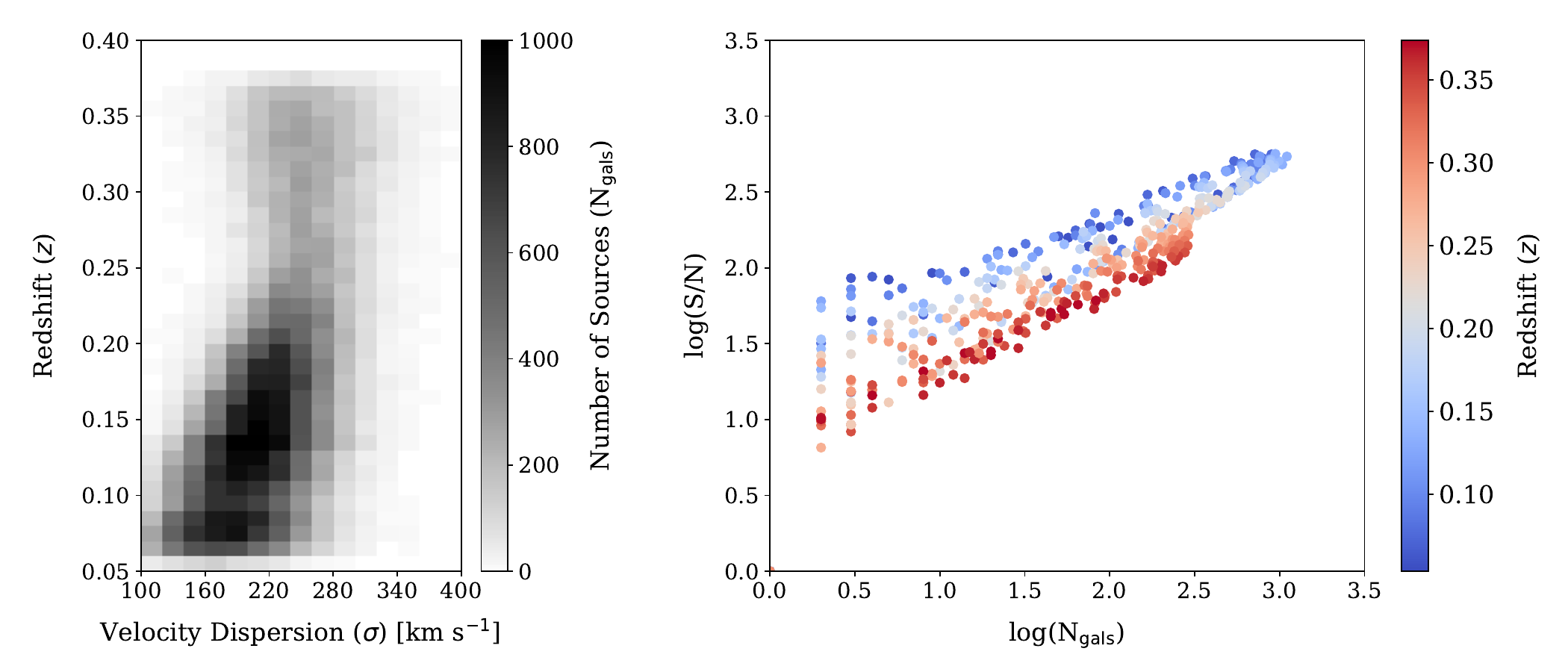}
    \caption{Properties of the ancillary stacks. Left panel: distribution of the number of galaxies included in the ancillary stacks defined by a linear constant spacing in both $\sigma$ and $z$. Right panel: relation between the S/N and the number of galaxies in each ancillary stack as a function of redshift.}
    \label{fig:SNNz}
\end{figure*}

Lick index measurements demand a high enough signal-to-noise ratio of the spectra, which is not guaranteed by individual galaxies and can compromise the reliability on the estimation of age, metallicity and $\alpha$-enhancement. Single galaxy spectra are thus shifted to the rest frame, normalised and then stacked. In this process, we followed \cite{borghi22a} to derive the composite spectra and their associated uncertainties. We chose a central stable window (rest frame) between $4200$ and $4400$ \AA \ and normalized all spectra as $\tilde{F}(\lambda) \equiv F(\lambda)/\langle F ([4200, 4400]) \rangle$, where $\langle\ldots\rangle$ denotes the median. Since the SDSS spectra are not sampled on the same wavelength array, they need to be resampled into a common grid and interpolated, for which we set a wavelength step of $0.1 $\AA. The flux of the stacked spectrum is computed through the median of the individual normalised spectra at every wavelength and the uncertainties are given by the normalised median absolute deviation as $\delta F_\text{stack}(\lambda) = 1.4826 \cdot \langle F_{i\in \text{stack}}(\lambda) - \langle F_{i \in \text{stack}}\rangle(\lambda)\rangle/ \sqrt{n}$, where $n$ is the number of objects in the stack and $F_i$ is each individual galaxy spectrum \citep{HOAGLIN83}.

To determine which sources we should stack together, we require that the final (stacked) spectra all have a similar signal-to-noise, a quantity that clearly depends on the number of sources included in each stack and their redshift. In other words, we wish to estimate the function $\text{N}(z, \langle \text{S}/\text{N} \rangle)$ that tells us the number of sources N needed at a given redshift z to obtain a median signal-to-noise ratio $\langle \text{S}/\text{N} \rangle$. For this reason, we had to build an auxiliary set of stacks through a two-dimensional grid in velocity dispersion and redshift. We set step sizes of $20 \text{ km}\,$s$^{-1}$ (from $100$ to $400 \text{ km}/\text{s}$) and $0.01$ (from $0.05$ to $0.40$) respectively. This is shown in the left panel of Fig. \ref{fig:SNNz}, where each pixel (auxiliary stack) is coloured according to the number of galaxies it contains. To sample $\text{N}(z, \langle \text{S}/\text{N} \rangle)$. For each auxiliary stack, we measured its signal-to-noise and kept track of the number of single galaxies included within. The result is plotted in the right panel of Fig. \ref{fig:SNNz}, where each point corresponds to a different pixel. As expected, stacks need more galaxies at a higher redshift to attain a given $\langle \text{S}/\text{N} \rangle$ and, at high redshift, it is uncommon to observe pixels with $\gtrsim 320$ sources.

Finally, we perform a linear fit of $\log(\langle \text{S}/\text{N} \rangle)$ as a function of $\log(\text{N})$ in bins of $z$ to sample N($z$) at a fixed $\langle \text{S}/\text{N} \rangle$, which we set to $300$ as the reference value. At this point we have a set of values of $\log(\text{N})$ that correspond to a set of values of $z$, with $\langle \text{S}/\text{N} \rangle$ fixed to $300$. The final N($z$) function is an interpolation of this set of data. The final stacks are then constructed taking into account the effect of velocity dispersion; for which we defined six groups logarithmically spaced from $10^2$ to $10^{2.6} \text{ km}\,$ s$^{-1}$. This choice roughly corresponds to setting the bin limits at $[100, 125, 160, 200, 250, 320, 400] \ \text{km}\,$s$^{-1}$. Then, the procedure to assemble the stacks in each group is the following:

\begin{enumerate}
\item We order galaxies in ascending redshift.
\item We "open" a new stack and include, one by one, galaxies in order of redshift (loop explained in steps 3-6).
\item After each inclusion, we compute the average redshift of the stack, $\langle z\rangle_\text{loc}$, and keep track of the number of galaxies included, $n_\text{loc}$.
\item We compute, using the N($z$) function we built from the auxiliary stacks, the number of sources that will be necessary to obtain $\langle \text{S}/\text{N} \rangle \sim 300$ at redshift $\langle z\rangle_\text{loc}$. This is, $\text{N}_\text{ref} \equiv  \text{N}(\langle z\rangle_\text{loc})$.
\item If $\text{N}_\text{ref}>n_\text{loc}$, we continue the loop with the following galaxy (step 3). If $\text{N}_\text{ref}<n_\text{loc}$ we "close" the stack (step 6).
\item When the stack is closed we can proceed to measure the spectral indices. Then, a new stack is "open" starting from the next galaxy, higher in redshift, to the last one included in the closed stack.

\end{enumerate}

The final stacks can be found in Figs. \ref{fig:cosmography_TMJ} and \ref{fig:cosmography_sMILES}, distributed in the plane $t-z$ according to the ages we estimated using the SPS models that we present in section \ref{sec:SPS}. Unless explicitly stated otherwise, any distribution of ages presented in the remainder of the article is computed on stacks built trying to match a median signal-to-noise of $300$.

From the final stacks, all those that include an internal distribution of sources that exceeds in width an average dispersion of $\Delta z = 0.01$ ($\sigma_z = 0.005$) are discarded, in order to guarantee that in each stack only the most similar individual galaxies are included. Eventually, it will be observed that some of those data points that cover vast redshift windows end up showing a behaviour typical of outliers (Figs. \ref{fig:cosmography_TMJ} and \ref{fig:cosmography_sMILES}). This final cut also makes it impossible to fit the highest velocity dispersion group ($320 < \sigma \ [\text{km s}^{-1}] < 400$) as we are left with no valid points. In any case, this group could be physically problematic as these velocity dispersions would be associated \citep{cappellari06, zahid16} to extremely massive galaxies of close to $10^{12} \ M_\odot$ in stellar mass.

\subsection{Lick index measurements\label{sec:Lick}}

The original set of Lick indices \citep{worthey94,trager98} includes 21 features identifying atomic and molecular absorptions typical of the light of SSPs of passively ageing galaxies, name-ordered in increasing wavelength from $\text{CN}_1$ to $\text{TiO}_2$. Then the Balmer $-\delta$ and $-\gamma$ indices were also included (\cite{worthey97}) given their sensitivity to stellar ages, and they were given two definitions, a wide band labelled with $_\text{A}$ and a narrower one labelled with $_\text{F}$. All indices, whether atomic or molecular, are defined by a central band ($\lambda_\text{cb}$, $\lambda_\text{cr}$) over which an integral is performed and two additional bands blueward ($\lambda_\text{bb}$, $\lambda_\text{br}$) and redward ($\lambda_{\text{rb}}$, $\lambda_\text{rr}$) of the central region. These are used to define the pseudocontinuum, estimated as a straight line joining the median flux over each of the two bands, that is,
\begin{equation}
    F_\text{p-c}(\lambda) = \frac{F_\text{r}\left(\lambda -\lambda_\text{b}\right) - F_\text{b}\left(\lambda - \lambda_\text{r}\right)}{\lambda_\text{r} - \lambda_\text{b}} \ \ \ ,
    \label{eq:methods_Fc}
\end{equation}
where $\lambda_\text{b}$, $\lambda_\text{r}$, $F_\text{b}$ and $F_\text{r}$ are the central wavelengths and the median fluxes of the pseudocontinuum regions. Then, indices are computed as
\begin{equation}
    I_\text{at.} = \int_{\lambda_\text{cb}}^{\lambda_\text{cr}} \left(1 - \frac{F(\lambda)}{F_\text{p-c}(\lambda)}\right) \text{d} \lambda , \ \ \
    \label{eq:methods_Iat}
\end{equation}
if they are atomic, or
\begin{equation}
    I_\text{mol.} = -2.5 \log\left(\frac{1}{\lambda_\text{cr} - \lambda_\text{cb}}\int_{\lambda_\text{cb}}^{\lambda_\text{cr}} \frac{F(\lambda)}{F_\text{p-c}(\lambda)}\text{d} \lambda\right) , \ \ \
    \label{eq:methods_Imol}
\end{equation}
if they are molecular. Atomic indices have units of wavelength and are defined in the same way as EWs, while molecular indices are dimensionless. With these definitions, any other atomic or molecular absorption (or emission) can also be defined. Indeed, we used the atomic definition for the computation of the emission in $\text{[OII]}\lambda 3727$, $\text{[OIII]}\lambda 5007$ and $\text{H}\alpha$ for the spectroscopic part of the selection of sources, while $\text{H}\beta$ is already included in the original group of Lick indices. All index measurements are performed using the {\fontfamily{qcr}\selectfont pyLick} code developed by \cite{borghi22a}. By default, the code also predict other features beyond the set of $25$ Lick indices; namely the two $\text{CaII}$ (both $\text{K}$ and $\text{H}$) the $\text{D}4000$ (\cite{hamilton85}) and the $\text{D}4000_\text{n}$ (\cite{balogh99}). The errors associated to the indices are computed through a Monte Carlo resampling of the spectra using the uncertainties of the fluxes.

\subsection{Spectral resolution and velocity dispersion\label{sec:velDisp}}

The index predictions from the models have an intrinsic resolution equal to that of the stellar library on which they are based. For this work, they are both built from the MILES library, which has an rms wavelength resolution of $\lambda_\text{IR}\approx1.06$\AA. For a fair comparison, index measurements should be at the same resolution as the models. However,  SDSS spectra have an effective resolution arising from the convolution of the intrinsic SDSS instrumental resolution, and then galaxy spectra suffer from the intrinsic effect of velocity dispersion of the stars in the galaxy. Therefore, in principle, the data and/or model spectra need to be corrected so that resolutions are matched. In particular, given a flux density $F_{\text{R}_0}$ at a resolution $\text{R}_0$, the downgraded spectrum at a resolution $\text{R}$ is given by the convolution
\begin{equation}
    F_\text{R}(\lambda) = \frac{1}{\sqrt{2\pi \sigma_\text{match}^2}} \int \text{d}\text{k} \ F_{\text{R}_0}(\text{k}) \ \exp{\left(-\frac{1}{2} \left(\frac{\lambda - \text{k}}{\sigma_\text{match}}\right)^2\right)} \ \ \ ,
    \label{eq:methods_convolution}
\end{equation}
where $\sigma_\text{match}$ is the Gaussian filter rms that needs to be applied in order to get the desired resolution. The value of this quantity can be obtained as
\begin{equation}
    \sigma_\text{R}^2 = \sigma_{\text{R}_0}^2 + \sigma_\text{match} ^2 = \sigma_\text{VD} ^2 + \sigma_\text{IR} ^2 + \sigma_\text{match} ^2
    \label{eq:methods_sigmasum}
\end{equation}
where $\sigma_\text{R}$ is the aimed final rms resolution, and we have identified the present resolution $\sigma_{\text{R}_0}$ of a (data) spectrum with the quadratic sum of both the instrumental resolution and velocity dispersion resolution before the degradation is performed. The latter is linear in wavelength and takes the form $\sigma_\text{VD} = \sigma_\text{VD}(\sigma, \lambda) = \lambda \cdot (\sigma/c)$, where $\sigma$ is the velocity dispersion. In particular, if we wanted to take a spectrum with a certain velocity dispersion $\sigma_0$ and instrumental resolution $\sigma_{\text{IR}_0}$ to a larger velocity dispersion $\sigma_f$ and instrumental resolution $\sigma_{\text{IR}_f}$, one should perform the convolution (\ref{eq:methods_convolution}) with a filter
\begin{equation}
    \begin{split}
    \sigma_\text{match} &= \sqrt{\sigma_{\text{IR}_f} ^2 + \sigma_\text{VD} ^2(\sigma_f,\lambda) - \sigma_{\text{IR}_0} ^2 - \sigma_\text{VD} ^2 (\sigma_0,\lambda) } \\ &=  \sqrt{\sigma_{\text{IR}_f} ^2 - \sigma_{\text{IR}_0} ^2 + \lambda ^2 \frac{\sigma_f ^2 - \sigma_0 ^2}{c^2}}   
    \end{split}
    \label{eq:methods_sigmaexample}
\end{equation}

\begin{figure}
	\includegraphics[width=\columnwidth]{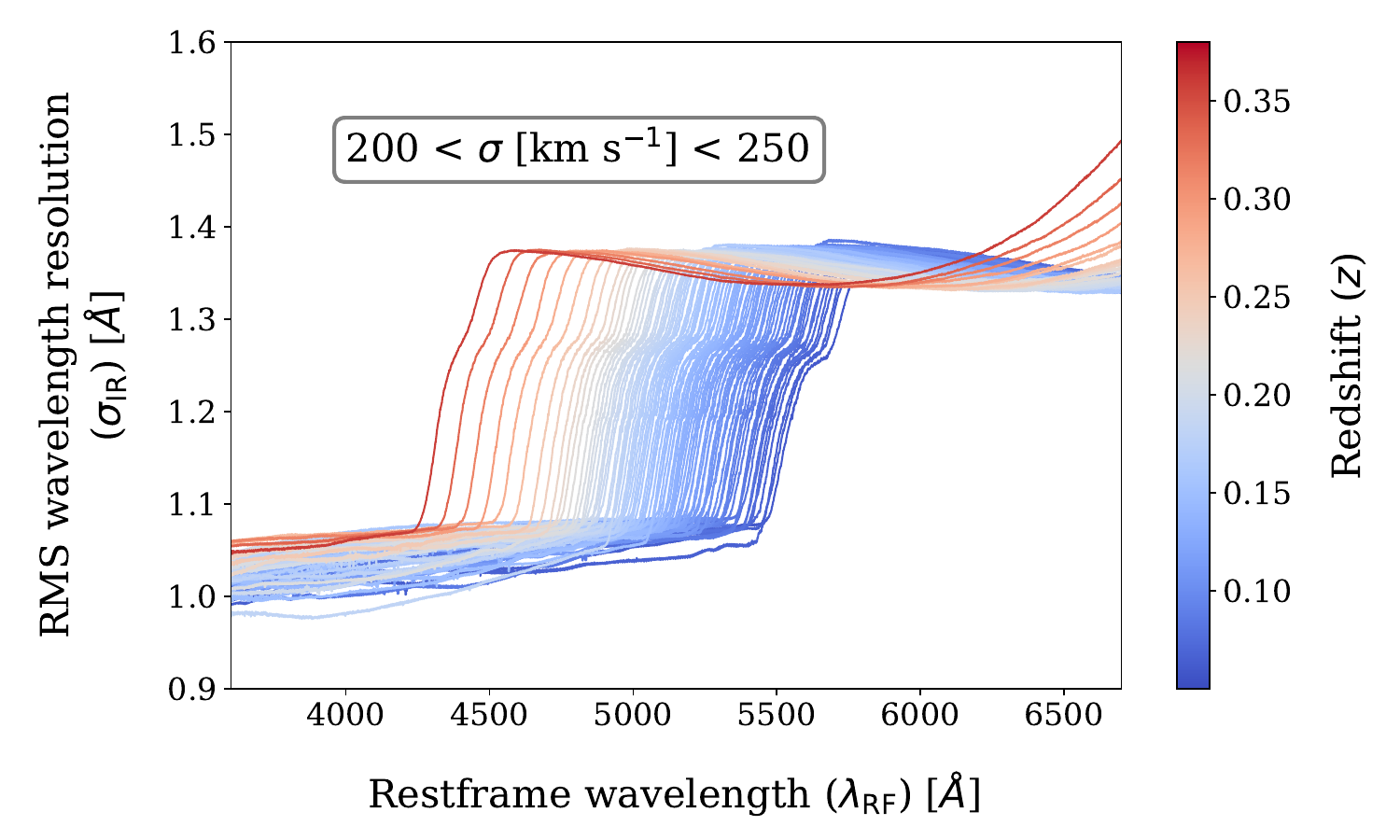}
    \caption{RMS wavelength resolution ($\sigma_\text{R}$) of SDSS stacked spectra as a function of rest-frame wavelength for different redshifts.}
    \label{fig:resolution_vd200250}
\end{figure}

Fig. \ref{fig:resolution_vd200250} shows the instrumental resolution of SDSS for sources at different redshifts (for the particular case of $200<\sigma \ [\text{km}/\text{s}] < 250$), which is a non-trivial function of wavelength, characterised by a sharp step-function-like jump at around $5500$ \AA \ in observed wavelength. In particular, the rms instrumental resolution of the data increases from $\sim 1.0$ \AA \ on the blue side to $\gtrsim 1.35$ \AA \ on the red side. It should be noted that, at very low redshifts, this implies that the instrumental resolution of the data is similar as that of the MILES stellar library within the range of interest for Lick indices. While this would mean that no instrumental resolution correction would be needed for these sources, the situation changes with redshift and the effect becomes apparent, as the reddest indices progressively enter the lower resolution regime. Although the effect of this correction is only minor, it is appropriate to take it into account. Therefore, all science (data) spectra will be degraded to a common instrumental resolution of $1.5$ \AA; that is, with $\sigma_\text{match}=(1.5^2 - \sigma^2_{\text{IR}_0}(\lambda))^{1/2}$.

Nevertheless, the procedure to then fit Lick indices with the SPS models depends on the model used. When fitting with the Knowles model we will leverage the provided model spectra by degrading them to match the velocity dispersion and instrumental resolution of the object we fit in each case. In other words, we will create an index-strength model in which velocity dispersion is just a configuration parameter. The transit from the (appropriately smoothed) spectra to Lick index predictions is always performed with the pyLick code \citep{borghi22a}. On the other hand, when using the TMJ model, an intermediate correction of the data needs to be performed. This model is built in as a direct prediction of index strengths at MILES ($1.06$\AA) resolution, for which we need to derive a correction (covering both a degraded instrumental resolution of $1.5$\AA and velocity dispersion) that is applied to the indices once they have been measured. The goal is to find a function $C_I(\sigma)$ that takes an index $I$ measured from a spectrum of a galaxy with velocity dispersion $\sigma$ and intrinsic instrumental resolution $\sigma_\text{IR}$ to the value $I_0$ it would be expected to have if measured at a MILES instrumental resolution and without the effect of velocity dispersion, mirroring the output of the TMJ model. Depending on whether the index is atomic or molecular the correction is written as $I_0 = I(\sigma) \cdot C_I(\sigma)$ or $I_0 = I(\sigma) + C_I(\sigma)$, respectively. In other words,
\begin{equation}
    C_I(\sigma, \sigma_\text{IR}) = \left\{\begin{matrix} 
    I(0, \sigma_\text{MILES})/I(\sigma, \sigma_\text{IR}), & \text{if atomic,}\\
    I(0, \sigma_\text{MILES})-I(\sigma, \sigma_\text{IR}), & \text{if molecular,}\\
    \end{matrix} \right.
    \label{eq:correction}
\end{equation}
where $\sigma_\text{IR} = 1.5$\AA. In order to compute the correction, several methods are present in the literature. Following \cite{carson10}, we take the SDSS spectra\footnote{SDSS velocity dispersion templates: \url{https://classic.sdss.org/dr7/algorithms/veldisp.php}} of giant stars in M67 (NGC2682) and perform smoothings of progressively higher velocity dispersions. Then, the change in the indices is averaged over all stars, and the sampling of $C_I(\sigma)$ is parametrised by a polynomial of order 3 for all indices. It should be noted that $I_0$ can be measured directly from the stars even though they have the SDSS resolution imprinted in their spectra, since, similarly to what happens with low redshift galaxy spectra, the entire set of Lick indices falls below the jump in resolution.

We highlight that the corrections $C(\sigma)$ carry an associated uncertainty $\delta C$ that we calculated in their derivation and take into account. These uncertainties are small compared to the statistical uncertainty of the data (index measurements) and are treated altogether as statistical uncertainty, despite the systematic nature of the former. For further information, we include Appendix \ref{sec:app_VD}, where we show the correction functions $C(\sigma)$ we used and give the parameters of the 3rd order polynomials. We also show for reference the difference in stellar age estimated for the set of data due to the use of the two methodologies presented to address the velocity dispersion effect. We remind that only the Knowles-modelled ages can be used for this particular study, as taking the TMJ model to different velocity dispersions is not possible.

\subsection{Stellar population models\label{sec:SPS}}

The measured indices are fit via an MCMC algorithm to both the TMJ and Knowles models. The underlying assumption is that the spectra can be approximated as a coeval stellar population born in a single burst, which is the rationale behind the use of cosmic chronometers. While recent studies fitting Lick indices have typically employed direct Lick index models like TMJ, we have decided to also test the sMILES-based SSP model to compare the results. As commented in the introduction, we advise that the choice of SPS model can alter significantly the results \citep{moresco20} in matter of both age and $H$ estimations. For what concerns the specificities of the models included, TMJ use the evolutionary synthesis code of \cite{maraston05}, the \citep{salpeter55} IMF, the MILES empirical library and \citep{cassisi97} stellar tracks. The Knowles model includes a larger semi-empirical library based on MILES, known as sMILES \citep{knowles21}. This library is built to correctly reproduce the MILES stellar spectra at the typical metallicities and $\alpha$-abundances of the stars in our neighbourhood, but using theoretical tools to go beyond those restrictions in the parameter space. They use the BaSTI isochrones \citep{pietrinferni04, pietrinferni06, pietrinferni21} and the model is available for five different IMFs, from which we have chosen the Chabrier IMF \citep{chabrier03}. These models can be interpreted as functions like

\begin{equation}
    I_i \equiv I_i(t, [\text{Z}/\text{H}], [\alpha / \text{Fe}]),
    \label{eq:model}
\end{equation}
where the subindex $_i$ denotes a given Lick index, which are a function of age ($t$), metallicity ($[\text{Z}/\text{H}]$) and abundance of $\alpha$ elements ($[\alpha/\text{Fe}]$). Each "base" model is sampled on a specific grid of parameters, which yields a discreet function that needs to be interpolated for the fit. Once this is done, the age, metallicity and $\alpha$-enhancement are found via an MCMC algorithm. In particular, we used \emph{emcee} \citep{FOREMAN13}, the Python implementation of the the Goodman \& Weare affine invariant MCMC ensemble sampler \citep{GOODMAN10}. The log-likelihood is written as
\begin{equation}
    \log{\mathcal{L}} = -\sum_i \log{\sqrt{2\pi \sigma_{I_i} ^2}} -\frac{1}{2}\sum_i \left(\frac{\hat{I}_i - I_i}{\sigma_{\hat{I}_i}}\right)^2
    \ \ \ ,
    \label{eq:loglike_Lick}
\end{equation}
where $\sigma_{\hat{I}_i}$ are the uncertainties associated to the Lick index strength measurements, $\hat{I}_i$, and $I_i$ are the model indices from (\ref{eq:model}). Once a particular model for the fit is selected, one should pay attention to two key aspects: the set of indices chosen for the fit, and the grid onto which the base model $I_i(t, [\text{Z}/\text{H}], [\alpha / \text{Fe}])$ is interpolated. Regarding the former, the recommendations provided by the authors should be followed not to bias the results. For the TMJ model, following their study using globular clusters without individual element variations, the use of certain indices is discouraged since they are not well reproduced by the data. In essence, our set comprises all of the indices of the baseline used by \cite{johansson12}, that is H$\delta_\text{A}$, H$\delta_\text{F}$, Fe$4383$, Fe$4531$, H$\beta$, Mgb, Fe$5270$, Fe$5335$, Fe$5406$; plus $\text{G}4300$, H$\gamma_\text{A}$, H$\gamma_\text{F}$ and the Mg$_2$, since they are well reproduced by globular cluster data in \cite{thomas11}. In turn, for the Knowles model, we discarded indices known to be poorly modelled, too broad or sensitive to the IMF, following \cite{knowles23}. The set of Lick indices is thus H$\delta_\text{A}$, H$\delta_\text{F}$, G$4300$, H$\gamma_\text{A}$, H$\gamma_\text{F}$, Fe$4383$, Ca$4455$, Fe$4531$, H$\beta$, Mgb, Fe$5270$, Fe$5335$, Fe$5406$, Fe$5709$,  Fe$5782$. We emphasise at this point that using a fixed set of indices across the entire dataset is strongly recommended, as \cite{borghi22b} found age estimates could vary by up to $\sim 2$ Gyr depending on the indices used.

In general, Balmer indices are sensitive to age, iron indices to metallicity and magnesium indices to $\alpha$-enhancement. \cite{knowles23} showed that the plane $H\beta_0$-[MgFe] could separate, almost orthogonally, ages and metallicities, where H$\beta_0$ is an optimised definition of H$\beta$. We saw, however, that for our selection of very passively evolving ETGs, the Balmer-$\beta$ feature alone produced extremely old ages, as \cite{vazdekis15} had already shown. Instead, the combination of the three Balmer lines ($\beta$, $\gamma$, $\delta$) returns much more reasonable ages. When it comes to metallicity and $\alpha$-enhancement, \cite{knowles23} showed that the plane Mg$_\text{b}$-Fe$5335$ or -Fe$5270$ separated well between different alpha enhancement groups. The three of them are included and recommended for their use in stellar population fits by themselves and \cite{thomas11}. Moreover, a combination of these three indices makes the [MgFe] index, which is sensitive to metallicity but insensitive to $\alpha$-enhancement.

Regarding the grid of the models, the trade-off between the computational cost and the precision when using an overly fine interpolating grid should be addressed. The base grid for the TMJ model features ages distributed in the interval $[0.1, \ 15.0]$ Gyr, linearly spaced with steps of $0.1$ and $0.2$ Gyr between $0.1$ and $1.0$ Gyr and with steps of $1.0$ Gyr until $15.0$ Gyr. Metallicity $([\text{Z}/\text{H}])$ is sampled in $6$ values, namely $\{-2.25, -1.35, -0.33, 0.00, 0.35, 0.67\}$. Finally, $\alpha$-enhancement $([\alpha/\text{Fe}])$ is sampled in four values: $\{-0.3, 0.0, 0.3, 0.5\}$. Regarding the Knowles model, the base grid features ages distributed in the interval $[0.03, 14.00]$ Gyr, with a varying step size that increases with age, namely 0.01, $0.05$, $0.1$, $0.25$ and $0.5$ Gyr. Metallicity is sampled over $10$ values not evenly spaced from $-1.79$ to $0.26$, meaning it does not reach strong metallicities beyond $0.26$; in contrast to TMJ, which goes up to $0.67$. $\alpha$-enhancement is equally the least thoroughly sampled parameter, with $5$ values: $\{-0.2, 0.0, 0.2, 0.4, 0.6\}$.

\begin{figure}
	\includegraphics[width=\columnwidth]{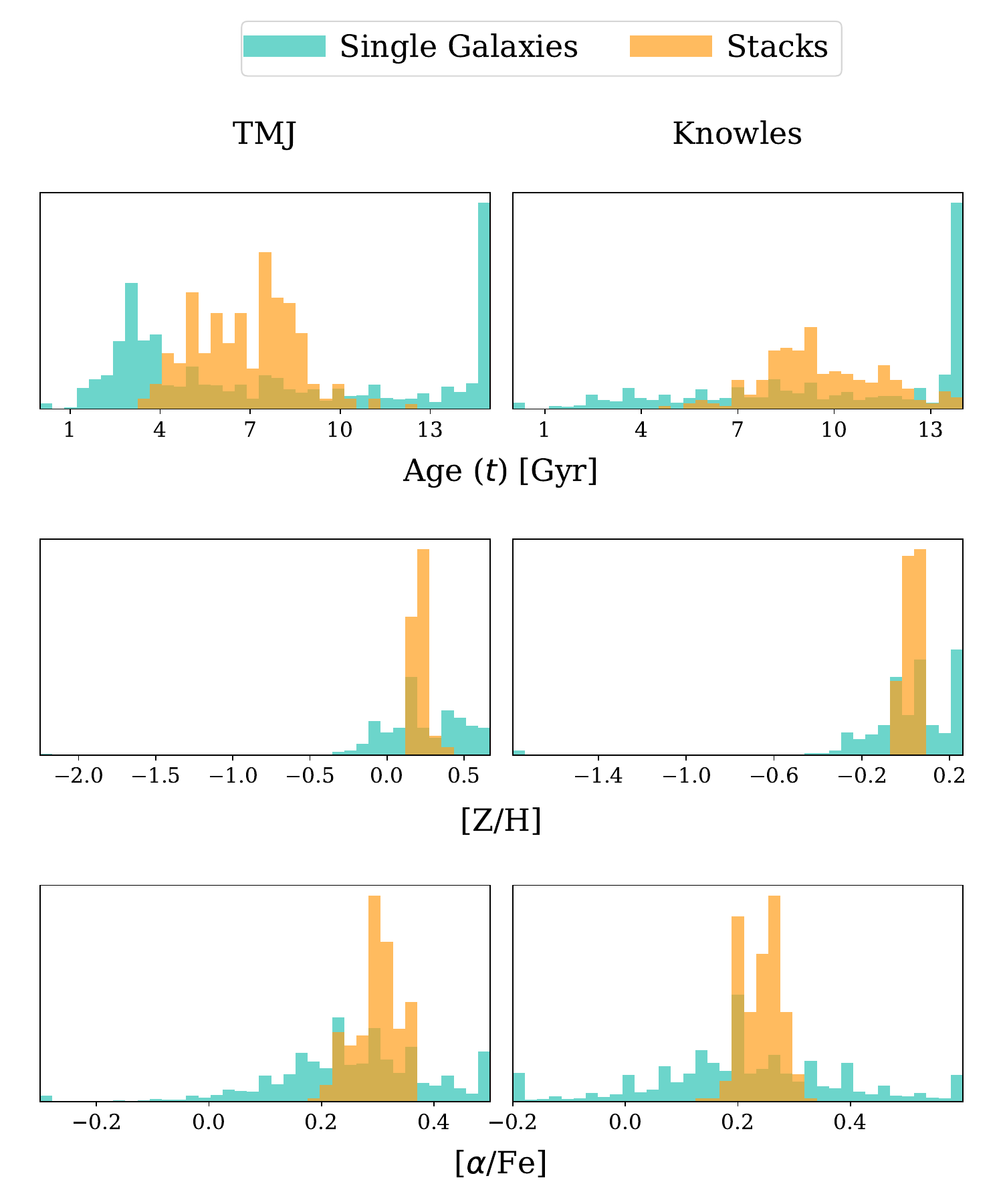}
    \caption{Age, metallicity and $\alpha$-enhancement distributions for the TMJ (left panels) and Knowles (right panels) models, interpolated to a grid of 50 points in each parameter within the original ranges. Results are shown for the Lick index fit to both single galaxies (in blue) and stacks (in yellow) of the final cosmic chronometer sample.}
    \label{fig:TMJKNW}
\end{figure}

Some of these values are typically not attained for massive passive galaxies \citep{parikh21,saracco23}. For example, metallicity is expected to range from slightly subsolar to supersolar metallicities, but not beyond $[\text{Z}/\text{H}] \sim 0.3$ \citep{clemens06,clemens09,thomas10,maiolino19}. Moreover, we have found that allowing for metallicities that go beyond the limit given by the Knowles model could bias a portion of the sample into smaller ages due to the age-metallicity relation and the poor sensitivity of the models at high stellar ages, which is not resolved by employing a thinner grid. For this reason, in addition to the computational cost, we decided to restrain the parameter limits of both models to $t$ $\in [0.1,\ 14.0]$ Gyr, $[\text{Z}/\text{H}] \in [-0.35, \ 0.26]$ and $[\alpha/\text{Fe}] \in [0.0, \ 0.4]$ and interpolate into a finer grid with step sizes of $\Delta t=0.1$ Gyr, $\Delta [\text{Z}/\text{H}]=0.01$ and $\Delta [\alpha/\text{Fe}]=0.01$. This choice of parameter limits is also supported by Fig. \ref{fig:TMJKNW}, where a third finer version of the base models (with 50 equally spaced points for each parameter in the original extent of the grids) was run on the galaxy stacks (shown in orange) as a preliminary test. As expected for massive ETGs, metallicities are either solar or supersolar and there is a significant abundance of $\alpha$-elements with respect to iron. For comparison, in blue, we show the result for the fit to the spectra of single galaxies. The distributions in this case are much more scattered and do no reflect the expected behaviour for massive passively evolving galaxies. Indeed, subsolar metallicities and iron enhancement with respect to $\alpha$-elements are found for a significant fraction of galaxies, especially when using the Knowles model. Additionally, extremely high values for the ages are found in a large number of galaxies, with an additional seemingly young population of $\sim 3$ Gyr obtained through the TMJ model. This underscores the importance of using high $\text{S}/\text{N}$ spectra in these studies, as results behave as expected when the fits are performed with the stacked data.

\subsection{Cosmographic fit\label{sec:COSMOmodel}}

The last step is to fit the stellar age-redshift relation in different velocity dispersion bins to derive the cosmographic parameters ($H_0$, $q_0$, $j_0$). This is performed using (\ref{eq:tdezdeHz}) and (\ref{eq:Hcosmographic_ord2}), which allow us to sample the Hubble parameter up to $z\lesssim 0.35$ without assuming a cosmological model. As anticipated at the end of Section \ref{sec:stacking}, we divided the sample into six logarithmic dispersion bins; namely [100,125), [125, 160), [160, 200), [200, 250), [250,320) and [320, 400) km$/\,$s$^{-1}$. In principle, this choice implies that six sets of age-redshift data should be fit jointly. Since every set characterises a different cosmic chronometer population defined by its mass, six distinct $t_0$ parameters need to be defined (i.e. $t_{0,100-125}$, $t_{0,125-160}$, $t_{0,160-200}$, $t_{0,200-250}$, $t_{0,250-320}$, $t_{0,320-400}$), which stand for the average stellar age of each ensemble at $z=0$. With all this, the log-likelihood reads
\begin{equation}
    \log{\mathcal{L}} = -\sum_{i, v} \frac{1}{2}\left(\frac{\hat{t}_{i, v} - t(z_{i, v}; t_{0, v}, H_0, q_0, j_0)}{\sigma_{\text{t}_{i, v}}}\right)^2
    \label{eq:loglike_Cosmo}
\end{equation}
where $i$ runs through the stacks in each velocity dispersion group, represented by $v$. For all parameters, we assign uniform prior probability distributions. The limits for these distributions were set between $0$ and $20$ Gyr for the zero-ages $(t_{0,v})$, $20$ and $120$ $\text{km s}^{-1} \text{ Mpc}^{-1}$ for the Hubble constant, $-10$ and $10$ for the deceleration parameter and $-100$ and $100$ for the jerk parameter. The choices for the cosmographic parameters were made in order to guarantee that the posterior distributions were constrained in the main cases (TMJ, Knowles), even if this allows the tails of the $q_0$ and $j_0$ distributions to attain values far from the conventional ones found in the literature. As will be seen, it is safer to do this due to the low sensitivity of the model on these kinematical parameters, particularly the jerk.

Additionally, we require the derivative of the Hubble parameter, $\text{d}H/\text{d}z$, to be positive, as suggested by the global trend observed from model-independent measurements of $H(z)$ \citep{simon05, stern10, zhang14, ratsimbazafy17, jiao22, tomasetti23, moresco12, moresco16a, borghi22a}. Physically, this translates into allowing the expansion of the Universe to accelerate but not too much, with a limit on $q(z)>-1$ at any redshift. In particular, the deceleration parameter today, $q_0$, has to follow this condition. This could be interpreted, under the introduction of a $w_0$CDM model, as the condition of reducing the amount of phantom dark energy states available. Specifically, the mathematical condition is $w_0 > 1/(\Omega_m -1)$, which allows for $w_0 < -1$ only when $\Omega_m \gg 0$. In particular, $w_0$ would present a lower limit in $\approx -1.43$ for $\Omega_m = 0.3$. Under the assumption of the $w_0$CDM model, we can combine our estimations for $q_0$ and $j_0$ \citep{capozziello12} to get the estimations of $\Omega_m$ and $w_0$ in those models, following
\begin{equation}
    \Omega_m (q_0, j_0) = \frac{2(j_0 - q_0 - 2q_0^2)}{1 + 2j_0 - 6q_0}
    \label{eq:Om_q0j0}
\end{equation}
and
\begin{equation}
    w_0 (q_0, j_0) = \frac{1+2j_0-6q_0}{-3+6q_0}
    \label{eq:w0_q0j0}
\end{equation}

Note that even if we apply these transformations, we can obtain values for $\Omega_m$ that exit the range $[0, 1]$. With these transformations, we can have a hint of the blind preference for values of $\Omega_m$ and $w_0$ aligned with the results from model-dependent probes.

\begin{figure*}
    \centering
    \includegraphics[width=\textwidth]{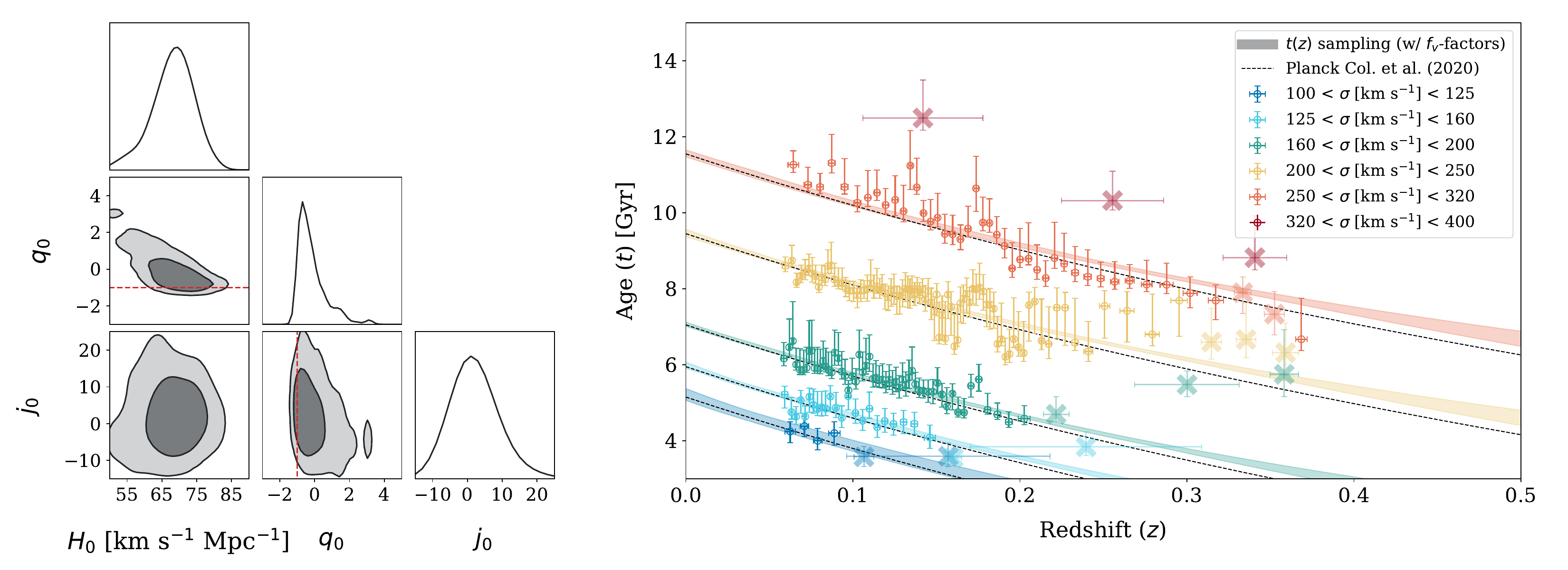}
    \caption{Joint cosmographic fit of $t(z)$ with ages estimated using the TMJ model. On the left, the corner-plot for the posterior probability distributions of $H_0$, $q_0$ and $j_0$ is presented, showing the $1\sigma$ and $2\sigma$ contour levels. On the right, the $t-z$ dispersion is shown. The empty circles in strong colour represent the data points we used for the fit, while the faded crosses represent the points that were the result of a set of galaxies with average dispersion in redshift greater than $0.01$. The shaded regions in the same colour as the data points are the $1\sigma$ of the sampling of the posterior distribution coming from the parameters on the left. The zero-ages of the velocity dispersion groups are not shown here, but we refer to the corner plot in Fig. \ref{fig:fullcosmoTMJ}; however, they are used for the sampling as well as to set a zero-age for the \protect\cite{planck18} evolution of $t(z)$, shown as black dashed lines. The dashed $q_0 = -1$ lines in the corner plots for the planes $q_0 - H_0$ and $j_0 - q_0$ represent the natural limit imposed by $dH/dz > 0$. This is the baseline $t-z$ relation in this work.}
    \label{fig:cosmography_TMJ}
\end{figure*}

\section{Results\label{sec:RESULTS}}

The present section covers the cosmographic analysis of the data, after the fit with both SPS models. Using TMJ-modelled ages we find a credible interval for $H_0$ aligned with other probes in the literature. Further discussion is provided for the other parameters included in the fit $(q_0, j_0)$, as well as for the cosmographic readout of the Knowles-modelled ages. Through the introduction of a $w_0$CDM model we can get some insight into the implications of our estimation for the parameter of the equation of state of dark energy, $w_0$. Finally, we examine the posterior sampling of our cosmographic fit in the Hubble diagram and compare our results with the point observations in the literature.

The right panels of Figs. \ref{fig:cosmography_TMJ} and \ref{fig:cosmography_sMILES} show the ages derived using the TMJ and Knowles models respectively, where each data point corresponds to a different stack. We found that more massive galaxies are older at every redshift and, at fixed velocity dispersion, the age decreases with redshift. This is in agreement with the cosmic chronometer ground principle: for a given velocity dispersion, we are statistically observing the same population of passive galaxies across cosmic time and thus the age evolution of the same galaxy population can be traced. The stratification in velocity dispersion groups is clear, and follows the downsizing scenario framework for the formation of ETGs. This aspect will be treated in detail in Section \ref{sec:stellararch}.

Before proceeding to the cosmographic fit of the ages, some points need to be addressed. First, the uncertainties in the ages derived from the Knowles model are systematically much smaller by a factor of $\sim6-7$. To investigate the potential reasons behind this, we examinated the differences in procedure when deriving ages with both models. The first difference is the set of Lick indices to be fit, as explained in Section \ref{sec:SPS}. However, after fitting a common set of indices for both models (defined by their overlap), no relevant changes are observed in ages or their uncertainties. The second difference lies on the treatment of the velocity dispersion effect. While for the TMJ model we have no other option than applying a correction -which introduces an extra systematic uncertainty associated to the computation of the correction-, for the Knowles model we can directly fit the data with the model indices, in which velocity dispersion is applied a priori to the synthetic spectra. However, using the TMJ methodology with the Knowles model is possible, allowing us to rule out the treatment of the velocity dispersion as the reason behind the disagreement on the value of uncertainties. In particular, we tried fitting the Knowles model at $0$ velocity dispersion with the indices (data) being corrected using the velocity dispersion corrections, analogously to the exact procedure of the TMJ model. The result were negligible changes in the ages, and the uncertainties remained the same size (further details are given in Appendix \ref{sec:app_VD}). Therefore, the only explanation is the nature of the models themselves, that is, the Knowles model seems to be much more precise. However, this will not necessarily play in our favour, due to the intrinsic scatter in the data.

The second point to address is the redshift dispersion within the stacks. As depicted by the horizontal error bars in the right panels of the aforementioned figures \ref{fig:cosmography_TMJ} and \ref{fig:cosmography_sMILES}, there is a (small) number of stacks (plotted with bold faded crosses) containing galaxies whose redshift covers a relatively wide range. Due to the stacking procedure, this typically occurs at the highest redshifts for a given velocity dispersion bin or for stacks with $320 < \sigma \ [ \text{km s}^{-1}] < 400$. Since these results are less reliable, we discard them from the start.

\begin{figure}
	\includegraphics[width=\columnwidth]{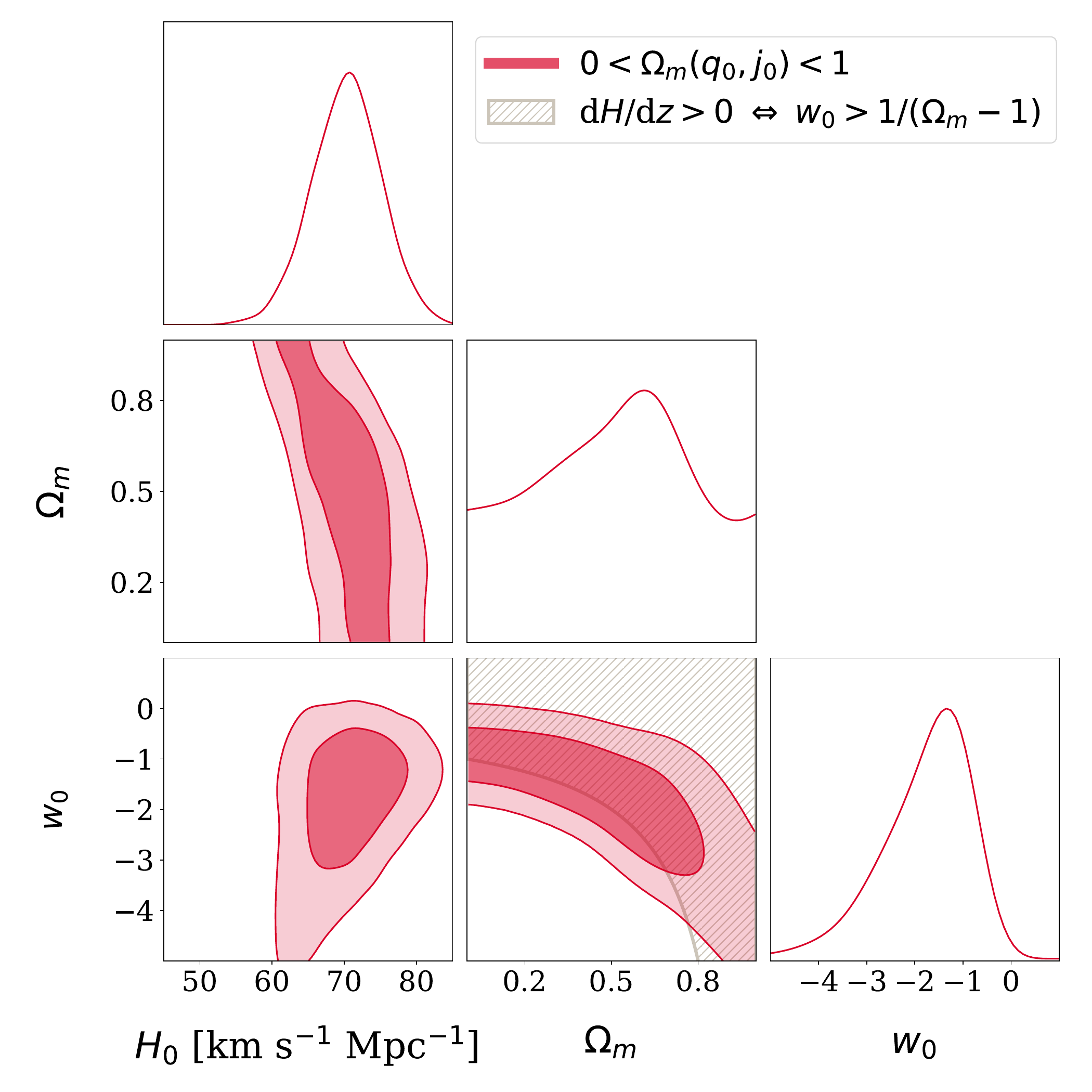}
    \caption{Posterior probability distribution for $H_0$, $\Omega_m$ and $w_0$. The sampled parameters were $q_0$ and $j_0$, then converted to $w_0$CDM parameters through equations (\ref{eq:Om_q0j0}) and (\ref{eq:w0_q0j0}). The sampling has been run under the $\Lambda$CDM-model condition of $0< \Omega_m(q_0, j_0) < 1$. The brown hatched region in the $\Omega_m-w_0$ plane represents the region allowed by $\text{d}H/\text{d}z > 0 \Leftrightarrow w_0 > 1/(\Omega_m -1) $.}
    \label{fig:corner_w0CDM}
\end{figure}

The last point to cover is related to a fluctuating pattern found in the ages with respect to redshift. Indeed, as seen more clearly for the TMJ model (Fig. \ref{fig:cosmography_TMJ}), the ages are found to oscillate at various redshifts, deviating from the expected smooth decline with redshift. This issue is particularly enhanced at $0.16 \lesssim z \lesssim 0.19$. This one particular "oscillation" is observed across all three velocity dispersion bins for which there are data at these redshifts. While we do not expect a large impact on the cosmographic results from the fit due to its symmetry with respect to the global trend, we found this issue of great relevance, and therefore decided to discuss it in depth in Section \ref{sec:indicestrends}.

\begin{figure*}
    \centering
    \includegraphics[width=\textwidth]{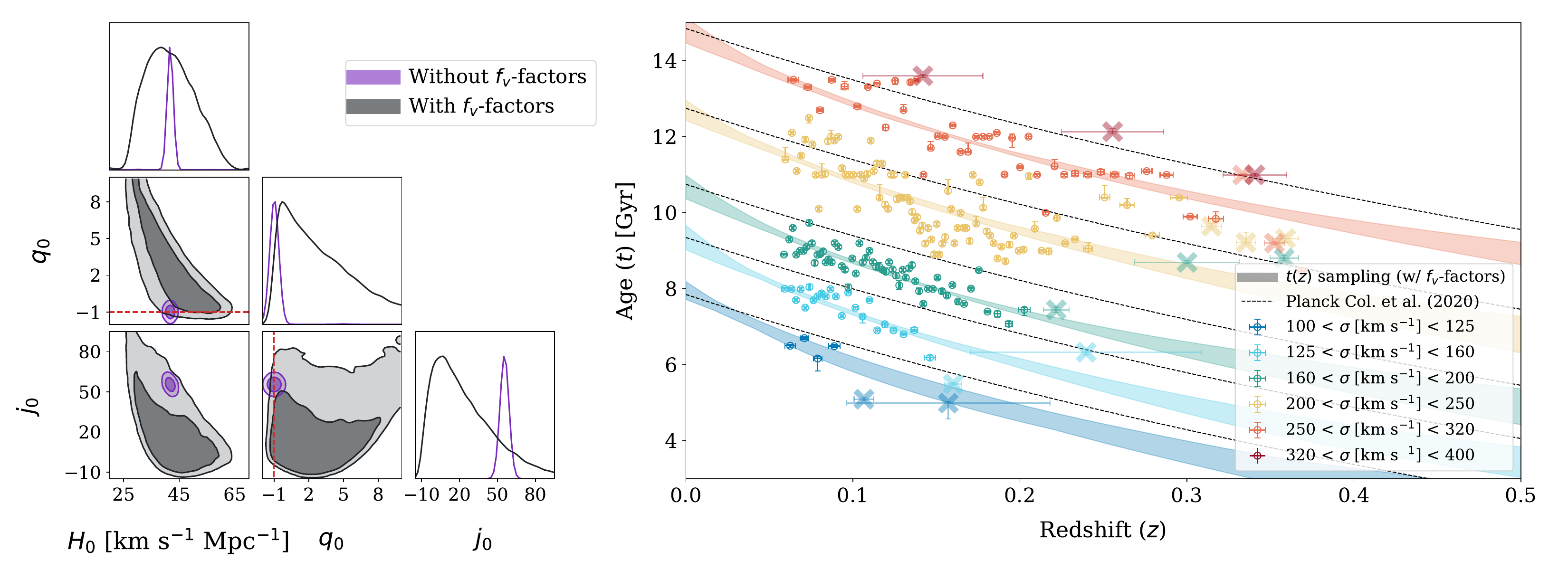}
    \caption{$t-z$ relations for the ages estimated using the Knowles model. The structure of this figure is the same as Fig. \ref{fig:cosmography_TMJ}. The grey contours represent the case with $f_{v}$ factors included, to account for the possible underestimation of age uncertainties. The $t-z$ samplings on the right panel refer to this case. The contours in purple represent the case where these factors have not been taken into account. The sampling of the associated posterior for $t-z$ is not represented on the right panel for this case.}
    \label{fig:cosmography_sMILES}
\end{figure*}

Once the distribution of ages has been individually assessed, we can turn to the cosmographic fit. In the right panel of Fig. \ref{fig:cosmography_TMJ}, apart from the $t(z)$ scatter of the data, we also show the $1\sigma$ sampling of the posterior distribution for $t(z)$, as obtained from the MCMC algorithm with the standard priors ($0<t_{0,i}<20 \text{ Gyr}$, $20 < H_0 < 120 \text{ km s}^{-1}\text{ Mpc}^{-1}$, $-10 < q_0 <10$, $-100 < j_0 <100$; as previously discussed in section \ref{sec:COSMOmodel}). For a comparative purpose, we also included the $\Lambda$CDM derived $t(z)$ following the results from \cite{planck18} as dashed black lines, taking as $t_{0,v}$ the peak of each posterior distribution on these parameters. Notice that the exclusion of widely $z$-extended data points eliminates all three belonging to the $320 < \sigma \ [\text{ km s}^{-1}] <400$ group, making it impossible for us to estimate $t_{0, 320-400}$.

\begin{figure*}
	\includegraphics[width=\textwidth]{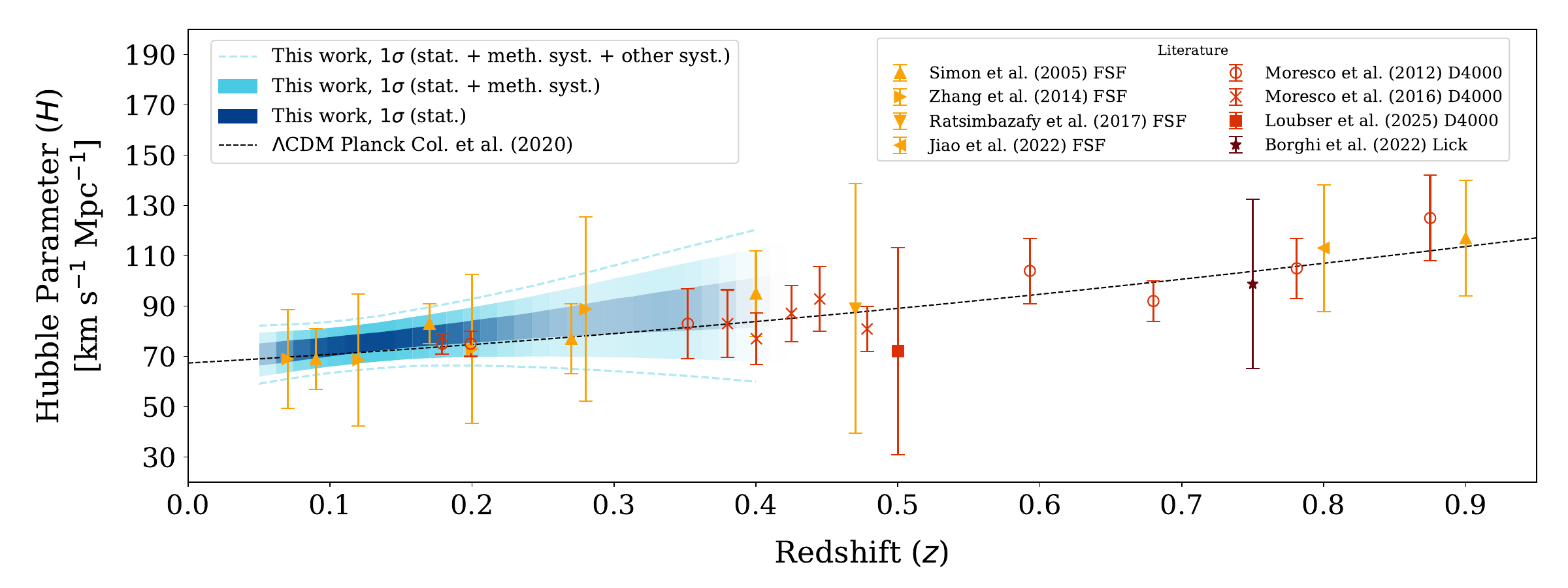}
    \caption{Reconstruction of the Hubble parameter. Our sampling is shown in shades of blue, corresponding to the posterior distribution obtained for $\text{H}_0$, $q_0$ and $j_0$ from Fig. \ref{fig:cosmography_TMJ}. The width of the tendency represents the range between the $16$th and $84$th percentiles of our the posterior sampling. The opacity of the colour shade represents the reliability of the $\text{H}(\text{z})$ measurement, which we based on the amount of sources (individual galaxies) at each $\text{z}$. The dashed black line represents the $\text{H}(\text{z})$ evolution according to \protect\cite{planck18}. The plot is extended up to $\text{z} = 0.9$ in order to show the result at lowest redshift obtained through Lick indices fit (dark red), by \protect\cite{borghi22b}. Other literature measurements are represented in yellow (obtained through Full Spectral Fitting) and orange (obtained through the measurement of D$4000$ of its narrow counterpart D$4000_\text{n}$).
    }
    \label{fig:Hfinal}
\end{figure*}

The corner plot corresponding to this sampling is shown in the left panel of the figure. The zero-redshift ages $(t_{0, v})$ of the individual cosmic chronometer samples are not plotted for the sake of a clear view of the other posteriors, but we anticipate they are extremely well constrained -at the exception of, obviously, $t_{0, 320-400}$-, as seen in Appendix C. We also note here that the estimations that we give throughout this discussion for the fit parameters include only statistical uncertainties. Systematic effects in both our methodology and from the SPS models will be addressed later (discussion around Fig. \ref{fig:Hfinal}).

The Hubble constant, $H_0$, is constrained to a mean value of $70.0^{+4.1}_{-7.6} \text{ km s}^{-1}\text{ Mpc}^{-1}$, which includes both the \cite{planck18} and \cite{RIESS22} estimations within the credible interval. The estimation of $q_0$ is relevant, clearly rejecting values below $-1$, having a seemingly strong cut-off at that value. For a direct view we have drawn this limit in the corner plots presenting the $q_0-H_0$ and $j_0-q_0$ planes. As we will see, this is due to the limitation on $q(z)$ (and thereby on $q_0$) imposed by $\text{d}H/\text{d}z > 0$. To inspect this in further detail, we examine the $16$th and $84$th percentiles and find that the modal value (maximum a posterior probability), $\text{mod}(q_0) = -0.95$, is at the left of the percentile $16$, $P_{16}(q_0) = -0.85$, while the percentile $84$ lies at $P_{84}(q_0) = 0.62$. It is clear that the posterior is skewed towards $q_0 = -1$ from its right. Indeed, when we inspect the MCMC chain thoroughly, we find that $99.91\%$ of the chain chose values greater or equal to $-1$, with only $0.09\%$ of samples remaining below, possibly due to the remainder in the approximation of $H(z)$. As previously said in Section \ref{sec:COSMOmodel}, we interpret this as a proof that the second order Taylor expansion of $H(z)$ is a good approximation in the redshift window we treat. Knowing this, we do not expect the region $q_0 < -1$ to be explored in any case by the MCMC chain, but we kept the lower limit of the prior in $-10$ as a consistency check. In order to assess how much this restriction on the derivative of $H$ means at a statistical level, we explored an extended parameter ($0< H_0 \ [\text{km s}^{-1} \text{ Mpc}^{-1}] < 300$, $-20 < q_0 < 20$, $-100 < j_0 < 300$) space after lifting the condition. We found out that marginals were much wider, in some cases beyond the limits set as standard, and a clear anti-correlation between $H_0$ and $q_0$ that effectively extends at $q_0 < -1$ (posterior shown in Appendix \ref{sec:corners}). The minimum of the $\chi^2$ in this case occurs at $H_0 \sim 220 \text{ km s}^{-1} \text{ Mpc}^{-1}$, but the typical values in this and all the regions explored clearly outside $\Lambda$ CDM ($H_0 \sim 70$ km s$^{-1}$ Mpc$^{-1}$, $q_0 \sim -0.5$, $j_0 \sim 1$) are similar to those found in the well-delimited posteriors in Fig. \ref{fig:cosmography_TMJ} (the baseline).

Finally, we assert that the cosmographic fit is highly insensitive to the jerk parameter, $j_0$, as the window needed in the prior to obtain a constrained posterior covers a range equivalent to more than $10$ times the value required by $\Lambda$CDM, $j_0 = 1$. Our estimation for the parameter is $j_0 = 0.9^{+5.5}_{-3.7}$, and the related corner plots show no degeneration of the jerk with either of the other two cosmographic parameters. However, we found, in the unrestricted fit without the limitation on the derivative of $H$, a seemingly functional relation between the preferred combinations of $H_0$ and $j_0$ that accepts $50<H_0 \ [\text{km s}^{-1} \text{ Mpc}^{-1}]<100$ only when $-50<j_0<50$.

Our assumption that the oscillations in redshift do not affect the global fit is supported a posteriori by an additional run of the cosmographic fit in which we have excluded the data points with redshifts in the range $0.15<z<0.20$. The marginal posterior distributions of the parameters are similar to those observed in Fig. \ref{fig:cosmography_TMJ}, with the median and minimum $\chi^2$ in the last $1000$ steps of the MCMC chain being roughly half the value found in the original case, the same order of magnitude. The marginals and corner plots for this case can be found in Appendix \ref{sec:corners}.

So far results remain model-independent in the framework of cosmology, but with the transformations in equations (\ref{eq:Om_q0j0}) and (\ref{eq:w0_q0j0}) we can take a look at the implications for a $w_0$CDM model. To this end, we decided to run an extra MCMC imposing $0 < \Omega_m(q_0, j_0) < 1$. We do so because, if we depended solely on the results for the sampling shown in Fig. \ref{fig:cosmography_TMJ}, the regions $\Omega_m > 1$ and $\Omega_m < 0$ would be populated, adding spurious probability to $w_0$ and $H_0$ coming from the sampling there. The probability density regions of the figure show the sampling when $0 < \Omega_m < 1$ is enforced, and the brown hatched area represents the region allowed by $w_0 > 1/(\Omega_m -1)$. This condition on the $\Omega_m-w_0$ plane is the result of having imposed $\text{d}H/\text{d}z > 0$. Due to the effect of the smoothing of the contour levels it might seem that there is a non-negligible amount of sampling at $w_0 < 1/(\Omega_m -1)$; however, when the steps of the MCMC in that region are counted, we find only $0.19\%$ of the total amount. This result is similar to the $0.09\%$ of samples that had $q_0 < -1$ in the paragraph above, and it can be attributed to the remainder in the approximation of $H(z)$ for its second order Taylor expansion. The newly constrained $H_0$ keeps almost the exact same posterior distribution as before, constrained now to $H_0 = 70.7^{+4.3}_{-5.1} \text{ km s}^{-1} \text{ Mpc}^{-1}$. $\Omega_m$ is mostly unconstrained, although central values seem to be preferred. Finally, we have a readout of $w_0$, with a posterior probability defined by $w_0 = -1.01 ^{+0.03}_{-1.75}$. This result is compatible with $\Lambda$CDM, although the fit seems to prefer the combinations of $w_0$ and $\Omega_m$ close to the limit $w_0 = 1/(\Omega_m-1)$, tightly limiting the right-hand tail of the posterior of $w_0$.

We now turn to the Knowles model. We apply the same conditions we did for the sampling in Fig. \ref{fig:cosmography_TMJ}. The direct application of the MCMC algorithm to the data yields the results shown in purple in the left panel of Fig. \ref{fig:cosmography_sMILES}. The posteriors constrain the cosmographic parameters to $H_0 = 41.7^{+0.6}_{-0.6}\text{ km s}^{-1} \text{ Mpc}^{-1}$, $q_0 = -0.99 ^{+0.08}_{-0.01}$ and $j_0 = 55.4^{+2.6}_{-2.4}$. The result for $q_0$ follows what we already observed in the TMJ case. The other two marginals, however, are well off and extremely thin, distancing from literature values by several $\sigma$. Such precise results for parameters so strongly deviating from their expected value prompts a careful reconsideration of all possible sources of error. First, we note that the narrowness of the posterior distributions is primarily driven by the size of the observational errors. Then, after analysing the last thousand iterations of the MCMC chain, we find a discrepancy of three orders of magnitude in the minimum $\chi^2$ compared to the fit in Fig. \ref{fig:cosmography_TMJ}. 

This suggests that the ages inferred using the Knowles model may not be useful to gain sensitivity on the cosmographic parameters in their current form. We conclude that the nature of the Knowles model (Appendix \ref{sec:app_VD}), affects the performance of the subsequent cosmographic fit by enlarging the scatter of the ages while reducing the value of their uncertainties. The overall effect is the restriction of the posteriors to clearly off values and worsening the overall $\chi^2$ values of the sampling. If there was an underestimation of the ages that we cannot control, it can be addressed by introducing a set of $f$ factors to the likelihood. We do so following the documentation of \emph{emcee}, by substituting
\begin{equation}
    \sigma_{i,v}^2 \to \sigma_{i,v}^2 + t_{i,v}^2(z_{i,v}; H_0, q_0, j_0)\cdot f_v^2
    \label{eq:factorf}
\end{equation}

In essence, an additional parameter $f_v$ is included in the MCMC for each velocity dispersion bin, with a view to capture the lacking variance at least to a certain degree. They are assigned wide uniform priors satisfying $-10 < \log f_v < 1$. The corresponding corner plot for the cosmographic parameters is shown in grey in the left panel of Fig. \ref{fig:cosmography_sMILES}, with the Hubble constant constrained to $38.2^{+12.9}_{-5.1}$ $\text{ km s}^{-1}\text{ Mpc}^{-1}$. The result is still incompatible with other probes at $2\sigma$, and the central value is lower than before. However, the width of the posterior allows for a $3\sigma$ agreement with $H_0 \sim 70$ $\text{ km s}^{-1}\text{ Mpc}^{-1}$. The posterior of $q_0$ is again skewed towards $-1$ ($\text{mod}(q_0) \sim -0.9$, $P_{16}(q_0) = -0.17$), although this time the positive tail of the sampling extends to the upper limit of the prior. If we observe the corner plot relative to the plane $q_0-H_0$ we can see a degeneracy that prefers higher $H_0$ values (still low in comparison to other probes) for smaller $q_0$ values. Finally, the model is still insensitive to $j_0$ as in the TMJ case, but we find a constraint ($j_0 = 5.8^{+42.6}_{-5.1}$), thanks to the introduction of $f_v$ factors, that is compatible with $j_0 = 1$ within $1\sigma$. We notice that this time the posterior of $j_0$ is right-skewed.

In the right panel of the figure, we omitted the $1\sigma$ sampling of $t(z)$ without $f_v$ factors. We did so to simplify the figure and avoid unnecessary visual noise, but they are tight and leave most of the data points out by several $\sigma$. For the rest, Fig. \ref{fig:cosmography_sMILES} contains the same elements as Fig. \ref{fig:cosmography_TMJ}.

It is clear that $f_v$ factors somewhat aligned the results with the expected values coming from other probes, even if they do not capture the full complexity of the underestimated variance. The power of these factors is clear, as the average value of the minimum $\chi^2$ over the last thousand steps of the MCMC chain is now of the order of magnitude obtained for the TMJ fit. As a final comment on the Knowles model relative to the $t-z$ relation, it could happen that the modelled ages are less sensitive to the index-oscillations observed in the TMJ case, as they are not as easily perceived on the right panel of Fig. \ref{fig:cosmography_sMILES}. This idea could be tested further if the intrinsic scatter of the data was reduced, and if true, would constitute a strong feature of the Knowles model while the fluctuating index-strength issue is understood. So far, we believe that caution is advisable when using the Knowles model for cosmographic purposes, as it seems to present some irregularities regarding the optimal tracking of the evolution of ETG ages across redshift. Some of the difference between models in the evolution of the differential age with redshift can be attributed to the effect of model variation when it comes to the estimation of the Hubble parameter \citep{moresco20}. For what is known in literature, the intra-model variation of the IMF is expected to be only a minor issue, but other age diagnostics or index selection could be inspected in future work.

\subsection{Assessment of systematic uncertainties}

So far, the uncertainties that have been included in the estimation of cosmographic parameters are mainly of statistical nature. Only the uncertainty associated with the velocity dispersion correction functions, which is added to the index measurements, is of a systematic nature. However, as discussed in Appendix \ref{sec:app_VD}, the contribution to the error budget is small. In the end, it is treated as part of the statistical uncertainties, as the stellar model fit makes it impossible to get them disentangled. Apart from these, there are other sources of systematic uncertainty that have been thoroughly treated in the literature. We distinguish between those that are external to this work (SPS models, their configuration: IMF, isochrones...), and those that arise from our methodology. We are not treating the former as we will perform the $H(z)$ estimation following only the TMJ ages, as we concluded that the Knowles model needs further inspection before deriving conclusions from it. For the rest of this section, all computations refer to ages estimated using the TMJ model.

Following the stream of our methodology, we can cite the following three sources of systematic uncertainty: the S/N level of the stacked spectra, the set of Lick indices used for the stellar parameter fit, the choice of the final reliable dataset for the cosmographic fit (effect of outliers and internal homogeneity of the stacks). We run a test on the S/N level and the redshift extension of the data points, to assess their contribution to the systematic error budget. Addressing the impact of the set of Lick indices chosen for the fit would require a complicated analysis in which several combinations of indices, selected to be sensitive enough to variations on the SPS model parameters, are included. Beyond the feasibility of the study, this work does not aim to test the performance of index combinations in the TMJ model, which was already done by \cite{johansson12}. Indeed, we chose our indices as the baseline described in that work, plus another four indices that \cite{thomas11} proved to be well calibrated with globular clusters. The most recent study by \cite{borghi22b} showed that the contribution of the set of indices to the systematic error budget in $H(z)$ would be of the order of that related to the binning (in our case the stacking). Finally, the exclusion of data points that averaged galaxies over a large redshift interval is addressed by performing the cosmographic fit both releasing and tightening the threshold for exclusion in $\sigma_z$.

In order to build the stacks we followed the methodology described in Sec. \ref{sec:stacking}, which leaves the selection of the S/N level as the sole arbitrary choice. We assess its contribution to the systematic uncertainty by creating $11$ groups of stacks, built up to have S/N levels from $200$ to $400$ in steps of $20$. The central group at $\text{S/N} \sim 300$ is the one that we used as reference throughout the paper. We obtain the posterior distributions from the cosmographic fit in each case, compute the median, over the last $1000$ steps of the MCMC chains, of the standard deviation between the $11$ groups at each redshift. This is,
\begin{equation}
    \sigma_{H, \text{syst}}(z) = \text{med}_k\left(\sqrt{   \text{Var}_i \left(H_{i} (z; H_{0k}, q_{0 k}, j_{0k} )\right) }\right),
    \label{eq:medianastdsyst}
\end{equation}
where the index $_k$ represents the sample from the MCMC chain and the index $_i$ represents each of the $11$ stacking groups. Equation (\ref{eq:medianastdsyst}) will also be used to evaluate the contribution to the systematic uncertainty introduced by the choice of the $\sigma_z$ threshold when excluding data points prior to the cosmographic fit. For this, we consider threshold values of $\sigma_z = \{\emptyset, 0.05, 0.02, 0.01, 0.005, 0.002\}$, where we use $\emptyset$ to denote the absence of any threshold (i.e. the inclusion of all stacks). The contribution of both sources of systematic uncertainty to the total is shown in Fig. \ref{fig:systematic}, in the following section, as it is presented in units of the statistical uncertainty of the posterior sampling of $H(z)$.

\subsection{The sampling of $H(z)$}

\begin{figure}
	\includegraphics[width=\columnwidth]{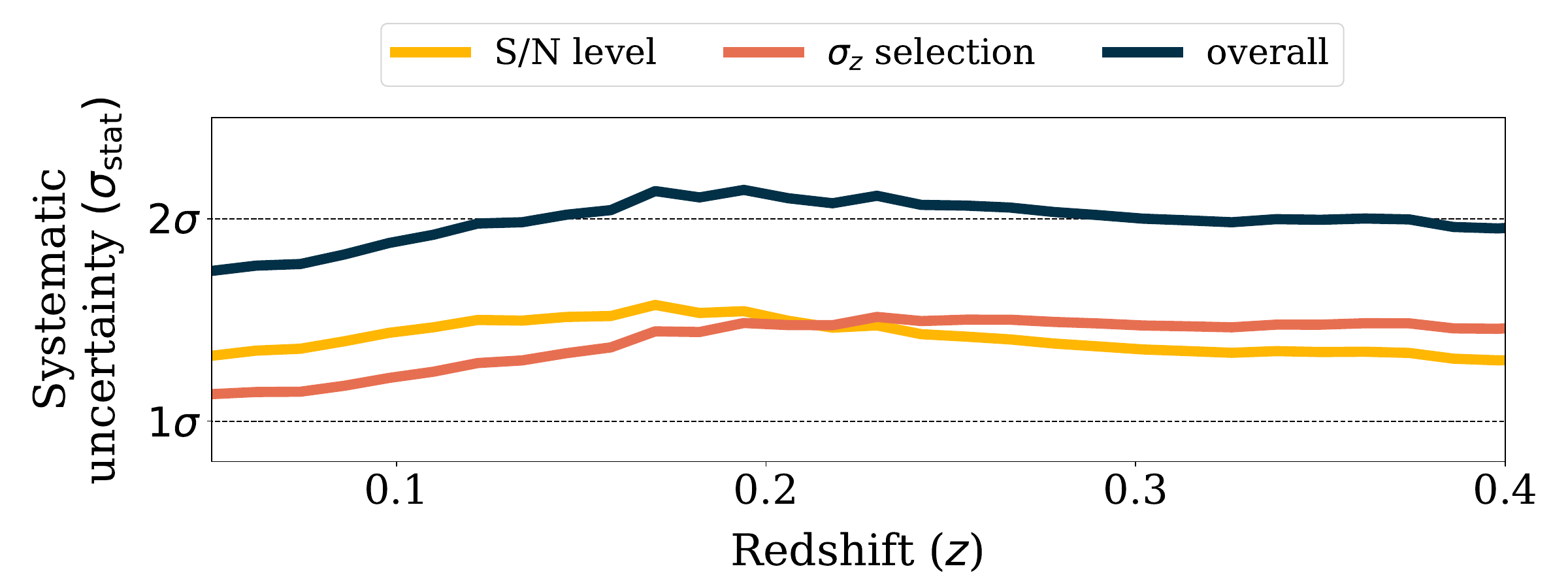}
    \caption{Systematic uncertainties arising from the methodology taken into account directly in this work. They are presented in units of $1\sigma$ of the statistical uncertainty $\sigma_\text{stat}$ at each redshift. The individual contributions of the level of S/N selected in the stacking procedure and the limit to the dispersion of sources in redshift in each stack are presented in yellow and dark orange respectively. In dark blue we present the overall systematic uncertainty.}
    \label{fig:systematic}
\end{figure}

As a final result we present in Fig. \ref{fig:Hfinal} the sampling of $H(z)$, using the posteriors from the baseline, in which we fit the TMJ-modelled ages imposing $\text{d}H/\text{d}z > 0$; this is, the MCMC run shown and discussed in Fig. \ref{fig:cosmography_TMJ}. At this point we introduce the systematic uncertainties that we discussed in the prior section. We do so following the typical quadratic sum: $\sigma_\text{total}^2= \sigma_\text{stat.}^2 + \sigma_\text{syst.}^2$. In Fig. \ref{fig:systematic} each of the contributions taken into account is presented in units of $1\sigma$ of the statistical uncertainty from the sampling of the MCMC run. Both are quite stable and slightly greater than the level of statistical uncertainty. The one associated to the S/N level chosen in the stacking procedure (yellow) dominates at low redshift, while the one related to the exclusion of data points (dark orange) is slightly greater at higher redshifts. Overall, the systematic uncertainty is stable and of the order of twice the statistical uncertainty.

In figure \ref{fig:Hfinal}, the final credible interval for $H(z)$ is presented in blue, where we have chosen to plot it in different shades according to the number of sources (individual galaxies) we had at each redshift. The light blue is chosen to represent the contribution of all sources of uncertainty that have been studied in this work, while the darker blue represents only the $1\sigma$ statistical uncertainty interval, computed as the region between the $16$th and $84$th percentiles of the sampling of the MCMC fit (parameters sampling in Fig. \ref{fig:cosmography_TMJ}). As a reference, we also included, as dashed light blue lines, the limits of the $1\sigma$ credible interval in case the main sources of systematic uncertainty not computed in this work (Lick index selection, SPS models) were collectively the same size \citep{borghi22b} as the systematic uncertainty taken into account.

For the sake of comparing the potential of this work with literature, we have included in yellow, orange and dark red the local model-independent measurements of $H(z)$ available to date up to $z = 1$. The colour of the observations depends on the methodology used for the analysis in each case. Full spectral fitting \citep{simon05, zhang14, ratsimbazafy17, jiao22}, in yellow, shows greater uncertainties than ages derived using the age-sensitive $\text{D}4000_\text{n}$ spectral feature \citep{moresco12, moresco16a, loubser25}, in orange. Our results extend along a redshift range $(z\lesssim 0.4)$ mostly populated by FSF measurements, and we see that our estimation is competitive in comparison. Within the same range, four/five local measurements through the $\text{D}4000_\text{n}$ feature show a relatively greater precision than ours. Finally, we present of the $\Lambda$CDM derived $H(z)$ from \cite{planck18}, that sits within the $1\sigma$ of our estimation. Naturally, our constrain power is stronger $H(z)$ when more sources are available, and it loses precision towards $z \gtrsim 0.3$. This pushes us to consider the extension of this methodology to higher redshift by the use of BOSS (Baryon Oscillation Spectroscopic Survey) and eBOSS (extended BOSS) data, using a Taylor expansion of $H(z)$ of appropriate order.

\section{Further Discussion\label{sec:otrosRESULTS}}

This section is included to fully exploit the results shown in the previous section. In the first place, we perform the linear fit of the scatter of $\log{(t)}$, $[\text{Z}/\text{H}]$ and $[\alpha/\text{Fe}]$ with $\log{(\sigma)}$, known as scaling relations. These functions try to give a quantitative guideline for the status of the currently well established downsizing scenario, by which more massive ETGs tend to have formed earlier in cosmic history, be more metallic and $\alpha$-enhanced. We obtain these relations for both the SPS models we consider in this work, using the low redshift ($z < 0.1$) end of our sample of sources. The second part of this section includes the discussion around the oscillations observed in the $t-z$ relation in Fig. \ref{fig:cosmography_TMJ}. In particular, we focus in the coordinated fluctuation, in all central and statistically significant velocity dispersion groups, at $0.16<z<0.19$. We unveil the tight correlation of this oscillation with a similar pattern on the index strength of some indices ($\text{H}\gamma_\text{F}$ and Fe$4383$) in that redshift range. A deeper inspection of the global trends of index strengths with redshift unveils an apparently overlooked behaviour that was already present in previous measurements of these Lick indices (i.e. \cite{tremonti04}).

\subsection{Stellar archaeology and scaling relations\label{sec:stellararch}}

\begin{figure*}
    \centering
    \includegraphics[width=0.49\textwidth]{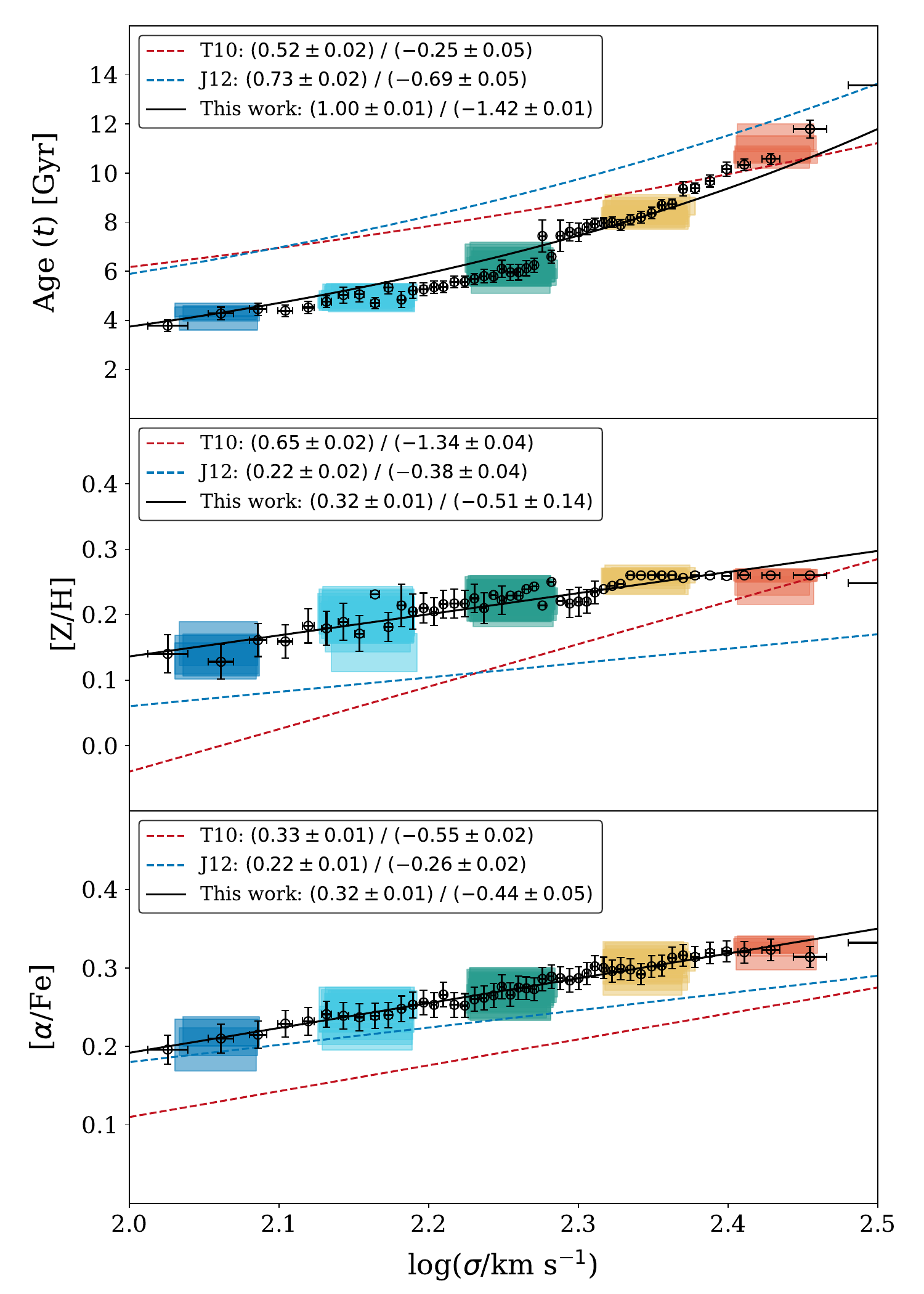}
    \includegraphics[width=0.49\textwidth]{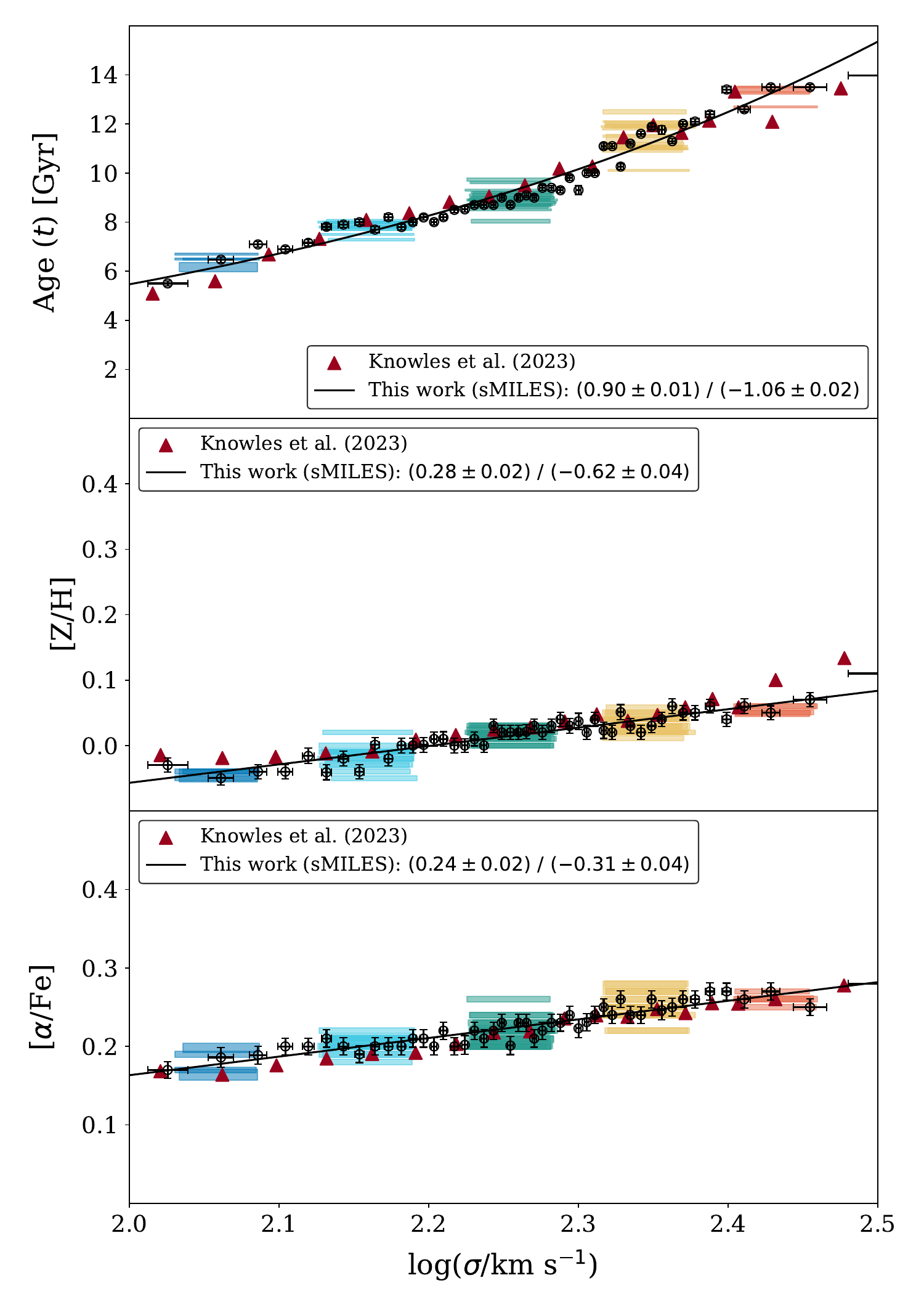}
    
    \caption{Left panel: Age, metallicity and $\alpha$-enhancement trends with velocity dispersion at low redshift ($0.05<z<0.10$). The black data points represent the results from stacks of $400$ galaxies. The red and blue dashed tendencies represent the \protect\cite{thomas10} and \protect\cite{johansson12} results (central values). The best log-log fit is plotted as a solid black line. We also depict the results from the "default" stacks as coloured rectangles representing the different velocity dispersion groups from dark blue ($100 < \sigma \ [\text{km s}^{-1}]<125$) to orange ($250 < \sigma \ [\text{km s}^{-1}]<320$). Note that the colour code of the velocity dispersion groups corresponds to the one used in Figs. \ref{fig:cosmography_TMJ} and \ref{fig:cosmography_sMILES}. Right panel: Same as left panel but with the Knowles model. Red triangles are the results by \protect\cite{knowles23} for stacks of massive ETGs, previously performed by \protect\cite{labarbera13}, using the same model.}
    \label{fig:archaeology_comparison}
\end{figure*}

We perform the study on the scaling relations by selecting all galaxies with redshifts $0.05 < z < 0.10$, $18955$ objects, which makes up around a $21\%$ of the whole sample. Given that the default stacks described in Section \ref{sec:stacking} are divided in six velocity dispersion bins (but we aim to derive a smooth relation from $\sigma\sim 100$ to $\sigma \sim 350 \text{ km s}^{-1}$), we construct another set of stacks for the purpose of this subsection. We stacked the galaxies in the subsample in groups of $400$ (ordered by velocity dispersion), since that number should ensure a high enough $\text{S}/\text{N}$ at this redshift according to Fig. \ref{fig:SNNz}. Lick indices are then measured and fit using both the TMJ and Knowles models.

Fig. \ref{fig:archaeology_comparison} shows the stellar population parameter trends in terms of velocity dispersion for the low-redshift sample. In particular, the results using the TMJ model are presented in the left panel, shown as black data points. The corresponding log-log linear fit is shown as a solid black line and, for comparison, the trends found by \cite{thomas10} and \cite{johansson12}\footnote{\cite{johansson12} performed an estimation of age, metallicity and $\alpha$-enhancement (represented by [O/Fe]) using the TMJ model with varying element abundances. The relations that are plotted are the ones that appear in their figures 6, 7 and 11 (upper panel).} are plotted with red and blue dotted lines, respectively. As a test of robustness, the coloured rectangles represent the results for the default stacks used for the cosmic chronometer analysis (Figs. \ref{fig:cosmography_TMJ} and \ref{fig:cosmography_sMILES}) that had an average redshift below $0.1$. The running of all three parameters behaves as expected with respect to velocity dispersion: the more massive the (passive) elliptical galaxy, the older, the more metal-rich and the more $\alpha$-enhanced it is. However, there is a caveat regarding the metallicities for $\log{\sigma} \gtrsim 2.3$: the posterior distributions saturates at the high end of the prior, indicating a preference for values higher than $0.26$, the limit set as explained in section \ref{sec:SPS}. While the $\text{Z}-\sigma$ fit omitted these data points, it should be noticed that relaxing this upper bound and allowing metallicities to extend up to $0.67$ breaks the smooth increase of age with velocity dispersion. Indeed, due to age-metallicity degeneracy, the values of $[\text{Z}/\text{H}]$ over $0.3$ cause the ages to decrease and produce an unstable trend towards the massive end for certain cases. As a consequence, we consider it safer to retain the hard prior on the maximum value of the metallicity.

\captionsetup[table]{skip=12pt}
\begin{table*}
    \centering
    \resizebox{\textwidth}{!}{
    \begin{tabular}{|l|c|c|}
        \hline
        & TMJ Model \rule{0pt}{15pt} &Knowles Model \rule{0pt}{15pt} \\ [1ex]
        \hline
        \multirow{3}{*}{\rotatebox{90}{\shortstack{Scaling\\Relations}}} & \ 
        $\log{(t)} = \left(-1.42 \pm 0.01\right) + \left(1.00 \pm 0.01\right)\log{(\sigma)}$ \rule{0pt}{4ex} \ & \
        $\log{(t)} = \left(-1.06 \pm 0.02\right) + \left(0.90 \pm 0.01\right)\log{(\sigma)}$ \ \ \  \\

        & \  $[\text{Z}/\text{H}] = \left(-0.51 \pm 0.14\right) + \left(0.32 \pm 0.01\right)\log{(\sigma)}$ \ \ \ \rule{0pt}{4ex} & \ \ \ 
        $[\text{Z}/\text{H}] = \left(-0.62 \pm 0.04\right) + \left(0.28 \pm 0.02\right)\log{(\sigma)}$ \ \ \  \\

        & \ \ \  $[\alpha/\text{Fe}] = \left(-0.44 \pm 0.05\right) + \left(0.32 \pm 0.01\right)\log{(\sigma)}$ \ \ \  \rule{0pt}{4ex} & \ \ \ 
        $[\alpha/\text{Fe}] = \left(-0.31 \pm 0.04\right) + \left(0.24 \pm 0.02\right)\log{(\sigma)}$ \ \ \  \\[1ex]
        \hline
        \multirow{3}{*}{\rotatebox{90}{\shortstack{Dispersion\\Relations}}} & 
        \ \ \ $\sigma_{\log{(t)}} = \left(0.54 \pm 0.15\right) + \left(-0.10 \pm 0.06\right)\log{(\sigma)}$ \ \ \  \rule{0pt}{4ex} &  \ \ \ 
        $\sigma_{\log{(t)}} = \left(1.82 \pm 0.01\right) + \left(-0.73 \pm 0.01\right)\log{(\sigma)}$ \ \ \  \\

        & \ \ \  $\sigma_{[\text{Z}/\text{H}]} = \left(0.87 \pm 0.07\right) + \left(-0.28 \pm 0.03\right)\log{(\sigma)}$ \ \ \  \rule{0pt}{4ex} & 
        \ \ \ $\sigma_{[\text{Z}/\text{H}]} = \left(1.12 \pm 0.03\right) + \left(-0.43 \pm 0.01\right)\log{(\sigma)}$ \ \ \  \\

        & \ \ \  $\sigma_{[\alpha/\text{Fe}]}= \left(0.52 \pm 0.03\right) + \left(-0.19 \pm 0.01\right)\log{(\sigma)}$ \ \ \ \rule{0pt}{4ex}  & 
        \ \ \  $\sigma_{[\alpha/\text{Fe}]} = \left(0.72 \pm 0.04\right) + \left(-0.27 \pm 0.02\right)\log{(\sigma)}$ \ \ \  \\[1ex]
        \hline
    \end{tabular}%
    }
    \caption{Scaling and dispersion relations for stellar population parameters with velocity dispersion for the TMJ (left) and Knowles (right) models.}
    \label{tab:scalingrels}
\end{table*}

Our results differ quantitatively from those in literature. In particular, we predict systematically lower ages than \cite{johansson12}, with an approximately constant offset of $\sim 2$ Gyr. We also observe lower ages compared to the \cite{thomas10} results in lower velocity dispersion data. Our results also indicate that galaxies tend to be more metal-rich and $\alpha$-enhanced. There are a number of reasons for this quantitative discrepancy. First, the Lick index model used in \cite{thomas10} is based on the \cite{thomas03} and \cite{thomas04} models, the precursors of the TMJ model (used both in \cite{johansson12} and in this work) that were not flux-calibrated. Second, both \cite{thomas10} and \cite{johansson12} performed an analysis of the spectra of single galaxies from the SDSS that were only morphologically selected to be ETGs. On the other hand, we work with high $\langle \text{S}/\text{N}\rangle$ stacked spectra of galaxies and aim at a selection of the most passively evolving sources. Third, the choice of Lick indices and the fitting procedure are different, since we fix a priori a set of indices for all galaxies based on the quality of their calibration with respect to globular cluster data as indicated in \cite{thomas11}. However, the qualitative physical picture remains unchanged and in line with the downsizing scenario.

The right panel of Fig. \ref{fig:archaeology_comparison} shows the corresponding results using the Knowles model, depicted in black. For comparison, we plot the results from \cite{knowles23} as red triangles, obtained via the same model and Lick index choice we used, taking the set of ETG stacks from \cite{labarbera13}. The age, metallicity and $\alpha$-enhancement for the default stacks used in the cosmic chronometer analysis are also shown as coloured rectangles. The agreement among the three sets of results is remarkable and serves as a further consistency check that the stacks have been constructed in a reliable manner. With regard to the galactic parameters, the qualitative trends remain unchanged and are consistent with a downsizing scenario. However, the Knowles model predicts systematically higher ages than TMJ, with an approximately constant offset of $\sim 2$ Gyr across all velocity dispersions. This translates directly into much smaller metallicities that reach even subsolar values for $\sigma \lesssim 160 \text{ km s}^{-1}$. Lastly, the sample is clearly $\alpha$-enhanced, although the values are slightly smaller than those of the TMJ model across all masses.

The average ages, metallicities and $\alpha$-enhancements obtained via a log-log linear fit of the above data can be regarded as the mean values of these quantities at the population level of massive quiescent galaxies. Even if the overall shape of the distributions for the single galaxy estimation is underwhelming, the mean of this distribution uncannily aligns with the results obtained using the more reliable stacking procedure. This is clearly seen in the upper-left panel of Fig. \ref{fig:bootst_TMJandK}, where the ages derived from Lick index fitting of single galaxies (plotted as density contours) are compared to the best fit for the stacking analysis with the TMJ model, depicted as a solid black line). While the median of single-galaxy ages (plotted as a thick blue line) seems to align with the values obtained from the stacks, a very large number of single massive galaxies are assigned very low or extremely high ages, as portrayed by the high density (darker blue contour) group (lobe) of individual galaxies in between $2$ and $5$ Gyr in the panel. This is also true for the metallicity and $\alpha$-enhancement trends, which are shown in the other two panels of the figure. This is a further hint that the $\text{S}/\text{N}$ of the spectra needs to be high enough to derive stellar population parameters in a reliable way.
\begin{figure*}
	\includegraphics[width=\textwidth]{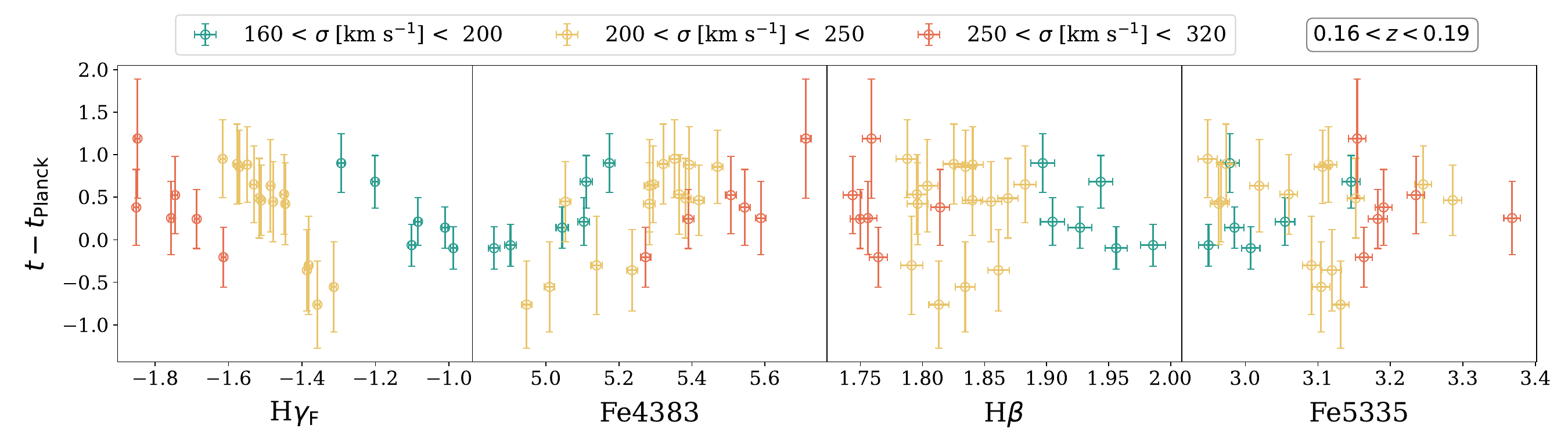}
    \caption{Difference between our estimated ages and the corresponding \protect\cite{planck18} trends (as shown in the form of dashed black lines in figure \ref{fig:cosmography_TMJ}) with respect to the index strength of H$\gamma_\text{F}$ (left panel) and Fe$4383$ (right panel). For the sake of clarity only the intermediate velocity dispersion groups, with enough data in the interest region $0.15<\text{z}<0.21$, are represented.
    }
    \label{fig:agediff_4indices}
\end{figure*}

However, we can exploit the distribution of the single-galaxy stellar population parameters to derive a mean dispersion at the population level for the age, metallicity and $\alpha$-enhancement of massive quiescent galaxies at low redshift, that is, $\sigma_{\log t}$, $\sigma_{[\text{Z}/\text{H}]}$ and $\sigma_{[\alpha/\text{Fe}]}$. Indeed, for every velocity dispersion bin, the standard deviation of the stellar population parameters with respect to their mean value can be computed. Via a Bootstrap resampling procedure, we can assign an uncertainty to the standard deviation and perform a simple linear fit with respect to velocity dispersion, as shown in the lower panels of Fig. \ref{fig:bootst_TMJandK}. As expected, the intrinsic dispersion in the ages (and thus in the formation times) decreases with velocity dispersion, in agreement with the popular view that the more massive the local quiescent galaxies, the more synchronous the formation of their progenitors was. The procedure is then repeated for the Knowles model, yielding the same qualitative behaviour, as shown in the right panels. However, the scatter in the ages for this model drops abruptly for large velocity dispersions due to the very high ages that it systematically assigns to the most massive galaxies. Through this approach, we provide an analytical recipe for the age, metallicity and $\alpha$-enhancement distribution of local quiescent galaxies and, in particular, for the formation time of their progenitors. These scaling relations can be used as inputs, for example, in studies relying on the formation time distribution of the progenitors of local quiescent galaxies \citep{bosi25}. The results are summarized in the left column of Table \ref{tab:scalingrels}.

\subsection{An oscillating pattern in the index-redshift relations\label{sec:indicestrends}}

When analysing the scatter of ages with redshift in Fig. \ref{fig:cosmography_TMJ}, we revealed some synchronous oscillations at various redshift intervals. These effects seem to break the global tendency in a manner that natural statistical fluctuations cannot explain. The effect is most clearly seen for the TMJ model and around $z\sim0.17$, although other small maxima could be identified at $z\sim0.08$ and $z\sim0.13$. After making sure that the fluctuations were independent of the $\text{S}/\text{N}$ of the stacks, we decided to look into the index measurements, aiming to assess whether there were specific indices responsible for the oscillatory pattern.

Fig. \ref{fig:agediff_4indices} shows the age difference with respect to a Planck cosmology for all stacks in the three central velocity dispersion groups\footnote{Note that, for clarity, we chose the same colour guide for velocity dispersion groups as in Figs. \ref{fig:cosmography_TMJ}, \ref{fig:cosmography_sMILES} and \ref{fig:archaeology_comparison}.} at redshifts $0.16<z<0.19$ (where the main oscillation is found) as a function of several Lick index strengths; namely H$\gamma_\text{F}$, Fe$4383$, H$\beta$ and Fe$5335$. It is clear that H$\gamma_\text{F}$ and Fe$4383$ portray a strong correlation with respect to the age deviation from Planck that is not present in the other two indices. It should be noted that this does not mean that H$\beta$ is not sensitive to stellar ages; in fact, it correlates very well when plotted over the whole sample. What the figure tells us is that certain indices are more responsible than others for the main fluctuation in the data irrespective of their intrinsic correlation with stellar ages.

\begin{figure}
	\includegraphics[width=\columnwidth]{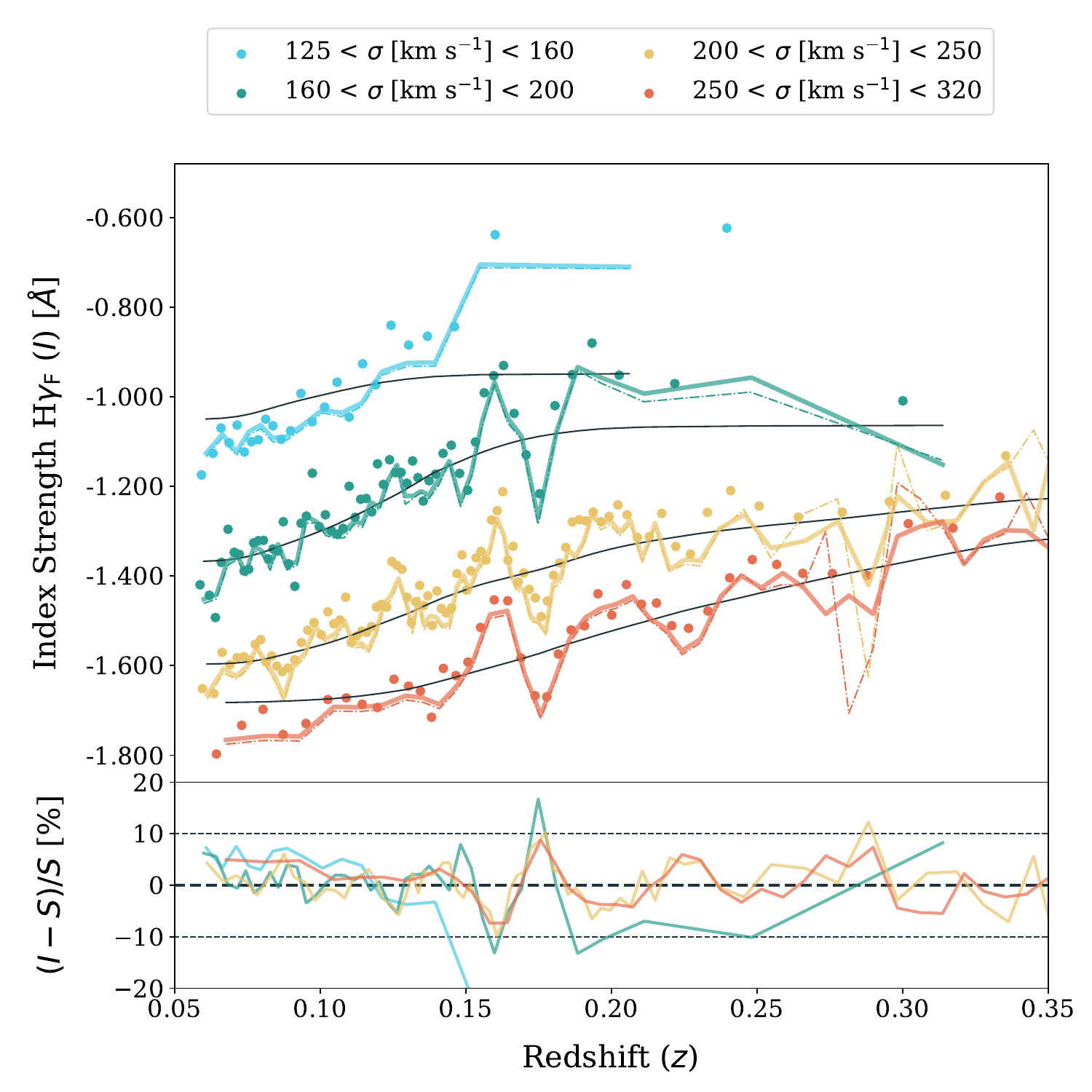}
    \caption{H$\gamma_\text{F}$ trend with redshift. The measurements from the stacks are shown as data points, coloured by the corresponding velocity dispersion bin accordingly to the colour code used in previous figures. The solid black lines represent the global trend via smoothed moving means, while the solid coloured lines are our measurements for single galaxies. The results from the SDSS database are depicted with coloured dashed lines.}
    \label{fig:HgF}
\end{figure}

As a consequence, we decided to look more closely at the index measurements and found an oscillatory behaviour with respect to redshift for a number of them. A paradigmatic example of this finding is shown in Fig. \ref{fig:HgF}, where the coloured dots are our measurements of the H$\gamma_\text{F}$ index for every galaxy stack in the four central velocity dispersion groups (again, the colour code for velocity dispersion is the same as in previous figures). As depicted in the bottom subpanel, a synchronous oscillation of $\sim 10\%$ with respect to the smoothed global trend (depicted as a solid black line) is clearly visible for three velocity dispersion bins around $z\sim 0.17$, where the main fluctuation feature is found for the ages. To determine if this could be related to the stacking procedure, we decided to measure the index strength for all single galaxies used in the stacks individually. The mean index value for every redshift is shown as a solid line, with the same colour code. As it is clearly observed, the single-galaxy trend follows perfectly that of the stacks, proving that the oscillatory phenomenon comes from a different nature. Once the stacking was discarded as the reason behind this effect, we could only turn to some flaw in the measurement of the index strengths. Fortunately, the SDSS archive offers measurements of Lick indices as well in their \texttt{galSpecIndx} table \citep{tremonti04, brinchmann04}. The mean for every redshift is shown with dashed thin lines, again with the same colours. The agreement is extremely perfect for $z<0.25$, to the point that the solid and dashed lines overlap almost at every point with our measurements coming from both single galaxies and stacks.

Since the trait is particularly enhanced at $0.16 \lesssim z \lesssim 0.19$, we can inspect our methodology to check for any particular divergence at that redshift. We have a strong non-linear behaviour in the SDSS instrumental resolution with wavelength at $\sim5500$\AA \ (rest frame) that is extended down to $\sim 4500$\AA \ at $z \sim 0.3$, as can be observed in figure \ref{fig:resolution_vd200250}. One could suspect of the treatment of the instrumental resolution as a driver of (or element that enhances) the fluctuation. At the aforementioned redshift window, all spectral features falling within $4810$\AA \ and $5350$\AA \ could be affected by the discontinuity in the instrumental resolution of SDSS. The indices that are included in that wavelength range, partially or totally, are H$\beta$, Fe$5015$, Mg$_1$, Mg$_2$, Mg$_\text{b}$, Fe$5270$ and Fe$5335$. From these, H$\beta$ and the bluest iron index were inspected in Fig. \ref{fig:agediff_4indices}, but showed no correlation in the age-index strength plane. The ones that present the clearest correlation in that plane, H$\gamma_\text{F}$ and Fe$4383$, are safely found in the region where $\sigma_\text{IR}$ is still $\sim 1.06$\AA, so for the moment we can say that the (non-linear wavelength dependent) instrumental resolution of SDSS is not the cause we are looking for.

The tests above show that there is an oscillatory trend in the H$\gamma_\text{F}$ index with respect to redshift that does not depend on galaxy mass or the $\text{S}/\text{N}$ of the spectrum. It is neither related to the stacking procedure nor the measurement itself, pointing to something rooted inside the individual galaxy spectra. Moreover, when we extend this analysis to other indices, we see that this strange behaviour is not only typical of H$\gamma_\text{F}$, nor it is in it that the trait is most enhanced. The pattern (although with different strengths and at different redshifts) is found as well, for instance, in the Fe$4383$, Fe$4531$, H$\delta_\text{A}$ and H$\beta$ index strengths. All of these are included in the fitting set, yet many others not included show as well oscillating patterns or fluctuations (Appendix \ref{sec:indexoscilappendix} is included to present the oscillations found in several other indices apart from H$\gamma_\text{F}$, shown in Fig. \ref{fig:HgF}), as happens with $\text{D}4000_\text{n}$ and $\text{C}_2 4668$. The latter is the most striking, with oscillations being enhanced, periodic and stable for a long redshift range, being more relevant than the index-velocity dispersion relation itself. The discontinuity feature suffers instead a significant drop at $z \sim 0.10$ (Fig. \ref{fig:otherindicesnofitD4000}), breaking the expected stable and monotonous decrease of the feature with redshift. We call attention to the fact that such behaviour could undermine the good work of the D$4000_\text{n}$ estimator at the referred redshift. Indeed, D$4000_\text{n}$ estimations of $H(z)$ at $z<0.4$ are relatively scarce (Fig. \ref{fig:Hfinal}), and \cite{moresco11} had already limited the work with this feature at $z > 0.15$, where it seems to gain stability.

Since the factors within our control do not seem to explain the observed oscillations, we believe that the only possibilities should be related to the SDSS data treatment. In particular, either the issue reflects a problem with the interpolation in the flux calibration procedure or is associated with atmospheric emission. As explained in \cite{stoughton02}, these effects are taken into account in the SDSS data reduction pipeline by means of the modelling of hot stars and subtraction of the empty sky. Regarding the former, the instrument response function is estimated using the observed spectra of hot young stars, whose spectra are well understood and calibrated. These spectra feature very deep Balmer lines, and modelling the continuum of these stars requires interpolating the flux at their wavelength region. This can introduce a bias in the response function at the rest-frame position of Balmer lines (since calibration stars are local) and, as a consequence, in the measurement of Lick absorption features that fall at the rest-frame wavelength of these Balmer lines. We take the H$\gamma_\text{F}$ index as an example. As shown in Fig. \ref{fig:HgF}, this index displays a strong oscillation between $z\sim0.16$ and $z\sim0.19$. Taking into account the definition of both the central and pseudocontinuum regions of this index, a quick computation shows that the rest-frame wavelength corresponding to this index for these redshifts lies well below $3800$\AA, far from the H$\delta$ Balmer index, the only one that could potentially be responsible for a bias in the measurement. While it is true that other smaller fluctuations in H$\gamma_\text{F}$, like the minimum around $z=0.08$, might fall at the right rest-frame wavelength for H$\delta$, we have checked that the interpolation issue could not account for the vast majority of oscillations in all indices. The second possibility would be related to telluric absorption or emission lines not correctly accounted for in the SDSS data reduction pipeline. Trying the wealth of existing atmospheric lines that could be responsible for the oscillating patterns is beyond the scope of this work, but preliminary tests seem to show that they cannot contribute to the bulk of this anomaly. Further work will address these findings.

\section{Conclusions\label{sec:conclusions}}

In this work we have employed a cosmographic approach in the context of cosmic chronometers to derive a continuous measurement of the Hubble parameter up to $z \sim 0.4$ via the fitting of Lick index absorption lines from massive and passively evolving galaxies. To this end, we carefully constructed a sample of such objects from the SDSS Legacy data, resulting in a sample of $90396$ objects meeting the expected physical requirements for these galaxies, in agreement with other samples in the literature. The spectra are reliably stacked in bins of both velocity dispersion and redshift to ensure a similar and high enough signal-to-noise ratio. After carefully considering instrumental resolution and velocity dispersion effects, Lick absorption line indices are measured and fit with two stellar population models, namely TMJ \citep{thomas11} and Knowles \citep{knowles23}, in order to derive stellar population parameters $(t, \text{[Z/H]},[\alpha/\text{Fe}])$. We emphasize the importance of using stacked spectra for the analysis, as demonstrated by the stellar population parameter distributions obtained from the fit single galaxy Lick indices, which feature a much less reliable and stable behaviour.

Results cover the cosmographic fit of the stellar ages, $t(z; H_0, q_0, j_0)$, through the integration of the inverse of the Hubble parameter, expanded in Taylor series of $z/(1+z)$ up to order $2$. The credible intervals given for the cosmographic parameters $\{H_0, q_0, j_0\}$ include only the statistical uncertainty associated to the scatter of the data and error propagation throughout the data treatment process. A brief summary of the systematic uncertainties due to our manipulation of data is included and applied directly to the final $H(z)$ estimation. Namely, we test the effect of the choice of the S/N level for the stacked spectra and the exclusion of stacks that average galaxies extended in redshift with $\sigma_z$ greater than a limit imposed by us (i.e. the baseline is $\sigma_z < 0.005$). We observed that these account for an impact similar to twice the statistical uncertainty. Other sources of systematic uncertainty related to the SPS models are not analysed in this work, but we included an additional margin in the estimation of $H(z)$ under the assumption that these could represent a similar amount to the ones we took into account. This is based on recent studies that assessed the effect of the SPS model and the Lick index combination on the $H(z)$ readout.

The TMJ model provides $\Lambda$CDM-aligned posteriors for our set of cosmographic parameters when the condition $dH/dz > 0$ is included. This yields a model-independent measurement of the Hubble constant, $H_0 = 70.0^{+4.1}_{-7.6}$ (stat.) km s$^{-1}$ Mpc$^{-1}$, that covers both late-time and early-time probes results. The deceleration parameter is skewed towards the limit imposed by the condition on the derivative of $H$, which translates into $q(z) > -1$; in particular $q_0 > -1$. The removal of this condition reveals the compatibility of the TMJ-modelled ages with combinations of cosmographic parameters in $q_0 < -1$ and $j_0 > 50$, which in those cases requires $H_0 > 100$. We give a cosmological readout of our results by introducing a $w_0$CDM model in which our estimated $q_0$ and $j_0$ (baseline) can be translated into $\Omega_m$ and $w_0$. By requiring that $0<\Omega_m(q_0, j_0)<1$, we found out that the condition $\text{d}H/\text{d}z>0$ only removes some phantom solutions from the possible results, while keeping $H_0$ virtually unchanged. The associated dark energy equation of state parameter $w_0$ is compatible with a cosmological constant, with some but not all phantom states allowed.

Our final result consists of drawing $H(z)$ using the MCMC chains from the baseline case (TMJ model with $dH/dz > 0$). The agreement with previous punctual measurements is good, and we conclude that our methodology is competitive with analyses based on FSF (whose uncertainties are somewhat greater than ours) and cover a region $(z < 0.15)$ where D$4000_\text{n}$ measurements from SDSS proved to be problematic. We associate the latter to the unveiled pattern of oscillations and fluctuations in the index-redshift relations for several Lick indices and other features, in particular D$4000_\text{n}$ and D$4000$.

Regarding the Knowles model, the distribution of stellar ages on the $t-z$ plane posed some problems for the cosmographic fit. Uncertainties in the derived ages were smaller by a factor of $\sim 7$ in comparison to the TMJ-modelled ages; being therefore much smaller than the intrinsic scatter in the data, which was also slightly higher than in the TMJ case. These issues were partially mitigated by the introduction of additional $f$-factor parameters in the likelihood. In our assessment, the overall age distribution remains less robust than that produced by the TMJ model. Hence, we advise some caution when using the Knowles-modelled ages for cosmographic applications, at least until the model has been more thoroughly examined. This includes evaluating its various assumptions, such as the choice of IMF, which is expected to contribute only marginally \citep{moresco20}, exploring alternative spectral diagnostics beyond those used in our fit, or even considering different methodologies other than spectral-feature fitting, taking advantage of the model being presented as full synthetic spectra for the given combinations of age, metallicity and $\alpha$-enhancement.

As a by-product, we give relations for the evolution of stellar population parameters with respect to velocity dispersion for both stellar population models. As expected, these scaling relations are in agreement with the downsizing scenario, although with quantitative differences with respect to results in the literature. Beyond those, we also offer a quantitative description for the dispersion relations of these parameters.

We conclude the work presenting the finding of an unexpected oscillating pattern in the distribution of index strengths with respect to redshift for a number of Lick indices, in particular some used in the cosmographic fits. These were found when trying to understand the fluctuating behaviour of the age-redshift dispersion. We observed that some index strength oscillations were more correlated with age oscillations than others; in particular, the Balmer $\text{H}\gamma$ feature (both $_\text{A}$ and $_\text{F}$ definitions) and some iron indices (in particular Fe$4383$), which are the main drivers of the oscillation around $z \sim 0.17$. We investigated their potential origin and discarded a statistical nature due to the synchronized behaviour of the oscillations across velocity dispersion bins. A methodological origin was also ruled out, as fluctuations are found both in the stacked and single-galaxy spectra and are also present in the Lick index measurements provided by the SDSS \citep{tremonti04, brinchmann04}. Furthermore, flux calibration issues related to the interpolation of the continuum in the data reduction pipeline of the SDSS do not seem to be able to account for such strong features, neither does a preliminary analysis of telluric absorption lines. To our knowledge, such oscillations have never been reported before and will be further studied in future work. In particular, the presence or absence of such oscillations in spectroscopic data from the GAMA survey will inform us about the role of the SDSS spectrograph and data reduction on their origin.

The big picture for future works is that SPS models will remain a critical bottleneck and source of systematic uncertainty. Deep spectroscopic surveys (e.g., DESI, Euclid), as well as fully exploiting BOSS and eBOSS data, may provide the necessary leverage at $z > 0.5$. At higher redshift, the parameters of the cosmographic fit may be more sensitive to kinematical parameters, and a custom $H(z)$ expansion in the vicinity of a non-zero redshift is even possible, giving us the counterparts of our set of cosmographic parameters at higher redshift.

\begin{acknowledgements}
This work was partially funded from the projects: “Data Science
methods for MultiMessenger Astrophysics \& Multi-Survey Cosmology” funded by the Italian
Ministry of University and Research, Programmazione triennale 2021/2023 (DM n.2503 dd.
9 December 2019), Programma Congiunto Scuole; EU H2020-MSCA-ITN-2019 n. 860744
BiD4BESt: Big Data applications for black hole Evolution STudies; Italian Research Center
on High Performance Computing Big Data and Quantum Computing (ICSC), project funded
by European Union — NextGenerationEU — and National Recovery and Resilience Plan
(NRRP) — Mission 4 Component 2 within the activities of Spoke 3 (Astrophysics and
Cosmos Observations); European Union — NextGenerationEU under the PRIN MUR 2022
project n. 20224JR28W “Charting unexplored avenues in Dark Matter”; INAF Large Grant 2022 funding scheme with the project “MeerKAT and LOFAR Team up: a Unique Radio
Window on Galaxy/AGN co-Evolution; INAF GO-GTO Normal 2023 funding scheme with
the project “Serendipitous H-ATLAS-fields Observations of Radio Extragalactic Sources
(SHORES)”. We further want to mention the helpful contribution and counsel of Alexandre Vazdekis, Nicola Borghi, Michele Moresco and Christy A. Tremonti.
\end{acknowledgements}

\bibliographystyle{aa} 
\bibliography{main}

\begin{thebibliography}{125}
\expandafter\ifx\csname natexlab\endcsname\relax\def\natexlab#1{#1}\fi

\bibitem[{{Abazajian} {et~al.}(2009){Abazajian}, {Adelman-McCarthy}, {Ag{\"u}eros}, {Allam}, {Allende Prieto}, {An}, {Anderson}, {Anderson}, {Annis}, {Bahcall}, {Bailer-Jones}, {Barentine}, {Bassett}, {Becker}, {Beers}, {Bell}, {Belokurov}, {Berlind}, {Berman}, {Bernardi}, {Bickerton}, {Bizyaev}, {Blakeslee}, {Blanton}, {Bochanski}, {Boroski}, {Brewington}, {Brinchmann}, {Brinkmann}, {Brunner}, {Budav{\'a}ri}, {Carey}, {Carliles}, {Carr}, {Castander}, {Cinabro}, {Connolly}, {Csabai}, {Cunha}, {Czarapata}, {Davenport}, {de Haas}, {Dilday}, {Doi}, {Eisenstein}, {Evans}, {Evans}, {Fan}, {Friedman}, {Frieman}, {Fukugita}, {G{\"a}nsicke}, {Gates}, {Gillespie}, {Gilmore}, {Gonzalez}, {Gonzalez}, {Grebel}, {Gunn}, {Gy{\"o}ry}, {Hall}, {Harding}, {Harris}, {Harvanek}, {Hawley}, {Hayes}, {Heckman}, {Hendry}, {Hennessy}, {Hindsley}, {Hoblitt}, {Hogan}, {Hogg}, {Holtzman}, {Hyde}, {Ichikawa}, {Ichikawa}, {Im}, {Ivezi{\'c}}, {Jester}, {Jiang}, {Johnson}, {Jorgensen}, {Juri{\'c}}, {Kent}, {Kessler}, {Kleinman}, {Knapp},
  {Konishi}, {Kron}, {Krzesinski}, {Kuropatkin}, {Lampeitl}, {Lebedeva}, {Lee}, {Lee}, {French Leger}, {L{\'e}pine}, {Li}, {Lima}, {Lin}, {Long}, {Loomis}, {Loveday}, {Lupton}, {Magnier}, {Malanushenko}, {Malanushenko}, {Mandelbaum}, {Margon}, {Marriner}, {Mart{\'\i}nez-Delgado}, {Matsubara}, {McGehee}, {McKay}, {Meiksin}, {Morrison}, {Mullally}, {Munn}, {Murphy}, {Nash}, {Nebot}, {Neilsen}, {Newberg}, {Newman}, {Nichol}, {Nicinski}, {Nieto-Santisteban}, {Nitta}, {Okamura}, {Oravetz}, {Ostriker}, {Owen}, {Padmanabhan}, {Pan}, {Park}, {Pauls}, {Peoples}, {Percival}, {Pier}, {Pope}, {Pourbaix}, {Price}, {Purger}, {Quinn}, {Raddick}, {Re Fiorentin}, {Richards}, {Richmond}, {Riess}, {Rix}, {Rockosi}, {Sako}, {Schlegel}, {Schneider}, {Scholz}, {Schreiber}, {Schwope}, {Seljak}, {Sesar}, {Sheldon}, {Shimasaku}, {Sibley}, {Simmons}, {Sivarani}, {Allyn Smith}, {Smith}, {Smol{\v{c}}i{\'c}}, {Snedden}, {Stebbins}, {Steinmetz}, {Stoughton}, {Strauss}, {SubbaRao}, {Suto}, {Szalay}, {Szapudi}, {Szkody}, {Tanaka},
  {Tegmark}, {Teodoro}, {Thakar}, {Tremonti}, {Tucker}, {Uomoto}, {Vanden Berk}, {Vandenberg}, {Vidrih}, {Vogeley}, {Voges}, {Vogt}, {Wadadekar}, {Watters}, {Weinberg}, {West}, {White}, {Wilhite}, {Wonders}, {Yanny}, \& {Yocum}}]{ABAZAJIAN09}
{Abazajian}, K.~N., {Adelman-McCarthy}, J.~K., {Ag{\"u}eros}, M.~A., {et~al.} 2009, \apjs, 182, 543

\bibitem[{{Abbott} {et~al.}(2020){Abbott}, {Aguena}, {Alarcon}, {Allam}, {Allen}, {Annis}, {Avila}, {Bacon}, {Bechtol}, {Bermeo}, {Bernstein}, {Bertin}, {Bhargava}, {Bocquet}, {Brooks}, {Brout}, {Buckley-Geer}, {Burke}, {Carnero Rosell}, {Carrasco Kind}, {Carretero}, {Castander}, {Cawthon}, {Chang}, {Chen}, {Choi}, {Costanzi}, {Crocce}, {da Costa}, {Davis}, {De Vicente}, {DeRose}, {Desai}, {Diehl}, {Dietrich}, {Dodelson}, {Doel}, {Drlica-Wagner}, {Eckert}, {Eifler}, {Elvin-Poole}, {Estrada}, {Everett}, {Evrard}, {Farahi}, {Ferrero}, {Flaugher}, {Fosalba}, {Frieman}, {Garc{\'\i}a-Bellido}, {Gatti}, {Gaztanaga}, {Gerdes}, {Giannantonio}, {Giles}, {Grandis}, {Gruen}, {Gruendl}, {Gschwend}, {Gutierrez}, {Hartley}, {Hinton}, {Hollowood}, {Honscheid}, {Hoyle}, {Huterer}, {James}, {Jarvis}, {Jeltema}, {Johnson}, {Johnson}, {Kent}, {Krause}, {Kron}, {Kuehn}, {Kuropatkin}, {Lahav}, {Li}, {Lidman}, {Lima}, {Lin}, {MacCrann}, {Maia}, {Mantz}, {Marshall}, {Martini}, {Mayers}, {Melchior}, {Mena-Fern{\'a}ndez},
  {Menanteau}, {Miquel}, {Mohr}, {Nichol}, {Nord}, {Ogando}, {Palmese}, {Paz-Chinch{\'o}n}, {Plazas}, {Prat}, {Rau}, {Romer}, {Roodman}, {Rooney}, {Rozo}, {Rykoff}, {Sako}, {Samuroff}, {S{\'a}nchez}, {Sanchez}, {Saro}, {Scarpine}, {Schubnell}, {Scolnic}, {Serrano}, {Sevilla-Noarbe}, {Sheldon}, {Smith}, {Smith}, {Suchyta}, {Swanson}, {Tarle}, {Thomas}, {To}, {Troxel}, {Tucker}, {Varga}, {von der Linden}, {Walker}, {Wechsler}, {Weller}, {Wilkinson}, {Wu}, {Yanny}, {Zhang}, {Zhang}, {Zuntz}, \& {DES Collaboration}}]{ABBOTT20}
{Abbott}, T.~M.~C., {Aguena}, M., {Alarcon}, A., {et~al.} 2020, \prd, 102, 023509

\bibitem[{{Abbott} {et~al.}(2022){Abbott}, {Aguena}, {Alarcon}, {Allam}, {Alves}, {Amon}, {Andrade-Oliveira}, {Annis}, {Avila}, {Bacon}, {Baxter}, {Bechtol}, {Becker}, {Bernstein}, {Bhargava}, {Birrer}, {Blazek}, {Brandao-Souza}, {Bridle}, {Brooks}, {Buckley-Geer}, {Burke}, {Camacho}, {Campos}, {Carnero Rosell}, {Carrasco Kind}, {Carretero}, {Castander}, {Cawthon}, {Chang}, {Chen}, {Chen}, {Choi}, {Conselice}, {Cordero}, {Costanzi}, {Crocce}, {da Costa}, {da Silva Pereira}, {Davis}, {Davis}, {De Vicente}, {DeRose}, {Desai}, {Di Valentino}, {Diehl}, {Dietrich}, {Dodelson}, {Doel}, {Doux}, {Drlica-Wagner}, {Eckert}, {Eifler}, {Elsner}, {Elvin-Poole}, {Everett}, {Evrard}, {Fang}, {Farahi}, {Fernandez}, {Ferrero}, {Fert{\'e}}, {Fosalba}, {Friedrich}, {Frieman}, {Garc{\'\i}a-Bellido}, {Gatti}, {Gaztanaga}, {Gerdes}, {Giannantonio}, {Giannini}, {Gruen}, {Gruendl}, {Gschwend}, {Gutierrez}, {Harrison}, {Hartley}, {Herner}, {Hinton}, {Hollowood}, {Honscheid}, {Hoyle}, {Huff}, {Huterer}, {Jain}, {James}, {Jarvis},
  {Jeffrey}, {Jeltema}, {Kovacs}, {Krause}, {Kron}, {Kuehn}, {Kuropatkin}, {Lahav}, {Leget}, {Lemos}, {Liddle}, {Lidman}, {Lima}, {Lin}, {MacCrann}, {Maia}, {Marshall}, {Martini}, {McCullough}, {Melchior}, {Mena-Fern{\'a}ndez}, {Menanteau}, {Miquel}, {Mohr}, {Morgan}, {Muir}, {Myles}, {Nadathur}, {Navarro-Alsina}, {Nichol}, {Ogando}, {Omori}, {Palmese}, {Pandey}, {Park}, {Paz-Chinch{\'o}n}, {Petravick}, {Pieres}, {Plazas Malag{\'o}n}, {Porredon}, {Prat}, {Raveri}, {Rodriguez-Monroy}, {Rollins}, {Romer}, {Roodman}, {Rosenfeld}, {Ross}, {Rykoff}, {Samuroff}, {S{\'a}nchez}, {Sanchez}, {Sanchez}, {Sanchez Cid}, {Scarpine}, {Schubnell}, {Scolnic}, {Secco}, {Serrano}, {Sevilla-Noarbe}, {Sheldon}, {Shin}, {Smith}, {Soares-Santos}, {Suchyta}, {Swanson}, {Tabbutt}, {Tarle}, {Thomas}, {To}, {Troja}, {Troxel}, {Tucker}, {Tutusaus}, {Varga}, {Walker}, {Weaverdyck}, {Wechsler}, {Weller}, {Yanny}, {Yin}, {Zhang}, {Zuntz}, \& {DES Collaboration}}]{ABBOTT22}
{Abbott}, T.~M.~C., {Aguena}, M., {Alarcon}, A., {et~al.} 2022, \prd, 105, 023520

\bibitem[{{Abraham} {et~al.}(2004){Abraham}, {Glazebrook}, {McCarthy}, {Crampton}, {Murowinski}, {J{\o}rgensen}, {Roth}, {Hook}, {Savaglio}, {Chen}, {Marzke}, \& {Carlberg}}]{abraham04}
{Abraham}, R.~G., {Glazebrook}, K., {McCarthy}, P.~J., {et~al.} 2004, \aj, 127, 2455

\bibitem[{{Adame} {et~al.}(2025){Adame}, {Aguilar}, {Ahlen}, {Alam}, {Alexander}, {Alvarez}, {Alves}, {Anand}, {Andrade}, {Armengaud}, {Avila}, {Aviles}, {Awan}, {Bahr-Kalus}, {Bailey}, {Baltay}, {Bault}, {Behera}, {BenZvi}, {Bera}, {Beutler}, {Bianchi}, {Blake}, {Blum}, {Brieden}, {Brodzeller}, {Brooks}, {Buckley-Geer}, {Burtin}, {Calderon}, {Canning}, {Carnero Rosell}, {Cereskaite}, {Cervantes-Cota}, {Chabanier}, {Chaussidon}, {Chaves-Montero}, {Chen}, {Chen}, {Claybaugh}, {Cole}, {Cuceu}, {Davis}, {Dawson}, {de la Macorra}, {de Mattia}, {Deiosso}, {Dey}, {Dey}, {Ding}, {Doel}, {Edelstein}, {Eftekharzadeh}, {Eisenstein}, {Elliott}, {Fagrelius}, {Fanning}, {Ferraro}, {Ereza}, {Findlay}, {Flaugher}, {Font-Ribera}, {Forero-S{\'a}nchez}, {Forero-Romero}, {Frenk}, {Garcia-Quintero}, {Gazta{\~n}aga}, {Gil-Mar{\'\i}n}, {Gontcho a Gontcho}, {Gonzalez-Morales}, {Gonzalez-Perez}, {Gordon}, {Green}, {Gruen}, {Gsponer}, {Gutierrez}, {Guy}, {Hadzhiyska}, {Hahn}, {Hanif}, {Herrera-Alcantar}, {Honscheid}, {Howlett},
  {Huterer}, {Ir{\v{s}}i{\v{c}}}, {Ishak}, {Juneau}, {Kara{\c{c}}ayl{\i}}, {Kehoe}, {Kent}, {Kirkby}, {Kremin}, {Krolewski}, {Lai}, {Lan}, {Landriau}, {Lang}, {Lasker}, {Le Goff}, {Le Guillou}, {Leauthaud}, {Levi}, {Li}, {Linder}, {Lodha}, {Magneville}, {Manera}, {Margala}, {Martini}, {Maus}, {McDonald}, {Medina-Varela}, {Meisner}, {Mena-Fern{\'a}ndez}, {Miquel}, {Moon}, {Moore}, {Moustakas}, {Mueller}, {Mu{\~n}oz-Guti{\'e}rrez}, {Myers}, {Nadathur}, {Napolitano}, {Neveux}, {Newman}, {Nguyen}, {Nie}, {Niz}, {Noriega}, {Padmanabhan}, {Paillas}, {Palanque-Delabrouille}, {Pan}, {Penmetsa}, {Percival}, {Pieri}, {Pinon}, {Poppett}, {Porredon}, {Prada}, {P{\'e}rez-Fern{\'a}ndez}, {P{\'e}rez-R{\`a}fols}, {Rabinowitz}, {Raichoor}, {Ram{\'\i}rez-P{\'e}rez}, {Ramirez-Solano}, {Rashkovetskyi}, {Ravoux}, {Rezaie}, {Rich}, {Rocher}, {Rockosi}, {Roe}, {Rosado-Marin}, {Ross}, {Rossi}, {Ruggeri}, {Ruhlmann-Kleider}, {Samushia}, {Sanchez}, {Saulder}, {Schlafly}, {Schlegel}, {Schubnell}, {Seo}, {Shafieloo}, {Sharples},
  {Silber}, {Slosar}, {Smith}, {Sprayberry}, {Tan}, {Tarl{\'e}}, {Taylor}, {Trusov}, {Ure{\~n}a-L{\'o}pez}, {Vaisakh}, {Valcin}, {Valdes}, {Vargas-Maga{\~n}a}, {Verde}, {Walther}, {Wang}, {Wang}, {Weaver}, {Weaverdyck}, {Wechsler}, {Weinberg}, {White}, {Yu}, {Yu}, {Yuan}, {Y{\`e}che}, {Zaborowski}, {Zarrouk}, {Zhang}, {Zhao}, {Zhao}, {Zhou}, \& {Zhuang}}]{ADAME25}
{Adame}, A.~G., {Aguilar}, J., {Ahlen}, S., {et~al.} 2025, \jcap, 2025, 021

\bibitem[{{Aiola} {et~al.}(2020){Aiola}, {Calabrese}, {Maurin}, {Naess}, {Schmitt}, {Abitbol}, {Addison}, {Ade}, {Alonso}, {Amiri}, {Amodeo}, {Angile}, {Austermann}, {Baildon}, {Battaglia}, {Beall}, {Bean}, {Becker}, {Bond}, {Bruno}, {Calafut}, {Campusano}, {Carrero}, {Chesmore}, {Cho}, {Choi}, {Clark}, {Cothard}, {Crichton}, {Crowley}, {Darwish}, {Datta}, {Denison}, {Devlin}, {Duell}, {Duff}, {Duivenvoorden}, {Dunkley}, {D{\"u}nner}, {Essinger-Hileman}, {Fankhanel}, {Ferraro}, {Fox}, {Fuzia}, {Gallardo}, {Gluscevic}, {Golec}, {Grace}, {Gralla}, {Guan}, {Hall}, {Halpern}, {Han}, {Hargrave}, {Hasselfield}, {Helton}, {Henderson}, {Hensley}, {Hill}, {Hilton}, {Hilton}, {Hincks}, {Hlo{\v{z}}ek}, {Ho}, {Hubmayr}, {Huffenberger}, {Hughes}, {Infante}, {Irwin}, {Jackson}, {Klein}, {Knowles}, {Koopman}, {Kosowsky}, {Lakey}, {Li}, {Li}, {Li}, {Lokken}, {Louis}, {Lungu}, {MacInnis}, {Madhavacheril}, {Maldonado}, {Mallaby-Kay}, {Marsden}, {McMahon}, {Menanteau}, {Moodley}, {Morton}, {Namikawa}, {Nati}, {Newburgh},
  {Nibarger}, {Nicola}, {Niemack}, {Nolta}, {Orlowski-Sherer}, {Page}, {Pappas}, {Partridge}, {Phakathi}, {Pisano}, {Prince}, {Puddu}, {Qu}, {Rivera}, {Robertson}, {Rojas}, {Salatino}, {Schaan}, {Schillaci}, {Sehgal}, {Sherwin}, {Sierra}, {Sievers}, {Sifon}, {Sikhosana}, {Simon}, {Spergel}, {Staggs}, {Stevens}, {Storer}, {Sunder}, {Switzer}, {Thorne}, {Thornton}, {Trac}, {Treu}, {Tucker}, {Vale}, {Van Engelen}, {Van Lanen}, {Vavagiakis}, {Wagoner}, {Wang}, {Ward}, {Wollack}, {Xu}, {Zago}, \& {Zhu}}]{aiola20}
{Aiola}, S., {Calabrese}, E., {Maurin}, L., {et~al.} 2020, \jcap, 2020, 047

\bibitem[{Balogh {et~al.}(1999)Balogh, Morris, Yee, Carlberg, \& Ellingson}]{balogh99}
Balogh, M.~L., Morris, S.~L., Yee, H. K.~C., Carlberg, R.~G., \& Ellingson, E. 1999, ApJ, 527, 54

\bibitem[{{Beutler} {et~al.}(2011){Beutler}, {Blake}, {Colless}, {Jones}, {Staveley-Smith}, {Campbell}, {Parker}, {Saunders}, \& {Watson}}]{BEUTLER11}
{Beutler}, F., {Blake}, C., {Colless}, M., {et~al.} 2011, \mnras, 416, 3017

\bibitem[{Borghi {et~al.}(2022{\natexlab{a}})Borghi, Moresco, \& Cimatti}]{borghi22b}
Borghi, N., Moresco, M., \& Cimatti, A. 2022{\natexlab{a}}, ApJL, 928, L4

\bibitem[{Borghi {et~al.}(2022{\natexlab{b}})Borghi, Moresco, Cimatti, Huchet, Quai, \& Pozzetti}]{borghi22a}
Borghi, N., Moresco, M., Cimatti, A., {et~al.} 2022{\natexlab{b}}, ApJ, 927, 164

\bibitem[{{Bosi} {et~al.}(2025){Bosi}, {Lapi}, {Boco}, {Alonso-Alvarez}, {Muniz-Cueli}, {Antinozzi}, {Behiri}, {Giulietti}, {Massardi}, {Spera}, {Bressan}, {Baccigalupi}, \& {Danese}}]{bosi25}
{Bosi}, M., {Lapi}, A., {Boco}, L., {et~al.} 2025, arXiv e-prints, arXiv:2503.22543

\bibitem[{Brinchmann {et~al.}(2004)Brinchmann, Charlot, White, Tremonti, Kauffmann, Heckman, \& Brinkmann}]{brinchmann04}
Brinchmann, J., Charlot, S., White, S. D.~M., {et~al.} 2004, MNRAS, 351, 1151

\bibitem[{{Brout} {et~al.}(2022){Brout}, {Scolnic}, {Popovic}, {Riess}, {Carr}, {Zuntz}, {Kessler}, {Davis}, {Hinton}, {Jones}, {Kenworthy}, {Peterson}, {Said}, {Taylor}, {Ali}, {Armstrong}, {Charvu}, {Dwomoh}, {Meldorf}, {Palmese}, {Qu}, {Rose}, {Sanchez}, {Stubbs}, {Vincenzi}, {Wood}, {Brown}, {Chen}, {Chambers}, {Coulter}, {Dai}, {Dimitriadis}, {Filippenko}, {Foley}, {Jha}, {Kelsey}, {Kirshner}, {M{\"o}ller}, {Muir}, {Nadathur}, {Pan}, {Rest}, {Rojas-Bravo}, {Sako}, {Siebert}, {Smith}, {Stahl}, \& {Wiseman}}]{BROUT22}
{Brout}, D., {Scolnic}, D., {Popovic}, B., {et~al.} 2022, \apj, 938, 110

\bibitem[{{Bruzual} \& {Charlot}(2003)}]{BRUZUAL03}
{Bruzual}, G. \& {Charlot}, S. 2003, \mnras, 344, 1000

\bibitem[{Capozziello {et~al.}(2011)Capozziello, Lazkoz, \& Salzano}]{capozziello12}
Capozziello, S., Lazkoz, R., \& Salzano, V. 2011, Phys. Rev. D, 84, 124061

\bibitem[{{Cappellari} {et~al.}(2006){Cappellari}, {Bacon}, {Bureau}, {Damen}, {Davies}, {de Zeeuw}, {Emsellem}, {Falc{\'o}n-Barroso}, {Krajnovi{\'c}}, {Kuntschner}, {McDermid}, {Peletier}, {Sarzi}, {van den Bosch}, \& {van de Ven}}]{cappellari06}
{Cappellari}, M., {Bacon}, R., {Bureau}, M., {et~al.} 2006, \mnras, 366, 1126

\bibitem[{{Carnall} {et~al.}(2018){Carnall}, {McLure}, {Dunlop}, \& {Dav{\'e}}}]{CARNALL18}
{Carnall}, A.~C., {McLure}, R.~J., {Dunlop}, J.~S., \& {Dav{\'e}}, R. 2018, \mnras, 480, 4379

\bibitem[{{Carollo} \& {Danziger}(1994)}]{carollo94}
{Carollo}, C.~M. \& {Danziger}, I.~J. 1994, \mnras, 270, 523

\bibitem[{Carretero {et~al.}(2007)Carretero, Vazdekis, \& Beckman}]{carretero07}
Carretero, C., Vazdekis, A., \& Beckman, J.~E. 2007, MNRAS, 375, 1025

\bibitem[{Carson \& Nichol(2010)}]{carson10}
Carson, D.~P. \& Nichol, R.~C. 2010, MNRAS, 408, 213

\bibitem[{Cassisi {et~al.}(1997)Cassisi, Castellani, \& Castellani}]{cassisi97}
Cassisi, S., Castellani, M., \& Castellani, V. 1997, A\&A, 317, 108

\bibitem[{Chabrier(2003)}]{chabrier03}
Chabrier, G. 2003, PASP, 115, 763

\bibitem[{{Chevallard} \& {Charlot}(2016)}]{CHEVALLARD16}
{Chevallard}, J. \& {Charlot}, S. 2016, \mnras, 462, 1415

\bibitem[{Chilingarian {et~al.}(2010)Chilingarian, Melchior, \& Zolotukhin}]{chilingarian10}
Chilingarian, I.~V., Melchior, A.-L., \& Zolotukhin, I.~Y. 2010, MNRAS, 405, 1409

\bibitem[{{Chung} {et~al.}(2013){Chung}, {Yoon}, {Lee}, \& {Lee}}]{chung13}
{Chung}, C., {Yoon}, S.-J., {Lee}, S.-Y., \& {Lee}, Y.-W. 2013, \apjs, 204, 3

\bibitem[{{Cid Fernandes} {et~al.}(2005){Cid Fernandes}, {Mateus}, {Sodr{\'e}}, {Stasi{\'n}ska}, \& {Gomes}}]{CID05}
{Cid Fernandes}, R., {Mateus}, A., {Sodr{\'e}}, L., {Stasi{\'n}ska}, G., \& {Gomes}, J.~M. 2005, \mnras, 358, 363

\bibitem[{Cimatti {et~al.}(2006)Cimatti, Daddi, \& Renzini}]{cimatti06}
Cimatti, A., Daddi, E., \& Renzini, A. 2006, A\&A, 453, L29

\bibitem[{Cincunegui {et~al.}(2007)Cincunegui, Díaz, \& Mauas}]{cincunegui07}
Cincunegui, C., Díaz, R., \& Mauas, P. 2007, A\&A, 469, 309

\bibitem[{Clemens {et~al.}(2006)Clemens, Bressan, Nikolic, Alexander, \& Annibali}]{clemens06}
Clemens, M.~S., Bressan, A., Nikolic, B., Alexander, P., \& Annibali, F. 2006, MNRAS, 702

\bibitem[{{Clemens} {et~al.}(2009){Clemens}, {Bressan}, {Nikolic}, \& {Rampazzo}}]{clemens09}
{Clemens}, M.~S., {Bressan}, A., {Nikolic}, B., \& {Rampazzo}, R. 2009, MNRAS, 392, L35

\bibitem[{{Conroy} {et~al.}(2014){Conroy}, {Graves}, \& {van Dokkum}}]{CONROY14}
{Conroy}, C., {Graves}, G.~J., \& {van Dokkum}, P.~G. 2014, \apj, 780, 33

\bibitem[{{Conroy} \& {van Dokkum}(2012)}]{conroy12}
{Conroy}, C. \& {van Dokkum}, P. 2012, \apj, 747, 69

\bibitem[{{Conroy} {et~al.}(2018){Conroy}, {Villaume}, {van Dokkum}, \& {Lind}}]{conroy18}
{Conroy}, C., {Villaume}, A., {van Dokkum}, P.~G., \& {Lind}, K. 2018, \apj, 854, 139

\bibitem[{{Cowie} {et~al.}(1996){Cowie}, {Songaila}, {Hu}, \& {Cohen}}]{COWIE96}
{Cowie}, L.~L., {Songaila}, A., {Hu}, E.~M., \& {Cohen}, J.~G. 1996, \aj, 112, 839

\bibitem[{{Dark Energy Survey and Kilo-Degree Survey Collaboration} {et~al.}(2023){Dark Energy Survey and Kilo-Degree Survey Collaboration}, {Abbott}, {Aguena}, {Alarcon}, {Alves}, {Amon}, {Andrade-Oliveira}, {Asgari}, {Avila}, {Bacon}, {Bechtol}, {Becker}, {Bernstein}, {Bertin}, {Bilicki}, {Blazek}, {Bocquet}, {Brooks}, {Burger}, {Burke}, {Camacho}, {Campos}, {Carnero Rosell}, {Carrasco Kind}, {Carretero}, {Castander}, {Cawthon}, {Chang}, {Chen}, {Choi}, {Conselice}, {Cordero}, {Crocce}, {da Costa}, {da Silva Pereira}, {Dalal}, {Davis}, {de Jong}, {DeRose}, {Desai}, {Diehl}, {Dodelson}, {Doel}, {Doux}, {Drlica-Wagner}, {Dvornik}, {Eckert}, {Eifler}, {Elvin-Poole}, {Everett}, {Fang}, {Ferrero}, {Fert{\'e}}, {Flaugher}, {Friedrich}, {Frieman}, {Garc{\'\i}a-Bellido}, {Gatti}, {Giannini}, {Giblin}, {Gruen}, {Gruendl}, {Gutierrez}, {Harrison}, {Hartley}, {Herner}, {Heymans}, {Hildebrandt}, {Hinton}, {Hoekstra}, {Hollowood}, {Honscheid}, {Huang}, {Huff}, {Huterer}, {James}, {Jarvis}, {Jeffrey}, {Jeltema},
  {Joachimi}, {Joudaki}, {Kannawadi}, {Krause}, {Kuehn}, {Kuijken}, {Kuropatkin}, {Lahav}, {Leget}, {Lemos}, {Li}, {Li}, {Liddle}, {Lima}, {Lin}, {Lin}, {MacCrann}, {Mahony}, {Marshall}, {McCullough}, {Mena-Fern{\'a}ndez}, {Menanteau}, {Miquel}, {Mohr}, {Muir}, {Myles}, {Napolitano}, {Navarro-Alsina}, {Ogando}, {Palmese}, {Pandey}, {Park}, {Paterno}, {Peacock}, {Petravick}, {Pieres}, {Plazas Malag{\'o}n}, {Porredon}, {Prat}, {Radovich}, {Raveri}, {Reischke}, {Robertson}, {Rollins}, {Romer}, {Roodman}, {Rykoff}, {Samuroff}, {S{\'a}nchez}, {Sanchez}, {Sanchez}, {Schneider}, {Secco}, {Sevilla-Noarbe}, {Shan}, {Sheldon}, {Shin}, {Sif{\'o}n}, {Smith}, {Soares-Santos}, {St{\"o}lzner}, {Suchyta}, {Swanson}, {Tarle}, {Thomas}, {To}, {Troxel}, {Tr{\"o}ster}, {Tutusaus}, {van den Busch}, {Varga}, {Walker}, {Weaverdyck}, {Wechsler}, {Weller}, {Wiseman}, {Wright}, {Yanny}, {Yin}, {Yoon}, {Zhang}, \& {Zuntz}}]{DES+KIDS}
{Dark Energy Survey and Kilo-Degree Survey Collaboration}, {Abbott}, T.~M.~C., {Aguena}, M., {et~al.} 2023, The Open Journal of Astrophysics, 6, 36

\bibitem[{{Davies} {et~al.}(1993){Davies}, {Sadler}, \& {Peletier}}]{DAVIES93}
{Davies}, R.~L., {Sadler}, E.~M., \& {Peletier}, R.~F. 1993, \mnras, 262, 650

\bibitem[{{Di Teodoro} {et~al.}(2016){Di Teodoro}, {Fraternali}, \& {Miller}}]{diteodoro16}
{Di Teodoro}, E.~M., {Fraternali}, F., \& {Miller}, S.~H. 2016, \aap, 594, A77

\bibitem[{{Di Valentino} {et~al.}(2021){Di Valentino}, {Mena}, {Pan}, {Visinelli}, {Yang}, {Melchiorri}, {Mota}, {Riess}, \& {Silk}}]{DIVALENTINO21}
{Di Valentino}, E., {Mena}, O., {Pan}, S., {et~al.} 2021, Classical and Quantum Gravity, 38, 153001

\bibitem[{{Dutcher} {et~al.}(2021){Dutcher}, {Balkenhol}, {Ade}, {Ahmed}, {Anderes}, {Anderson}, {Archipley}, {Avva}, {Aylor}, {Barry}, {Basu Thakur}, {Benabed}, {Bender}, {Benson}, {Bianchini}, {Bleem}, {Bouchet}, {Bryant}, {Byrum}, {Carlstrom}, {Carter}, {Cecil}, {Chang}, {Chaubal}, {Chen}, {Cho}, {Chou}, {Cliche}, {Crawford}, {Cukierman}, {Daley}, {de Haan}, {Denison}, {Dibert}, {Ding}, {Dobbs}, {Everett}, {Feng}, {Ferguson}, {Foster}, {Fu}, {Galli}, {Gambrel}, {Gardner}, {Goeckner-Wald}, {Gualtieri}, {Guns}, {Gupta}, {Guyser}, {Halverson}, {Harke-Hosemann}, {Harrington}, {Henning}, {Hilton}, {Hivon}, {Holder}, {Holzapfel}, {Hood}, {Howe}, {Huang}, {Irwin}, {Jeong}, {Jonas}, {Jones}, {Khaire}, {Knox}, {Kofman}, {Korman}, {Kubik}, {Kuhlmann}, {Kuo}, {Lee}, {Leitch}, {Lowitz}, {Lu}, {Meyer}, {Michalik}, {Millea}, {Montgomery}, {Nadolski}, {Natoli}, {Nguyen}, {Noble}, {Novosad}, {Omori}, {Padin}, {Pan}, {Paschos}, {Pearson}, {Posada}, {Prabhu}, {Quan}, {Raghunathan}, {Rahlin}, {Reichardt}, {Riebel}, {Riedel},
  {Rouble}, {Ruhl}, {Sayre}, {Schiappucci}, {Shirokoff}, {Smecher}, {Sobrin}, {Stark}, {Stephen}, {Story}, {Suzuki}, {Thompson}, {Thorne}, {Tucker}, {Umilta}, {Vale}, {Vanderlinde}, {Vieira}, {Wang}, {Whitehorn}, {Wu}, {Yefremenko}, {Yoon}, {Young}, \& {SPT-3G Collaboration}}]{DUTCHER21}
{Dutcher}, D., {Balkenhol}, L., {Ade}, P.~A.~R., {et~al.} 2021, \prd, 104, 022003

\bibitem[{{Dvornik} {et~al.}(2023){Dvornik}, {Heymans}, {Asgari}, {Mahony}, {Joachimi}, {Bilicki}, {Chisari}, {Hildebrandt}, {Hoekstra}, {Johnston}, {Kuijken}, {Mead}, {Miyatake}, {Nishimichi}, {Reischke}, {Unruh}, \& {Wright}}]{DVORNIK23}
{Dvornik}, A., {Heymans}, C., {Asgari}, M., {et~al.} 2023, \aap, 675, A189

\bibitem[{Eisenstein {et~al.}(2011)Eisenstein, Weinberg, Agol, Aihara, Allende~Prieto, Anderson, Arns, Aubourg, Bailey, Balbinot, Barkhouser, Beers, Berlind, Bickerton, Bizyaev, Blanton, Bochanski, Bolton, Bosman, Bovy, Brandt, Breslauer, Brewington, Brinkmann, Brown, Brownstein, Burger, Busca, Campbell, Cargile, Carithers, Carlberg, Carr, Chang, Chen, Chiappini, Comparat, Connolly, Cortes, Croft, Cunha, da~Costa, Davenport, Dawson, De~Lee, Porto~de Mello, de~Simoni, Dean, Dhital, Ealet, Ebelke, Edmondson, Eiting, Escoffier, Esposito, Evans, Fan, Femenía~Castellá, Dutra~Ferreira, Fitzgerald, Fleming, Font-Ribera, Ford, Frinchaboy, García~Pérez, Gaudi, Ge, Ghezzi, Gillespie, Gilmore, Girardi, Gott, Gould, Grebel, Gunn, Hamilton, Harding, Harris, Hawley, Hearty, Hennawi, González~Hernández, Ho, Hogg, Holtzman, Honscheid, Inada, Ivans, Jiang, Jiang, Johnson, Jordan, Jordan, Kauffmann, Kazin, Kirkby, Klaene, Knapp, Kneib, Kochanek, Koesterke, Kollmeier, Kron, Lampeitl, Lang, Lawler, Le~Goff, Lee, Lee,
  Leisenring, Lin, Liu, Long, Loomis, Lucatello, Lundgren, Lupton, Ma, Ma, MacDonald, Mack, Mahadevan, Maia, Majewski, Makler, Malanushenko, Malanushenko, Mandelbaum, Maraston, Margala, Maseman, Masters, McBride, McDonald, McGreer, McMahon, Mena~Requejo, Ménard, Miralda-Escudé, Morrison, Mullally, Muna, Murayama, Myers, Naugle, Neto, Nguyen, Nichol, Nidever, O’Connell, Ogando, Olmstead, Oravetz, Padmanabhan, Paegert, Palanque-Delabrouille, Pan, Pandey, Parejko, Pâris, Pellegrini, Pepper, Percival, Petitjean, Pfaffenberger, Pforr, Phleps, Pichon, Pieri, Prada, Price-Whelan, Raddick, Ramos, Reid, Reyle, Rich, Richards, Rieke, Rieke, Rix, Robin, Rocha-Pinto, Rockosi, Roe, Rollinde, Ross, Ross, Rossetto, Sánchez, Santiago, Sayres, Schiavon, Schlegel, Schlesinger, Schmidt, Schneider, Sellgren, Shelden, Sheldon, Shetrone, Shu, Silverman, Simmerer, Simmons, Sivarani, Skrutskie, Slosar, Smee, Smith, Snedden, Stassun, Steele, Steinmetz, Stockett, Stollberg, Strauss, Szalay, Tanaka, Thakar, Thomas, Tinker,
  Tofflemire, Tojeiro, Tremonti, Vargas~Magaña, Verde, Vogt, Wake, Wan, Wang, Weaver, White, White, Wilson, Wisniewski, Wood-Vasey, Yanny, Yasuda, Yèche, York, Young, Zasowski, Zehavi, \& Zhao}]{eisenstein11}
Eisenstein, D.~J., Weinberg, D.~H., Agol, E., {et~al.} 2011, AJ, 142, 72

\bibitem[{{Eisenstein} {et~al.}(2005){Eisenstein}, {Zehavi}, {Hogg}, {Scoccimarro}, {Blanton}, {Nichol}, {Scranton}, {Seo}, {Tegmark}, {Zheng}, {Anderson}, {Annis}, {Bahcall}, {Brinkmann}, {Burles}, {Castander}, {Connolly}, {Csabai}, {Doi}, {Fukugita}, {Frieman}, {Glazebrook}, {Gunn}, {Hendry}, {Hennessy}, {Ivezi{\'c}}, {Kent}, {Knapp}, {Lin}, {Loh}, {Lupton}, {Margon}, {McKay}, {Meiksin}, {Munn}, {Pope}, {Richmond}, {Schlegel}, {Schneider}, {Shimasaku}, {Stoughton}, {Strauss}, {SubbaRao}, {Szalay}, {Szapudi}, {Tucker}, {Yanny}, \& {York}}]{EISENSTEIN05}
{Eisenstein}, D.~J., {Zehavi}, I., {Hogg}, D.~W., {et~al.} 2005, \apj, 633, 560

\bibitem[{Ferreras {et~al.}(2019)Ferreras, Scott, Barbera, Croom, van~de Sande, Hopkins, Colless, Barone, d′Eugenio, Bland-Hawthorn, Brough, Bryant, Konstantopoulos, Lagos, Lawrence, López-Sánchez, Medling, Owers, \& Richards}]{ferreras19}
Ferreras, I., Scott, N., Barbera, F.~L., {et~al.} 2019, MNRAS

\bibitem[{Foreman-Mackey {et~al.}(2013)Foreman-Mackey, Hogg, Lang, {et~al.}}]{FOREMAN13}
Foreman-Mackey, D., Hogg, D.~W., Lang, D., {et~al.} 2013, Publications of the Astronomical Society of the Pacific, 125, 306

\bibitem[{{Gallazzi} {et~al.}(2005){Gallazzi}, {Charlot}, {Brinchmann}, {White}, \& {Tremonti}}]{GALLAZZI05}
{Gallazzi}, A., {Charlot}, S., {Brinchmann}, J., {White}, S. D.~M., \& {Tremonti}, C.~A. 2005, \mnras, 362, 41

\bibitem[{Goodman \& Weare(2010)}]{GOODMAN10}
Goodman, J. \& Weare, J. 2010, Communications in Applied Mathematics and Computational Science, 5, 65

\bibitem[{Hamilton(1985)}]{hamilton85}
Hamilton, D. 1985, ApJ, 297, 371

\bibitem[{{Heavens} {et~al.}(2000){Heavens}, {Jimenez}, \& {Lahav}}]{HEAVENS00}
{Heavens}, A.~F., {Jimenez}, R., \& {Lahav}, O. 2000, \mnras, 317, 965

\bibitem[{{Hoaglin} {et~al.}(1983){Hoaglin}, {Mosteller}, \& {Tukey}}]{HOAGLIN83}
{Hoaglin}, D.~C., {Mosteller}, F., \& {Tukey}, J.~W. 1983, {Understanding robust and exploratory data anlysis}

\bibitem[{Holden {et~al.}(2012)Holden, van~der Wel, Rix, \& Franx}]{holden12}
Holden, B.~P., van~der Wel, A., Rix, H.-W., \& Franx, M. 2012, ApJ, 749, 96

\bibitem[{{Ilbert, O.} {et~al.}(2013){Ilbert, O.}, {McCracken, H. J.}, {Le Fèvre, O.}, {Capak, P.}, {Dunlop, J.}, {Karim, A.}, {Renzini, M. A.}, {Caputi, K.}, {Boissier, S.}, {Arnouts, S.}, {Aussel, H.}, {Comparat, J.}, {Guo, Q.}, {Hudelot, P.}, {Kartaltepe, J.}, {Kneib, J. P.}, {Krogager, J. K.}, {Le Floc’h, E.}, {Lilly, S.}, {Mellier, Y.}, {Milvang-Jensen, B.}, {Moutard, T.}, {Onodera, M.}, {Richard, J.}, {Salvato, M.}, {Sanders, D. B.}, {Scoville, N.}, {Silverman, J. D.}, {Taniguchi, Y.}, {Tasca, L.}, {Thomas, R.}, {Toft, S.}, {Tresse, L.}, {Vergani, D.}, {Wolk, M.}, \& {Zirm, A.}}]{ilbert13}
{Ilbert, O.}, {McCracken, H. J.}, {Le Fèvre, O.}, {et~al.} 2013, A\&A, 556, A55

\bibitem[{Jiao {et~al.}(2022)Jiao, Borghi, Moresco, \& T.-J.}]{jiao22}
Jiao, K., Borghi, N., Moresco, M., \& T.-J., Z. 2022, ApJS, 265, 48

\bibitem[{Jimenez \& Loeb(2002)}]{jimenez02}
Jimenez, R. \& Loeb, A. 2002, ApJ, 573, 37

\bibitem[{Johansson {et~al.}(2012)Johansson, Thomas, \& Maraston}]{johansson12}
Johansson, J., Thomas, D., \& Maraston, C. 2012, MNRAS, 421, 1908

\bibitem[{{J{\o}rgensen}(1999)}]{jorgensen99}
{J{\o}rgensen}, I. 1999, \mnras, 306, 607

\bibitem[{{Kauffmann} {et~al.}(2003){Kauffmann}, {Heckman}, {White}, {Charlot}, {Tremonti}, {Brinchmann}, {Bruzual}, {Peng}, {Seibert}, {Bernardi}, {Blanton}, {Brinkmann}, {Castander}, {Cs{\'a}bai}, {Fukugita}, {Ivezic}, {Munn}, {Nichol}, {Padmanabhan}, {Thakar}, {Weinberg}, \& {York}}]{KAUFFMANN03}
{Kauffmann}, G., {Heckman}, T.~M., {White}, S. D.~M., {et~al.} 2003, \mnras, 341, 33

\bibitem[{Khostovan {et~al.}(2020)Khostovan, Malhotra, Rhoads, Jiang, Wang, Wold, ya~Zheng, Barrientos, Coughlin, Harish, Hu, Infante, Perez, Pharo, Valdes, Walker, \& Yang}]{khostovan20}
Khostovan, A., Malhotra, S., Rhoads, J., {et~al.} 2020, MNRAS, 493, 3966

\bibitem[{Knowles {et~al.}(2021)Knowles, Sansom, Allende-Prieto, \& Vazdekis}]{knowles21}
Knowles, A.~T., Sansom, A.~E., Allende-Prieto, C., \& Vazdekis, A. 2021, MNRAS, 504, 2286

\bibitem[{{Knowles} {et~al.}(2023){Knowles}, {Sansom}, {Vazdekis}, \& {Allende Prieto}}]{knowles23}
{Knowles}, A.~T., {Sansom}, A.~E., {Vazdekis}, A., \& {Allende Prieto}, C. 2023, \mnras, 523, 3450

\bibitem[{La~Barbera {et~al.}(2013)La~Barbera, Ferreras, Vazdekis, de~la Rosa, de~Carvalho, Trevisan, Falcón-Barroso, \& Ricciardelli}]{labarbera13}
La~Barbera, F., Ferreras, I., Vazdekis, A., {et~al.} 2013, MNRAS, 433, 3017

\bibitem[{{Longhetti} {et~al.}(2000){Longhetti}, {Bressan}, {Chiosi}, \& {Rampazzo}}]{LONGHETTI00}
{Longhetti}, M., {Bressan}, A., {Chiosi}, C., \& {Rampazzo}, R. 2000, \aap, 353, 917

\bibitem[{{Loubser} {et~al.}(2025){Loubser}, {Alabi}, {Hilton}, {Ma}, {Tang}, {Hatamkhani}, {Cress}, {Skelton}, \& {Nkosi}}]{loubser25}
{Loubser}, S.~I., {Alabi}, A.~B., {Hilton}, M., {et~al.} 2025, \mnras, 540, 3135

\bibitem[{{Maiolino} \& {Mannucci}(2019)}]{maiolino19}
{Maiolino}, R. \& {Mannucci}, F. 2019, \aapr, 27, 3

\bibitem[{Maraston(2005)}]{maraston05}
Maraston, C. 2005, MNRAS, 362, 799

\bibitem[{Maraston {et~al.}(2013)Maraston, Pforr, Henriques, Thomas, Wake, Brownstein, Capozzi, Tinker, Bundy, Skibba, Beifiori, Nichol, Edmondson, Schneider, Chen, Masters, Steele, Bolton, York, Weaver, Higgs, Bizyaev, Brewington, Malanushenko, Malanushenko, Snedden, Oravetz, Pan, Shelden, \& Simmons}]{maraston13}
Maraston, C., Pforr, J., Henriques, B.~M., {et~al.} 2013, MNRAS, 435, 2764

\bibitem[{Masters {et~al.}(2011)Masters, Maraston, Nichol, Thomas, Beifiori, Bundy, Edmondson, Higgs, Leauthaud, Mandelbaum, Pforr, Ross, Ross, Schneider, Skibba, Tinker, Tojeiro, Wake, Brinkmann, \& Weaver}]{masters11}
Masters, K.~L., Maraston, C., Nichol, R.~C., {et~al.} 2011, MNRAS, 418, 1055

\bibitem[{McLure {et~al.}(2018)McLure, Pentericci, Cimatti, Dunlop, Elbaz, Fontana, Nandra, Amorin, Bolzonella, Bongiorno, Carnall, Castellano, Cirasuolo, Cucciati, Cullen, DeBarros, Finkelstein, Fontanot, Franzetti, Fumana, Gargiulo, Garilli, Guaita, Hartley, Iovino, Jarvis, Juneau, Karman, Maccagni, Marchi, Mármol-Queraltó, Pompei, Pozzetti, Scodeggio, Sommariva, Talia, Almaini, Balestra, Bardelli, Bell, Bourne, Bowler, Brusa, Buitrago, Caputi, Cassata, Charlot, Citro, Cresci, Cristiani, Curtis-Lake, Dickinson, Fazio, Ferguson, Fiore, Franco, Fynbo, Galametz, Georgakakis, Giavalisco, Grazian, Hathi, Jung, Kim, Koekemoer, Khusanova, LeFèvre, Lotz, Mannucci, Maltby, Matsuoka, McLeod, Mendez-Hernandez, Mendez-Abreu, Mignoli, Moresco, Mortlock, Nonino, Pannella, Papovich, Popesso, Rosario, Salvato, Santini, Schaerer, Schreiber, Stark, Tasca, Thomas, Treu, Vanzella, Wild, Williams, Zamorani, \& Zucca}]{mclure18}
McLure, R.~J., Pentericci, L., Cimatti, A., {et~al.} 2018, MNRAS, 479, 25

\bibitem[{{Mehlert} {et~al.}(1998){Mehlert}, {Saglia}, {Bender}, \& {Wegner}}]{MEHLERT8}
{Mehlert}, D., {Saglia}, R.~P., {Bender}, R., \& {Wegner}, G. 1998, \aap, 332, 33

\bibitem[{{Moresco} {et~al.}(2022){Moresco}, {Amati}, {Amendola}, {Birrer}, {Blakeslee}, {Cantiello}, {Cimatti}, {Darling}, {Della Valle}, {Fishbach}, {Grillo}, {Hamaus}, {Holz}, {Izzo}, {Jimenez}, {Lusso}, {Meneghetti}, {Piedipalumbo}, {Pisani}, {Pourtsidou}, {Pozzetti}, {Quartin}, {Risaliti}, {Rosati}, \& {Verde}}]{moresco22}
{Moresco}, M., {Amati}, L., {Amendola}, L., {et~al.} 2022, Living Reviews in Relativity, 25, 6

\bibitem[{Moresco {et~al.}(2011)Moresco, Cimatti, Jimenez, Pozzetti, Zamorani, Bolzonella, Dunlop, Lamareille, Mignoli, Pearce, Rosati, Stern, Verde, Zucca, Carollo, Contini, Kneib, Fèvre, Lilly, Mainieri, Renzini, Scodeggio, Balestra, Gobat, McLure, Bardelli, Bongiorno, Caputi, Cucciati, de~la Torre, de~Ravel, Franzetti, Garilli, Iovino, Kampczyk, Knobel, Kovač, Borgne, Brun, Maier, Pelló, Peng, Perez-Montero, Presotto, Silverman, Tanaka, Tasca, Tresse, Vergani, Almaini, Barnes, Bordoloi, Bradshaw, Cappi, Chuter, Cirasuolo, Coppa, Diener, Foucaud, Hartley, Kamionkowski, Koekemoer, López-Sanjuan, McCracken, Nair, Oesch, Stanford, \& Welikala}]{moresco11}
Moresco, M., Cimatti, A., Jimenez, R., {et~al.} 2011, JCAP, 2011, 045

\bibitem[{Moresco {et~al.}(2012)Moresco, Cimatti, Jimenez, Pozzetti, Zamorani, Bolzonella, Dunlop, Lamareille, Mignoli, Pearce, Rosati, Stern, Verde, Zucca, Carollo, Contini, Kneib, Fèvre, Lilly, Mainieri, Renzini, Scodeggio, Balestra, Gobat, McLure, Bardelli, Bongiorno, Caputi, Cucciati, de~la Torre, de~Ravel, Franzetti, Garilli, Iovino, Kampczyk, Knobel, Kovač, Borgne, Brun, Maier, Pelló, Peng, Perez-Montero, Presotto, Silverman, Tanaka, Tasca, Tresse, Vergani, Almaini, Barnes, Bordoloi, Bradshaw, Cappi, Chuter, Cirasuolo, Coppa, Diener, Foucaud, Hartley, Kamionkowski, Koekemoer, López-Sanjuan, McCracken, Nair, Oesch, Stanford, \& Welikala}]{moresco12}
Moresco, M., Cimatti, A., Jimenez, R., {et~al.} 2012, JCAP, 2012, 006

\bibitem[{{Moresco} {et~al.}(2020){Moresco}, {Jimenez}, {Verde}, {Cimatti}, \& {Pozzetti}}]{moresco20}
{Moresco}, M., {Jimenez}, R., {Verde}, L., {Cimatti}, A., \& {Pozzetti}, L. 2020, \apj, 898, 82

\bibitem[{Moresco {et~al.}(2018)Moresco, Jimenez, Verde, Pozzetti, Cimatti, \& Citro}]{moresco18}
Moresco, M., Jimenez, R., Verde, L., {et~al.} 2018, ApJ, 868

\bibitem[{Moresco {et~al.}(2016)Moresco, Pozzetti, Cimatti, Jimenez, Maraston, Verde, Thomas, Citro, Tojeiro, \& Wikinson}]{moresco16a}
Moresco, M., Pozzetti, L., Cimatti, A., {et~al.} 2016, JCAP, 2016, 014

\bibitem[{{O'Mill} {et~al.}(2011){O'Mill}, {Duplancic}, {Garc{\'\i}a Lambas}, \& {Sodr{\'e}}}]{omill11}
{O'Mill}, A.~L., {Duplancic}, F., {Garc{\'\i}a Lambas}, D., \& {Sodr{\'e}}, Jr., L. 2011, \mnras, 413, 1395

\bibitem[{Parikh {et~al.}(2021)Parikh, Thomas, Maraston, Westfall, Andrews, Boardman, Drory, \& Oyarzun}]{parikh21}
Parikh, T., Thomas, D., Maraston, C., {et~al.} 2021, MNRAS, 502, 5508

\bibitem[{{Perlmutter} {et~al.}(1999){Perlmutter}, {Aldering}, {Goldhaber}, {Knop}, {Nugent}, {Castro}, {Deustua}, {Fabbro}, {Goobar}, {Groom}, {Hook}, {Kim}, {Kim}, {Lee}, {Nunes}, {Pain}, {Pennypacker}, {Quimby}, {Lidman}, {Ellis}, {Irwin}, {McMahon}, {Ruiz-Lapuente}, {Walton}, {Schaefer}, {Boyle}, {Filippenko}, {Matheson}, {Fruchter}, {Panagia}, {Newberg}, {Couch}, \& {Project}}]{PERLMUTTER99}
{Perlmutter}, S., {Aldering}, G., {Goldhaber}, G., {et~al.} 1999, \apj, 517, 565

\bibitem[{{Pietrinferni} {et~al.}(2004){Pietrinferni}, {Cassisi}, {Salaris}, \& {Castelli}}]{pietrinferni04}
{Pietrinferni}, A., {Cassisi}, S., {Salaris}, M., \& {Castelli}, F. 2004, \apj, 612, 168

\bibitem[{Pietrinferni {et~al.}(2006)Pietrinferni, Cassisi, Salaris, \& Castelli}]{pietrinferni06}
Pietrinferni, A., Cassisi, S., Salaris, M., \& Castelli, F. 2006, The Astrophysical Journal, 642, 797

\bibitem[{Pietrinferni {et~al.}(2021)Pietrinferni, Hidalgo, Cassisi, Salaris, Savino, Mucciarelli, Verma, Aguirre, Aparicio, \& Ferguson}]{pietrinferni21}
Pietrinferni, A., Hidalgo, S., Cassisi, S., {et~al.} 2021, The Astrophysical Journal, 908, 102

\bibitem[{Pimbblet {et~al.}(2019)Pimbblet, Crossett, Crossett, \& Fraser-McKelvie}]{pimbblet19}
Pimbblet, K., Crossett, J., Crossett, J., \& Fraser-McKelvie, A. 2019, MNRAS

\bibitem[{{Planck Collaboration} {et~al.}(2014){Planck Collaboration}, {Ade}, {Aghanim}, {Armitage-Caplan}, {Arnaud}, {Ashdown}, {Atrio-Barandela}, {Aumont}, {Baccigalupi}, {Banday}, {Barreiro}, {Bartlett}, {Battaner}, {Benabed}, {Beno{\^\i}t}, {Benoit-L{\'e}vy}, {Bernard}, {Bersanelli}, {Bielewicz}, {Bobin}, {Bock}, {Bonaldi}, {Bond}, {Borrill}, {Bouchet}, {Bridges}, {Bucher}, {Burigana}, {Butler}, {Calabrese}, {Cappellini}, {Cardoso}, {Catalano}, {Challinor}, {Chamballu}, {Chary}, {Chen}, {Chiang}, {Chiang}, {Christensen}, {Church}, {Clements}, {Colombi}, {Colombo}, {Couchot}, {Coulais}, {Crill}, {Curto}, {Cuttaia}, {Danese}, {Davies}, {Davis}, {de Bernardis}, {de Rosa}, {de Zotti}, {Delabrouille}, {Delouis}, {D{\'e}sert}, {Dickinson}, {Diego}, {Dolag}, {Dole}, {Donzelli}, {Dor{\'e}}, {Douspis}, {Dunkley}, {Dupac}, {Efstathiou}, {Elsner}, {En{\ss}lin}, {Eriksen}, {Finelli}, {Forni}, {Frailis}, {Fraisse}, {Franceschi}, {Gaier}, {Galeotta}, {Galli}, {Ganga}, {Giard}, {Giardino}, {Giraud-H{\'e}raud},
  {Gjerl{\o}w}, {Gonz{\'a}lez-Nuevo}, {G{\'o}rski}, {Gratton}, {Gregorio}, {Gruppuso}, {Gudmundsson}, {Haissinski}, {Hamann}, {Hansen}, {Hanson}, {Harrison}, {Henrot-Versill{\'e}}, {Hern{\'a}ndez-Monteagudo}, {Herranz}, {Hildebrandt}, {Hivon}, {Hobson}, {Holmes}, {Hornstrup}, {Hou}, {Hovest}, {Huffenberger}, {Jaffe}, {Jaffe}, {Jewell}, {Jones}, {Juvela}, {Keih{\"a}nen}, {Keskitalo}, {Kisner}, {Kneissl}, {Knoche}, {Knox}, {Kunz}, {Kurki-Suonio}, {Lagache}, {L{\"a}hteenm{\"a}ki}, {Lamarre}, {Lasenby}, {Lattanzi}, {Laureijs}, {Lawrence}, {Leach}, {Leahy}, {Leonardi}, {Le{\'o}n-Tavares}, {Lesgourgues}, {Lewis}, {Liguori}, {Lilje}, {Linden-V{\o}rnle}, {L{\'o}pez-Caniego}, {Lubin}, {Mac{\'\i}as-P{\'e}rez}, {Maffei}, {Maino}, {Mandolesi}, {Maris}, {Marshall}, {Martin}, {Mart{\'\i}nez-Gonz{\'a}lez}, {Masi}, {Massardi}, {Matarrese}, {Matthai}, {Mazzotta}, {Meinhold}, {Melchiorri}, {Melin}, {Mendes}, {Menegoni}, {Mennella}, {Migliaccio}, {Millea}, {Mitra}, {Miville-Desch{\^e}nes}, {Moneti}, {Montier}, {Morgante},
  {Mortlock}, {Moss}, {Munshi}, {Murphy}, {Naselsky}, {Nati}, {Natoli}, {Netterfield}, {N{\o}rgaard-Nielsen}, {Noviello}, {Novikov}, {Novikov}, {O'Dwyer}, {Osborne}, {Oxborrow}, {Paci}, {Pagano}, {Pajot}, {Paladini}, {Paoletti}, {Partridge}, {Pasian}, {Patanchon}, {Pearson}, {Pearson}, {Peiris}, {Perdereau}, {Perotto}, {Perrotta}, {Pettorino}, {Piacentini}, {Piat}, {Pierpaoli}, {Pietrobon}, {Plaszczynski}, {Platania}, \& {Pointecouteau}}]{PLANCK14}
{Planck Collaboration}, {Ade}, P.~A.~R., {Aghanim}, N., {et~al.} 2014, \aap, 571, A16

\bibitem[{{Planck Collaboration} {et~al.}(2020){Planck Collaboration}, {Aghanim, N.}, {Akrami, Y.}, {Ashdown, M.}, {Aumont, J.}, {Baccigalupi, C.}, {Ballardini, M.}, {Banday, A. J.}, {Barreiro, R. B.}, {Bartolo, N.}, {Basak, S.}, {Battye, R.}, {Benabed, K.}, {Bernard, J.-P.}, {Bersanelli, M.}, {Bielewicz, P.}, {Bock, J. J.}, {Bond, J. R.}, {Borrill, J.}, {Bouchet, F. R.}, {Boulanger, F.}, {Bucher, M.}, {Burigana, C.}, {Butler, R. C.}, {Calabrese, E.}, {Cardoso, J.-F.}, {Carron, J.}, {Challinor, A.}, {Chiang, H. C.}, {Chluba, J.}, {Colombo, L. P. L.}, {Combet, C.}, {Contreras, D.}, {Crill, B. P.}, {Cuttaia, F.}, {de Bernardis, P.}, {de Zotti, G.}, {Delabrouille, J.}, {Delouis, J.-M.}, {Di Valentino, E.}, {Diego, J. M.}, {Doré, O.}, {Douspis, M.}, {Ducout, A.}, {Dupac, X.}, {Dusini, S.}, {Efstathiou, G.}, {Elsner, F.}, {Enßlin, T. A.}, {Eriksen, H. K.}, {Fantaye, Y.}, {Farhang, M.}, {Fergusson, J.}, {Fernandez-Cobos, R.}, {Finelli, F.}, {Forastieri, F.}, {Frailis, M.}, {Fraisse, A. A.}, {Franceschi, E.},
  {Frolov, A.}, {Galeotta, S.}, {Galli, S.}, {Ganga, K.}, {Génova-Santos, R. T.}, {Gerbino, M.}, {Ghosh, T.}, {González-Nuevo, J.}, {Górski, K. M.}, {Gratton, S.}, {Gruppuso, A.}, {Gudmundsson, J. E.}, {Hamann, J.}, {Handley, W.}, {Hansen, F. K.}, {Herranz, D.}, {Hildebrandt, S. R.}, {Hivon, E.}, {Huang, Z.}, {Jaffe, A. H.}, {Jones, W. C.}, {Karakci, A.}, {Keihänen, E.}, {Keskitalo, R.}, {Kiiveri, K.}, {Kim, J.}, {Kisner, T. S.}, {Knox, L.}, {Krachmalnicoff, N.}, {Kunz, M.}, {Kurki-Suonio, H.}, {Lagache, G.}, {Lamarre, J.-M.}, {Lasenby, A.}, {Lattanzi, M.}, {Lawrence, C. R.}, {Le Jeune, M.}, {Lemos, P.}, {Lesgourgues, J.}, {Levrier, F.}, {Lewis, A.}, {Liguori, M.}, {Lilje, P. B.}, {Lilley, M.}, {Lindholm, V.}, {López-Caniego, M.}, {Lubin, P. M.}, {Ma, Y.-Z.}, {Macías-Pérez, J. F.}, {Maggio, G.}, {Maino, D.}, {Mandolesi, N.}, {Mangilli, A.}, {Marcos-Caballero, A.}, {Maris, M.}, {Martin, P. G.}, {Martinelli, M.}, {Martínez-González, E.}, {Matarrese, S.}, {Mauri, N.}, {McEwen, J. D.}, {Meinhold, P. R.},
  {Melchiorri, A.}, {Mennella, A.}, {Migliaccio, M.}, {Millea, M.}, {Mitra, S.}, {Miville-Deschênes, M.-A.}, {Molinari, D.}, {Montier, L.}, {Morgante, G.}, {Moss, A.}, {Natoli, P.}, {Nørgaard-Nielsen, H. U.}, {Pagano, L.}, {Paoletti, D.}, {Partridge, B.}, {Patanchon, G.}, {Peiris, H. V.}, {Perrotta, F.}, {Pettorino, V.}, {Piacentini, F.}, {Polastri, L.}, {Polenta, G.}, {Puget, J.-L.}, {Rachen, J. P.}, {Reinecke, M.}, {Remazeilles, M.}, {Renzi, A.}, {Rocha, G.}, {Rosset, C.}, {Roudier, G.}, {Rubiño-Martín, J. A.}, {Ruiz-Granados, B.}, {Salvati, L.}, {Sandri, M.}, {Savelainen, M.}, {Scott, D.}, {Shellard, E. P. S.}, {Sirignano, C.}, {Sirri, G.}, {Spencer, L. D.}, {Sunyaev, R.}, {Suur-Uski, A.-S.}, {Tauber, J. A.}, {Tavagnacco, D.}, {Tenti, M.}, {Toffolatti, L.}, {Tomasi, M.}, {Trombetti, T.}, {Valenziano, L.}, {Valiviita, J.}, {Van Tent, B.}, {Vibert, L.}, {Vielva, P.}, {Villa, F.}, {Vittorio, N.}, {Wandelt, B. D.}, {Wehus, I. K.}, {White, M.}, {White, S. D. M.}, {Zacchei, A.}, \& {Zonca, A.}}]{planck18}
{Planck Collaboration}, {Aghanim, N.}, {Akrami, Y.}, {et~al.} 2020, A\&A, 641, A6

\bibitem[{Ratsimbazafy {et~al.}(2017)Ratsimbazafy, Loubser, Crawford, Cress, Bassett, Nichol, \& Väisänen}]{ratsimbazafy17}
Ratsimbazafy, A.~L., Loubser, S.~I., Crawford, S.~M., {et~al.} 2017, MNRAS, 467, 3239

\bibitem[{{Riess} {et~al.}(2022){Riess}, {Yuan}, {Macri}, {Scolnic}, {Brout}, {Casertano}, {Jones}, {Murakami}, {Anand}, {Breuval}, {Brink}, {Filippenko}, {Hoffmann}, {Jha}, {D'arcy Kenworthy}, {Mackenty}, {Stahl}, \& {Zheng}}]{RIESS22}
{Riess}, A.~G., {Yuan}, W., {Macri}, L.~M., {et~al.} 2022, \apjl, 934, L7

\bibitem[{{Ross} {et~al.}(2015){Ross}, {Samushia}, {Howlett}, {Percival}, {Burden}, \& {Manera}}]{ROSS15}
{Ross}, A.~J., {Samushia}, L., {Howlett}, C., {et~al.} 2015, \mnras, 449, 835

\bibitem[{Salpeter(1955)}]{salpeter55}
Salpeter, E.~E. 1955, ApJ, 121, 161

\bibitem[{Saracco {et~al.}(2023)Saracco, Barbera, de~Propris, Bevacqua, Marchesini, de~Lucia, Fontanot, Hirschmann, Nonino, Pasquali, Spiniello, \& Tortora}]{saracco23}
Saracco, P., Barbera, F.~L., de~Propris, R., {et~al.} 2023, MNRAS

\bibitem[{Sargent {et~al.}(2015)Sargent, Daddi, Bournaud, Onodera, Feruglio, Martig, Gobat, Dannerbauer, \& Schinnerer}]{sargent15}
Sargent, M.~T., Daddi, E., Bournaud, F., {et~al.} 2015, ApJL, 806, L20

\bibitem[{{Schiavon}(2007)}]{schiavon07}
{Schiavon}, R.~P. 2007, \apjs, 171, 146

\bibitem[{{Schiavon} {et~al.}(2002){Schiavon}, {Faber}, {Castilho}, \& {Rose}}]{schiavon02}
{Schiavon}, R.~P., {Faber}, S.~M., {Castilho}, B.~V., \& {Rose}, J.~A. 2002, \apj, 580, 850

\bibitem[{{Scolnic} {et~al.}(2018){Scolnic}, {Jones}, {Rest}, {Pan}, {Chornock}, {Foley}, {Huber}, {Kessler}, {Narayan}, {Riess}, {Rodney}, {Berger}, {Brout}, {Challis}, {Drout}, {Finkbeiner}, {Lunnan}, {Kirshner}, {Sanders}, {Schlafly}, {Smartt}, {Stubbs}, {Tonry}, {Wood-Vasey}, {Foley}, {Hand}, {Johnson}, {Burgett}, {Chambers}, {Draper}, {Hodapp}, {Kaiser}, {Kudritzki}, {Magnier}, {Metcalfe}, {Bresolin}, {Gall}, {Kotak}, {McCrum}, \& {Smith}}]{SCOLNIC18}
{Scolnic}, D.~M., {Jones}, D.~O., {Rest}, A., {et~al.} 2018, \apj, 859, 101

\bibitem[{Shimasaku {et~al.}(2001)Shimasaku, Fukugita, Doi, Hamabe, Ichikawa, Okamura, Sekiguchi, Yasuda, Brinkmann, Csabai, Ichikawa, Ivezić, Kunszt, Schneider, Szokoly, Watanabe, \& York}]{shimasaku01}
Shimasaku, K., Fukugita, M., Doi, M., {et~al.} 2001, ApJ, 122, 1238

\bibitem[{Simon {et~al.}(2005)Simon, Verde, \& Jimenez}]{simon05}
Simon, J., Verde, L., \& Jimenez, R. 2005, Phys. Rev. D, 71, 123001

\bibitem[{Sobral {et~al.}(2015)Sobral, Matthee, Best, Smail, Khostovan, Milvang-Jensen, Kim, Stott, Calhau, Nayyeri, \& Mobasher}]{sobral15}
Sobral, D., Matthee, J., Best, P.~N., {et~al.} 2015, MNRAS, 451, 2303

\bibitem[{Stern {et~al.}(2010)Stern, Jimenez, Verde, Kamionkowski, \& Stanford}]{stern10}
Stern, D., Jimenez, R., Verde, L., Kamionkowski, M., \& Stanford, S.~A. 2010, JCAP, 2010, 008

\bibitem[{{Stoughton} {et~al.}(2002){Stoughton}, {Lupton}, {Bernardi}, {Blanton}, {Burles}, {Castander}, {Connolly}, {Eisenstein}, {Frieman}, {Hennessy}, {Hindsley}, {Ivezi{\'c}}, {Kent}, {Kunszt}, {Lee}, {Meiksin}, {Munn}, {Newberg}, {Nichol}, {Nicinski}, {Pier}, {Richards}, {Richmond}, {Schlegel}, {Smith}, {Strauss}, {SubbaRao}, {Szalay}, {Thakar}, {Tucker}, {Vanden Berk}, {Yanny}, {Adelman}, {Anderson}, {Anderson}, {Annis}, {Bahcall}, {Bakken}, {Bartelmann}, {Bastian}, {Bauer}, {Berman}, {B{\"o}hringer}, {Boroski}, {Bracker}, {Briegel}, {Briggs}, {Brinkmann}, {Brunner}, {Carey}, {Carr}, {Chen}, {Christian}, {Colestock}, {Crocker}, {Csabai}, {Czarapata}, {Dalcanton}, {Davidsen}, {Davis}, {Dehnen}, {Dodelson}, {Doi}, {Dombeck}, {Donahue}, {Ellman}, {Elms}, {Evans}, {Eyer}, {Fan}, {Federwitz}, {Friedman}, {Fukugita}, {Gal}, {Gillespie}, {Glazebrook}, {Gray}, {Grebel}, {Greenawalt}, {Greene}, {Gunn}, {de Haas}, {Haiman}, {Haldeman}, {Hall}, {Hamabe}, {Hansen}, {Harris}, {Harris}, {Harvanek}, {Hawley}, {Hayes},
  {Heckman}, {Helmi}, {Henden}, {Hogan}, {Hogg}, {Holmgren}, {Holtzman}, {Huang}, {Hull}, {Ichikawa}, {Ichikawa}, {Johnston}, {Kauffmann}, {Kim}, {Kimball}, {Kinney}, {Klaene}, {Kleinman}, {Klypin}, {Knapp}, {Korienek}, {Krolik}, {Kron}, {Krzesi{\'n}ski}, {Lamb}, {Leger}, {Limmongkol}, {Lindenmeyer}, {Long}, {Loomis}, {Loveday}, {MacKinnon}, {Mannery}, {Mantsch}, {Margon}, {McGehee}, {McKay}, {McLean}, {Menou}, {Merelli}, {Mo}, {Monet}, {Nakamura}, {Narayanan}, {Nash}, {Neilsen}, {Newman}, {Nitta}, {Odenkirchen}, {Okada}, {Okamura}, {Ostriker}, {Owen}, {Pauls}, {Peoples}, {Peterson}, {Petravick}, {Pope}, {Pordes}, {Postman}, {Prosapio}, {Quinn}, {Rechenmacher}, {Rivetta}, {Rix}, {Rockosi}, {Rosner}, {Ruthmansdorfer}, {Sandford}, {Schneider}, {Scranton}, {Sekiguchi}, {Sergey}, {Sheth}, {Shimasaku}, {Smee}, {Snedden}, {Stebbins}, {Stubbs}, {Szapudi}, {Szkody}, {Szokoly}, {Tabachnik}, {Tsvetanov}, {Uomoto}, {Vogeley}, {Voges}, {Waddell}, {Walterbos}, {Wang}, {Watanabe}, {Weinberg}, {White}, {White}, {Wilhite},
  {Wolfe}, {Yasuda}, {York}, {Zehavi}, \& {Zheng}}]{stoughton02}
{Stoughton}, C., {Lupton}, R.~H., {Bernardi}, M., {et~al.} 2002, \aj, 123, 485

\bibitem[{Strateva {et~al.}(2001)Strateva, Željko Ivezić, Knapp, Narayanan, Strauss, Gunn, Lupton, Schlegel, Bahcall, Brinkmann, Brunner, Budavári, Csabai, Castander, Doi, Fukugita, Győry, Hamabe, Hennessy, Ichikawa, Kunszt, Lamb, McKay, Okamura, Racusin, Sekiguchi, Schneider, Shimasaku, \& York}]{strateva01}
Strateva, I., Željko Ivezić, Knapp, G.~R., {et~al.} 2001, AJ, 122, 1861

\bibitem[{Suzuki {et~al.}(2016)Suzuki, Kodama, Sobral, Khostovan, Hayashi, Shimakawa, Koyama, Tadaki, Tanaka, Minowa, Yamamoto, Smail, \& Best}]{suzuki16}
Suzuki, T., Kodama, T., Sobral, D., {et~al.} 2016, MNRAS, 462, 181

\bibitem[{Sánchez {et~al.}(2020)Sánchez, Sánchez, Bernardi, Nikakhtar, Margalef-Bentabol, \& Sheth}]{sanchez20}
Sánchez, H.~D., Sánchez, H.~D., Bernardi, M., {et~al.} 2020, MNRAS, 495, 2894

\bibitem[{{Tantalo} \& {Chiosi}(2004)}]{tantalo04}
{Tantalo}, R. \& {Chiosi}, C. 2004, \mnras, 353, 917

\bibitem[{{Thomas} {et~al.}(2003){Thomas}, {Maraston}, \& {Bender}}]{thomas03}
{Thomas}, D., {Maraston}, C., \& {Bender}, R. 2003, MNRAS, 339, 897

\bibitem[{Thomas {et~al.}(2005)Thomas, {Maraston}, {Bender}, \& {Mendes de Oliveira}}]{thomas05}
Thomas, D., {Maraston}, C., {Bender}, R., \& {Mendes de Oliveira}, C. 2005, \apj, 621, 673

\bibitem[{Thomas {et~al.}(2011)Thomas, Maraston, \& Johansson}]{thomas11}
Thomas, D., Maraston, C., \& Johansson, J. 2011, MNRAS, 412, 2183–2198

\bibitem[{{Thomas} {et~al.}(2004){Thomas}, {Maraston}, \& {Korn}}]{thomas04}
{Thomas}, D., {Maraston}, C., \& {Korn}, A. 2004, MNRAS, 351, L19

\bibitem[{Thomas {et~al.}(2010)Thomas, Maraston, Schawinski, Sarzi, \& Silk}]{thomas10}
Thomas, D., Maraston, C., Schawinski, K., Sarzi, M., \& Silk, J. 2010, MNRAS, 404, 1775

\bibitem[{{Tojeiro} {et~al.}(2007){Tojeiro}, {Heavens}, {Jimenez}, \& {Panter}}]{TOJEIRO07}
{Tojeiro}, R., {Heavens}, A.~F., {Jimenez}, R., \& {Panter}, B. 2007, \mnras, 381, 1252

\bibitem[{{Tomasetti, E.} {et~al.}(2023){Tomasetti, E.}, {Moresco, M.}, {Borghi, N.}, {Jiao, K.}, {Cimatti, A.}, {Pozzetti, L.}, {Carnall, A. C.}, {McLure, R. J.}, \& {Pentericci, L.}}]{tomasetti23}
{Tomasetti, E.}, {Moresco, M.}, {Borghi, N.}, {et~al.} 2023, A\&A, 679, A96

\bibitem[{Trager {et~al.}(1998)Trager, Worthey, Faber, Burstein, \& González}]{trager98}
Trager, S.~C., Worthey, G., Faber, S.~M., Burstein, D., \& González, J.~J. 1998, ApJ, 116, 1

\bibitem[{Tremonti {et~al.}(2004)Tremonti, Heckman, Kauffmann, Brinchmann, Charlot, White, Seibert, Peng, Schlegel, Uomoto, Fukugita, \& Brinkmann}]{tremonti04}
Tremonti, C.~A., Heckman, T.~M., Kauffmann, G., {et~al.} 2004, ApJ, 613, 898

\bibitem[{{van der Wel} {et~al.}(2016){van der Wel}, {Noeske}, {Bezanson}, {Pacifici}, {Gallazzi}, {Franx}, {Mu{\~n}oz-Mateos}, {Bell}, {Brammer}, {Charlot}, {Chauk{\'e}}, {Labb{\'e}}, {Maseda}, {Muzzin}, {Rix}, {Sobral}, {van de Sande}, {van Dokkum}, {Wild}, \& {Wolf}}]{vanderwel16}
{van der Wel}, A., {Noeske}, K., {Bezanson}, R., {et~al.} 2016, \apjs, 223, 29

\bibitem[{{Vazdekis} {et~al.}(2015){Vazdekis}, {Coelho}, {Cassisi}, {Ricciardelli}, {Falc{\'o}n-Barroso}, {S{\'a}nchez-Bl{\'a}zquez}, {La Barbera}, {Beasley}, \& {Pietrinferni}}]{vazdekis15}
{Vazdekis}, A., {Coelho}, P., {Cassisi}, S., {et~al.} 2015, \mnras, 449, 1177

\bibitem[{{Vazdekis} {et~al.}(2010){Vazdekis}, {S{\'a}nchez-Bl{\'a}zquez}, {Falc{\'o}n-Barroso}, {Cenarro}, {Beasley}, {Cardiel}, {Gorgas}, \& {Peletier}}]{vazdekis10}
{Vazdekis}, A., {S{\'a}nchez-Bl{\'a}zquez}, P., {Falc{\'o}n-Barroso}, J., {et~al.} 2010, \mnras, 404, 1639

\bibitem[{Veale {et~al.}(2017)Veale, Ma, Greene, Thomas, Blakeslee, Walsh, \& Ito}]{veale17}
Veale, M., Ma, C.-P., Greene, J., {et~al.} 2017, MNRAS, 473, 5446

\bibitem[{Werle {et~al.}(2020)Werle, Fernandes, Asari, Coelho, Bruzual, Charlot, de~Carvalho, Herpich, Oliveira, Sodr'e, Dutra, Amorim, \& Sampaio}]{werle20}
Werle, A., Fernandes, R.~C., Asari, N.~V., {et~al.} 2020, MNRAS, 497, 3251

\bibitem[{White {et~al.}(2011)White, Blanton, Bolton, Schlegel, Tinker, Berlind, da~Costa, Kazin, Lin, Maia, McBride, Padmanabhan, Parejko, Percival, Prada, Ramos, Sheldon, de~Simoni, Skibba, Thomas, Wake, Zehavi, Zheng, Nichol, Schneider, Strauss, Weaver, \& Weinberg}]{white11}
White, M., Blanton, M., Bolton, A., {et~al.} 2011, ApJ, 728, 126

\bibitem[{{Wilkinson} {et~al.}(2017){Wilkinson}, {Maraston}, {Goddard}, {Thomas}, \& {Parikh}}]{WILKINSON17}
{Wilkinson}, D.~M., {Maraston}, C., {Goddard}, D., {Thomas}, D., \& {Parikh}, T. 2017, \mnras, 472, 4297

\bibitem[{{Worthey} {et~al.}(1992){Worthey}, {Faber}, \& {Gonzalez}}]{WORTHEY92}
{Worthey}, G., {Faber}, S.~M., \& {Gonzalez}, J.~J. 1992, \apj, 398, 69

\bibitem[{{Worthey} {et~al.}(1994){Worthey}, {Faber}, {Gonzalez}, \& {Burstein}}]{worthey94}
{Worthey}, G., {Faber}, S.~M., {Gonzalez}, J.~J., \& {Burstein}, D. 1994, ApJS, 94, 687

\bibitem[{Worthey \& Ottaviani(1997)}]{worthey97}
Worthey, G. \& Ottaviani, D.~L. 1997, ApJS, 111, 377

\bibitem[{Yildirim {et~al.}(2017)Yildirim, van~den Bosch, van~de Ven, Martín-Navarro, Walsh, Husemann, Gültekin, \& Gebhardt}]{yildirim17}
Yildirim, A., van~den Bosch, R. C.~E., van~de Ven, G., {et~al.} 2017, MNRAS, 468, 4216

\bibitem[{{York} {et~al.}(2000){York}, {Adelman}, {Anderson}, {Anderson}, {Annis}, {Bahcall}, {Bakken}, {Barkhouser}, {Bastian}, {Berman}, {Boroski}, {Bracker}, {Briegel}, {Briggs}, {Brinkmann}, {Brunner}, {Burles}, {Carey}, {Carr}, {Castander}, {Chen}, {Colestock}, {Connolly}, {Crocker}, {Csabai}, {Czarapata}, {Davis}, {Doi}, {Dombeck}, {Eisenstein}, {Ellman}, {Elms}, {Evans}, {Fan}, {Federwitz}, {Fiscelli}, {Friedman}, {Frieman}, {Fukugita}, {Gillespie}, {Gunn}, {Gurbani}, {de Haas}, {Haldeman}, {Harris}, {Hayes}, {Heckman}, {Hennessy}, {Hindsley}, {Holm}, {Holmgren}, {Huang}, {Hull}, {Husby}, {Ichikawa}, {Ichikawa}, {Ivezi{\'c}}, {Kent}, {Kim}, {Kinney}, {Klaene}, {Kleinman}, {Kleinman}, {Knapp}, {Korienek}, {Kron}, {Kunszt}, {Lamb}, {Lee}, {Leger}, {Limmongkol}, {Lindenmeyer}, {Long}, {Loomis}, {Loveday}, {Lucinio}, {Lupton}, {MacKinnon}, {Mannery}, {Mantsch}, {Margon}, {McGehee}, {McKay}, {Meiksin}, {Merelli}, {Monet}, {Munn}, {Narayanan}, {Nash}, {Neilsen}, {Neswold}, {Newberg}, {Nichol}, {Nicinski},
  {Nonino}, {Okada}, {Okamura}, {Ostriker}, {Owen}, {Pauls}, {Peoples}, {Peterson}, {Petravick}, {Pier}, {Pope}, {Pordes}, {Prosapio}, {Rechenmacher}, {Quinn}, {Richards}, {Richmond}, {Rivetta}, {Rockosi}, {Ruthmansdorfer}, {Sandford}, {Schlegel}, {Schneider}, {Sekiguchi}, {Sergey}, {Shimasaku}, {Siegmund}, {Smee}, {Smith}, {Snedden}, {Stone}, {Stoughton}, {Strauss}, {Stubbs}, {SubbaRao}, {Szalay}, {Szapudi}, {Szokoly}, {Thakar}, {Tremonti}, {Tucker}, {Uomoto}, {Vanden Berk}, {Vogeley}, {Waddell}, {Wang}, {Watanabe}, {Weinberg}, {Yanny}, {Yasuda}, \& {SDSS Collaboration}}]{YORK00}
{York}, D.~G., {Adelman}, J., {Anderson}, Jr., J.~E., {et~al.} 2000, \aj, 120, 1579

\bibitem[{{Zahid} {et~al.}(2016){Zahid}, {Geller}, {Fabricant}, \& {Hwang}}]{zahid16}
{Zahid}, H.~J., {Geller}, M.~J., {Fabricant}, D.~G., \& {Hwang}, H.~S. 2016, \apj, 832, 203

\bibitem[{Zhang {et~al.}(2014)Zhang, Han, Shuo, Siqi, Tong-Jie, \& Yan-Chun}]{zhang14}
Zhang, C., Han, Z., Shuo, Y., {et~al.} 2014, RAA, 14, 1221

\bibitem[{{Zhao} {et~al.}(2022){Zhao}, {Variu}, {He}, {Forero-S{\'a}nchez}, {Tamone}, {Chuang}, {Kitaura}, {Tao}, {Yu}, {Kneib}, {Percival}, {Shan}, {Zhao}, {Burtin}, {Dawson}, {Rossi}, {Schneider}, \& {de la Macorra}}]{ZHAO22}
{Zhao}, C., {Variu}, A., {He}, M., {et~al.} 2022, \mnras, 511, 5492

\end{thebibliography}

\begin{appendix}

\section{Correction of the velocity dispersion effect}
\label{sec:app_VD}

The process to transform the raw measurements of Lick indices (or in general spectral features) to data ready for the cosmographic fit was outlined in Sec. \ref{sec:velDisp}. In this Appendix we provide with the functional form of the corrections that need to be applied in order to take data at $\sigma$ $[\text{km s}^{-1}]$ velocity dispersion and instrumental spectral resolution of $1.5$\AA (root mean squared wavelength resolution) to zero velocity dispersion and $\sigma_\text{IR} = 1.06$\AA.

\begin{table}[H]
    \centering
    \captionsetup[table]{skip=10pt}
    \caption{Coefficients of the correction function for each index: $C_I(\sigma, \sigma_\text{IR}) = a_0 + a_1 \sigma + a_2 \sigma^2 + a_3 \sigma^3$.}
    \label{tab:Cfunctions}

    \resizebox{0.45\textwidth}{!}{%
    \tiny
    \begin{tabular}{|l|cccc|}
    \hline
    Index & $a_0$ & $a_1$ $(\times 10^{-4})$ & $a_2$ $(\times 10^{-6})$ & $a_3$ $(\times 10^{-8})$ \\ \hline
    H$\delta_\text{A}$ & 1.007 & -0.297 &  1.893 &  0.137 \\
    H$\delta_\text{F}$ & 0.938 & -9.164 & -9.720 &  1.396 \\
    CN$_1            $ & 0.004 &  0.380 &  0.164 & -0.029 \\
    CN$_2            $ & 0.003 & -0.020 &  0.455 & -0.051 \\
    Ca$4227          $ & 1.011 &  2.267 & -1.146 & -1.621 \\
    G$4300           $ & 1.002 & -0.254 &  1.653 & -0.172 \\
    H$\gamma_\text{A}$ & 0.996 & -0.263 & -0.951 &  0.221 \\
    H$\gamma_\text{F}$ & 1.010 & -0.641 &  2.561 & -0.362 \\
    Fe$4383          $ & 1.002 & -2.635 &  3.658 & -0.202 \\
    Ca$4455          $ & 1.036 &  0.737 &  7.826 & -0.191 \\
    Fe$4531          $ & 1.015 &  1.501 &  2.133 & -0.083 \\
    C$_2 4668        $ & 0.998 & -1.769 &  1.087 &  0.031 \\
    H$\beta          $ & 1.004 &  1.498 &  0.347 &  0.023 \\
    Fe$5015          $ & 1.021 &  1.398 &  5.114 & -0.491 \\
    Mg$_1            $ & 0.000 &  0.026 &  0.078 & -0.010 \\
    Mg$_2            $ & 0.000 &  0.077 &  0.068 & -0.008 \\
    Mg$_\text{b}     $ & 0.997 & -2.877 &  2.714 & -0.023 \\
    Fe$5270          $ & 1.010 &  0.100 &  3.823 & -0.309 \\
    Fe$5335          $ & 1.020 &  0.661 &  6.097 &  0.083 \\
    Fe$5406          $ & 1.004 & -2.339 &  4.860 &  0.341 \\
    Fe$5709          $ & 1.008 &  2.145 &  1.200 &  0.300 \\
    Fe$5782          $ & 1.002 & -1.793 &  4.609 &  0.654 \\
    NaD$             $ & 1.002 &  1.619 &  0.061 &  0.182 \\
    TiO$_1           $ & 0.000 & -0.026 &  0.009 &  0.000 \\
    TiO$_2           $ & 0.000 &  0.031 & -0.013 &  0.002 \\ \hline
    \end{tabular}}
\end{table}

Figure \ref{fig:VDCORS} shows the correction functions for all $25$ basic Lick indices. The choice of colour depends on whether the feature has been used in the fit with each of the models. All the corrections are computed as $3$rd order polynomials in $\sigma$ $[\text{km s}^{-1}]$, with central values for the parameters $a_i$, $C_I(\sigma, \sigma_\text{IR}) = \sum_i a_i \sigma^{i}$, shown in table \ref{tab:Cfunctions}. For the atomic indices the correction takes the form $C(\sigma, \sigma_{\text{IR}}) = I(0, \sigma_\text{MILES}) / I(\sigma, \sigma_\text{IR})$, while for the molecular indices (CN$_1$, CN$_2$, Mg$_1$, Mg$_2$, TiO$_1$, TiO$_2$) it is computed as $C(\sigma, \sigma_{\text{IR}}) = I(0, \sigma_\text{MILES}) - I(\sigma, \sigma_\text{IR})$. $\sigma_\text{IR} = 1.5$\AA \ and $\sigma_\text{MILES} = 1.06$\AA. The uncertainty associated to the correction $\delta C$ is small in all cases, due to the little scatter (over the sample of SDSS giant stars) and good approximation of the average computed $I_0/I(\sigma)$ to a third order polynomial. Both $H\delta$ features show a thicker line, representing the $1\sigma$ exceeding the thickness of the line indicating the fiducial value.

\begin{figure}[H]
	\includegraphics[width=\columnwidth]{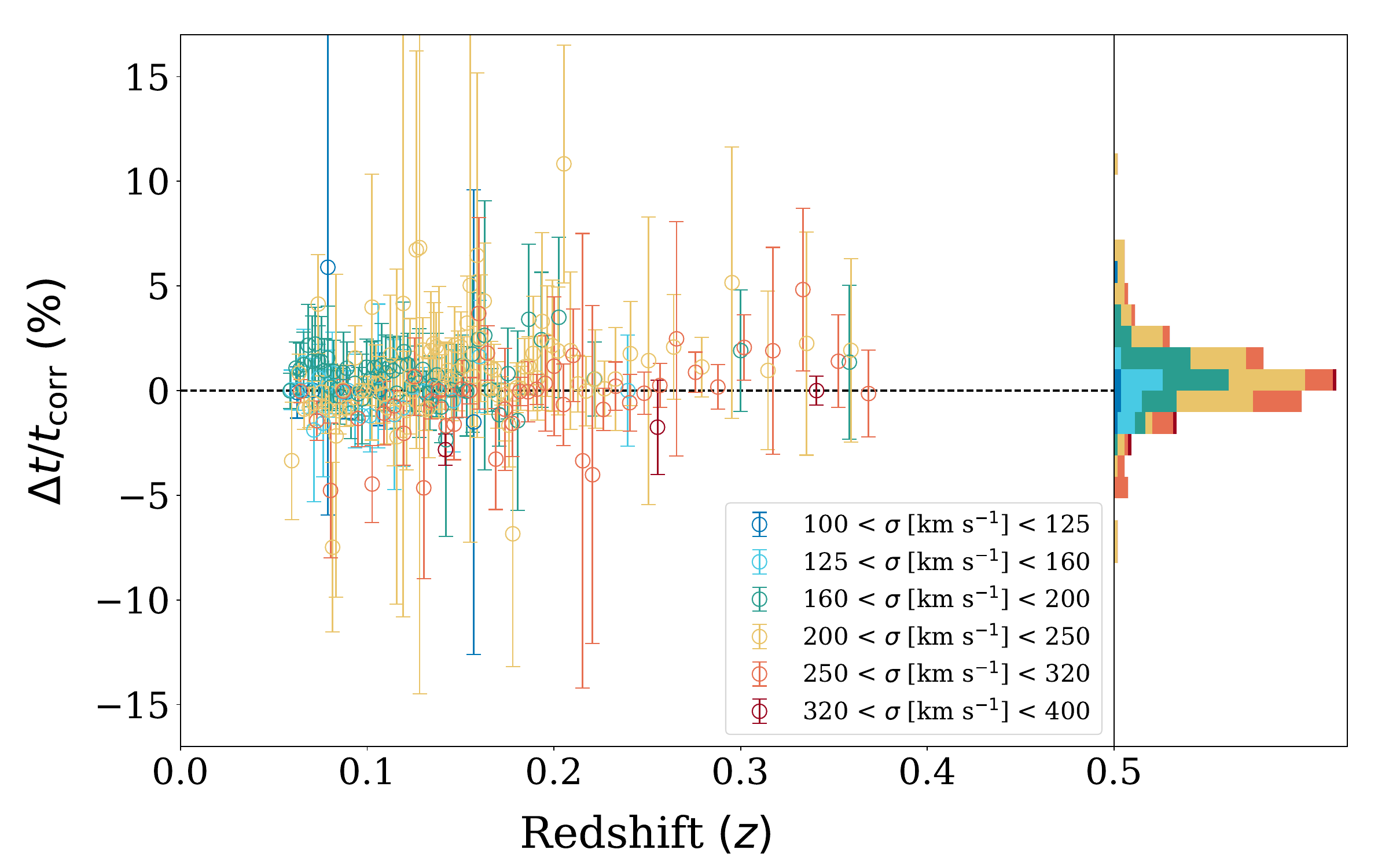}
    \caption{Effect on the modelled age (Knowles model) of varying the methodology to account for the velocity dispersion effect. Colours chosen to represent the velocity dispersion groups in Fig. \ref{fig:cosmography_sMILES}.}
    \label{fig:inmodelvoutmodel}
\end{figure}

In this work we have presented two different ways to address the correction of the velocity dispersion effect imprinted in galaxy spectra. The common way is the aforementioned one, through the correction functions, and applied a posteriori. The other one, which we have presented for the Knowles model, consists in correcting the SPS model to make velocity dispersion become a fine tuning configuration. However, this can only be applied to Lick index models that come from direct integration over spectra, which is not the case of the TMJ model. Even if, as discussed in the text (Sec. \ref{sec:RESULTS}), the ages derived with the Knowles model have to be taken with caution, we analyse the effect that these two different approaches have on the age distribution. To visualise the difference we present in Fig. \ref{fig:inmodelvoutmodel} the rate $\Delta t / t_\text{corr}$ between the age difference derived using both methodologies $\Delta t \equiv t_\text{in-model} - t_\text{corr}$, where $t_\text{in-model}$ is the age derived using the model adapted to different velocity dispersions and $t_\text{corr}$ is the age derived using the corrections.

\begin{figure}[H]
	\includegraphics[width=\columnwidth]{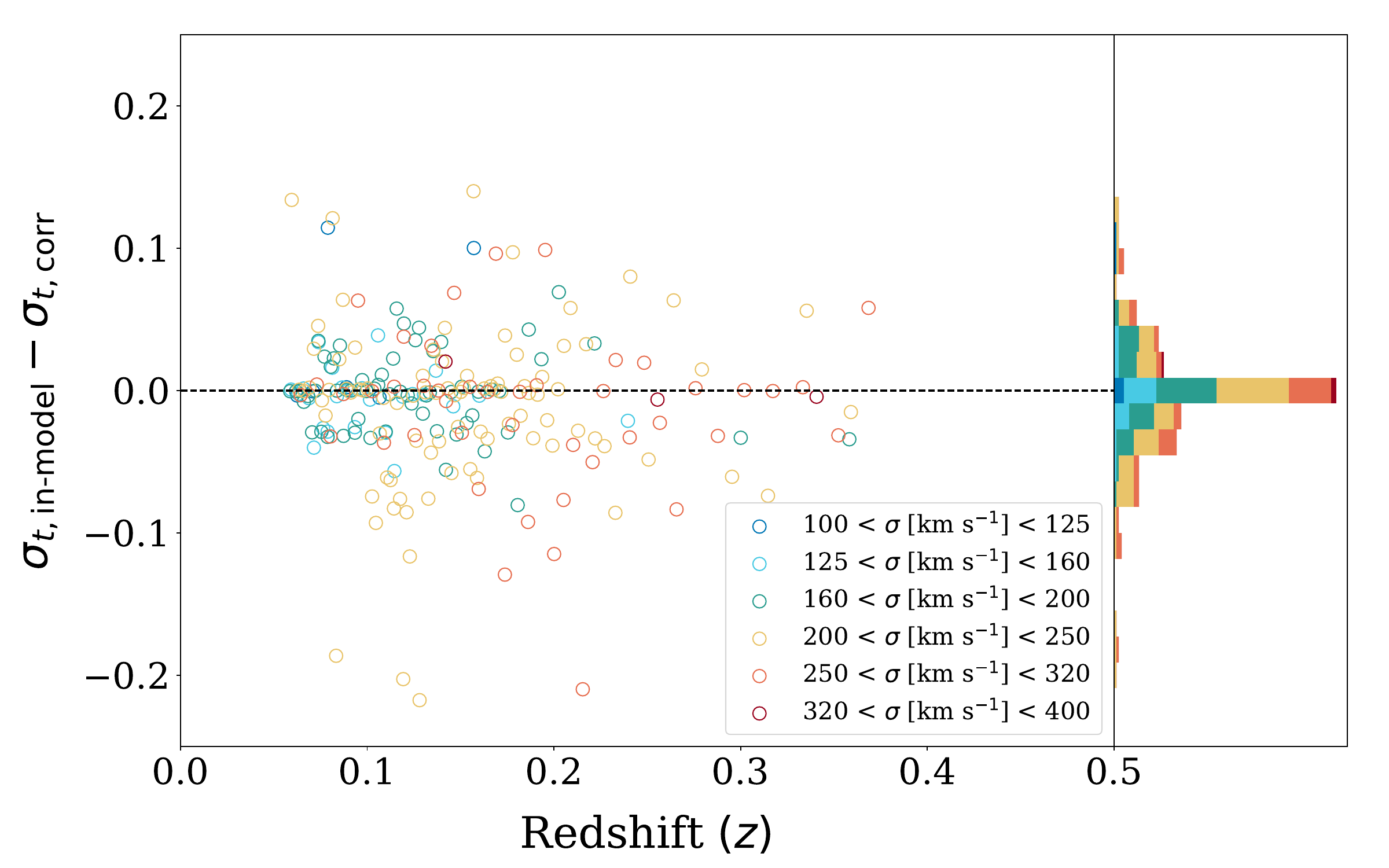}
    \caption{Comparison of the derived uncertainties in Knowles-modelled age coming after the two possible treatments of the velocity dispersion effect.}
    \label{fig:dtvdt}
\end{figure}

\begin{figure*}
    \centering
    \includegraphics[width=0.95\textwidth]{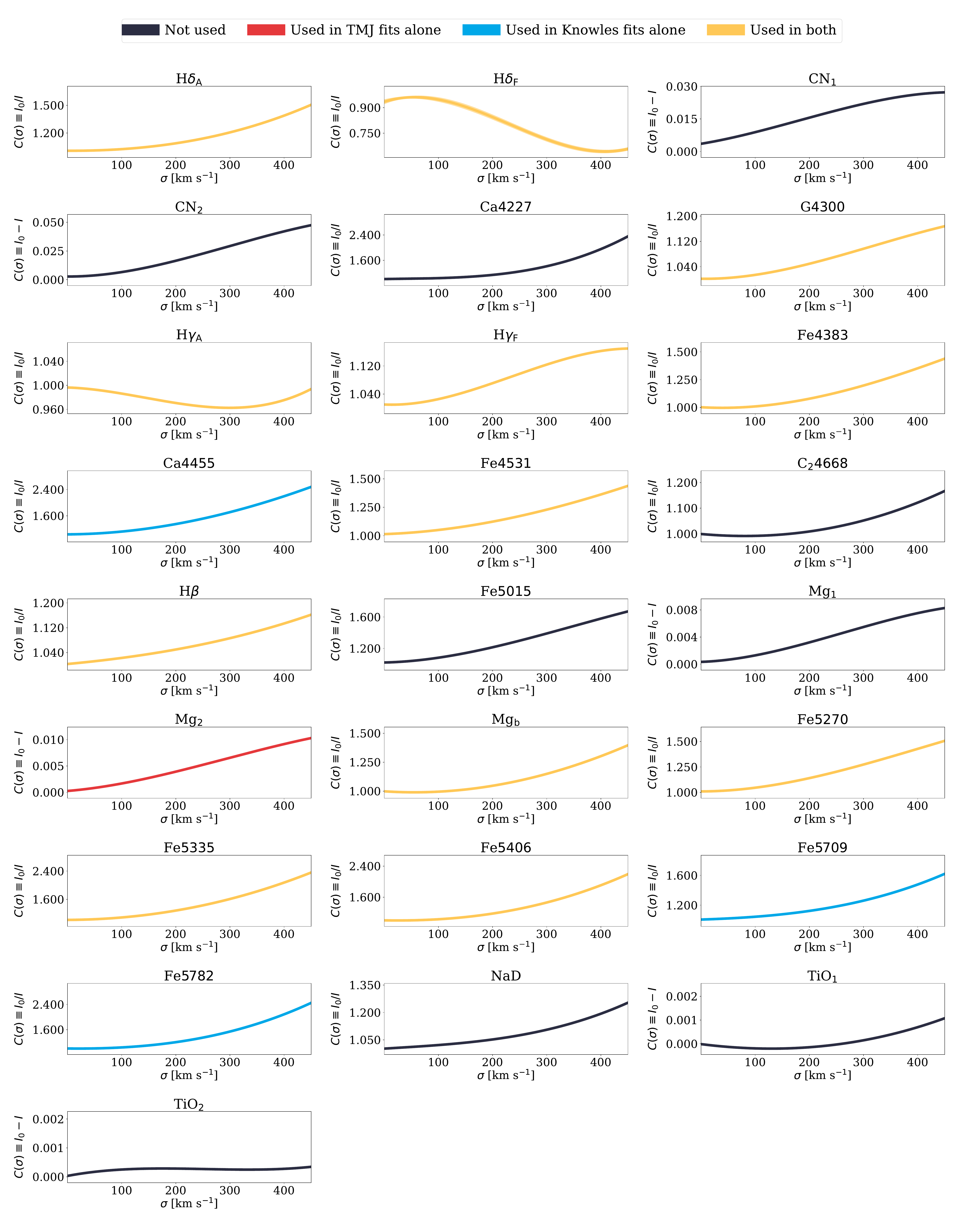}
    \caption{Velocity dispersion correction functions $C(\sigma, \sigma_\text{IR})$ for the $25$ Lick indices. Yellow lines were chosen to represent the indices that were used for the fit with both SPS models, while blue and red indicate those that have been only used with one of them (Knowles and TMJ respectively). Grey tendencies represent the corrections for the indices that have not been used with any of the models.}
    \label{fig:VDCORS}
\end{figure*}

The global distribution of $\Delta t / t_\text{corr}$ is Gaussian and with a mean close to $0$. Indeed, $P_{16}(\Delta t / t_\text{corr}) = -0.9\%$, $\text{Med}(\Delta t / t_\text{corr}) = 0.2\%$ and $P_{84}(\Delta t / t_\text{corr}) = 1.9\%$. We conclude that even if the ages from both methodologies seem highly compatible, some systematic uncertainty can be due to the treatment of the velocity dispersion effect. The uncertainty of the correction functions is already included as a part of the statistical uncertainty in the moment the galactic parameters fit is performed over the set of Lick index measurements. In general, this cannot be accounted for when the model is modified to include velocity dispersion as a fine-tuning parameter. Interested in assessing the impact of the different treatments of the velocity dispersion effect onto the uncertainties derived for the age estimation, we performed a fit of the corrected Lick indices with the Knowles model at zero velocity dispersion. The comparison of the difference in the uncertainties in age coming from both fits, $\sigma_{t, \text{in-model}} - \sigma_{t, \text{corr}}$ is presented in Fig. \ref{fig:dtvdt}.

The distributions for all the velocity dispersion groups, on the right panel in Fig. \ref{fig:dtvdt}, shows a perfect compatibility of the uncertainties in both cases. We see that the choice of methodology to treat the velocity dispersion effect is irrelevant to the age uncertainties, which supports the idea that the nature of each model is the root of the differences observed between Figs. \ref{fig:cosmography_TMJ} and \ref{fig:cosmography_sMILES}.

\section{Cosmographic fits}
\label{sec:corners}

\begin{figure}
    \includegraphics[width=0.49\textwidth]{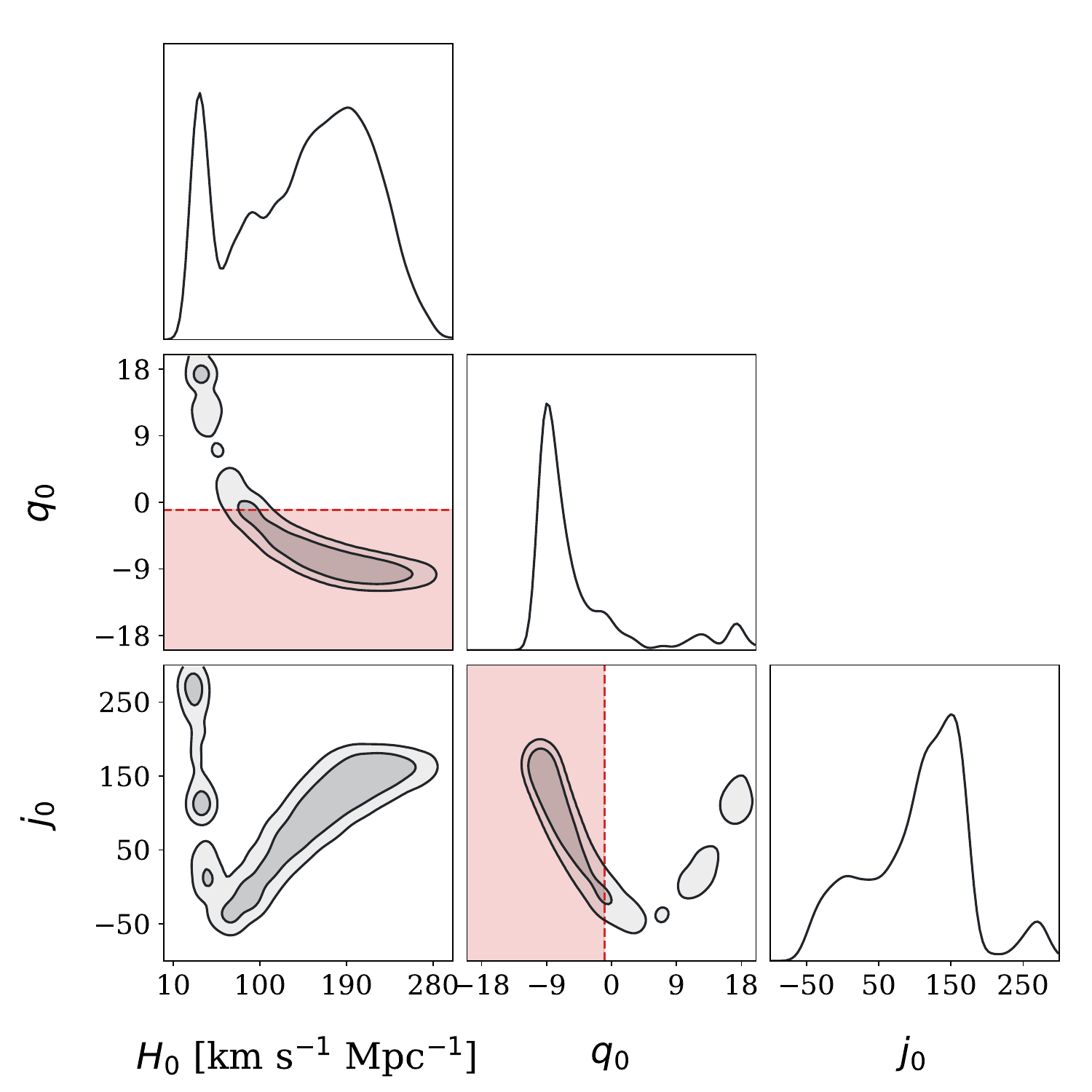}
    \includegraphics[width=0.49\textwidth]{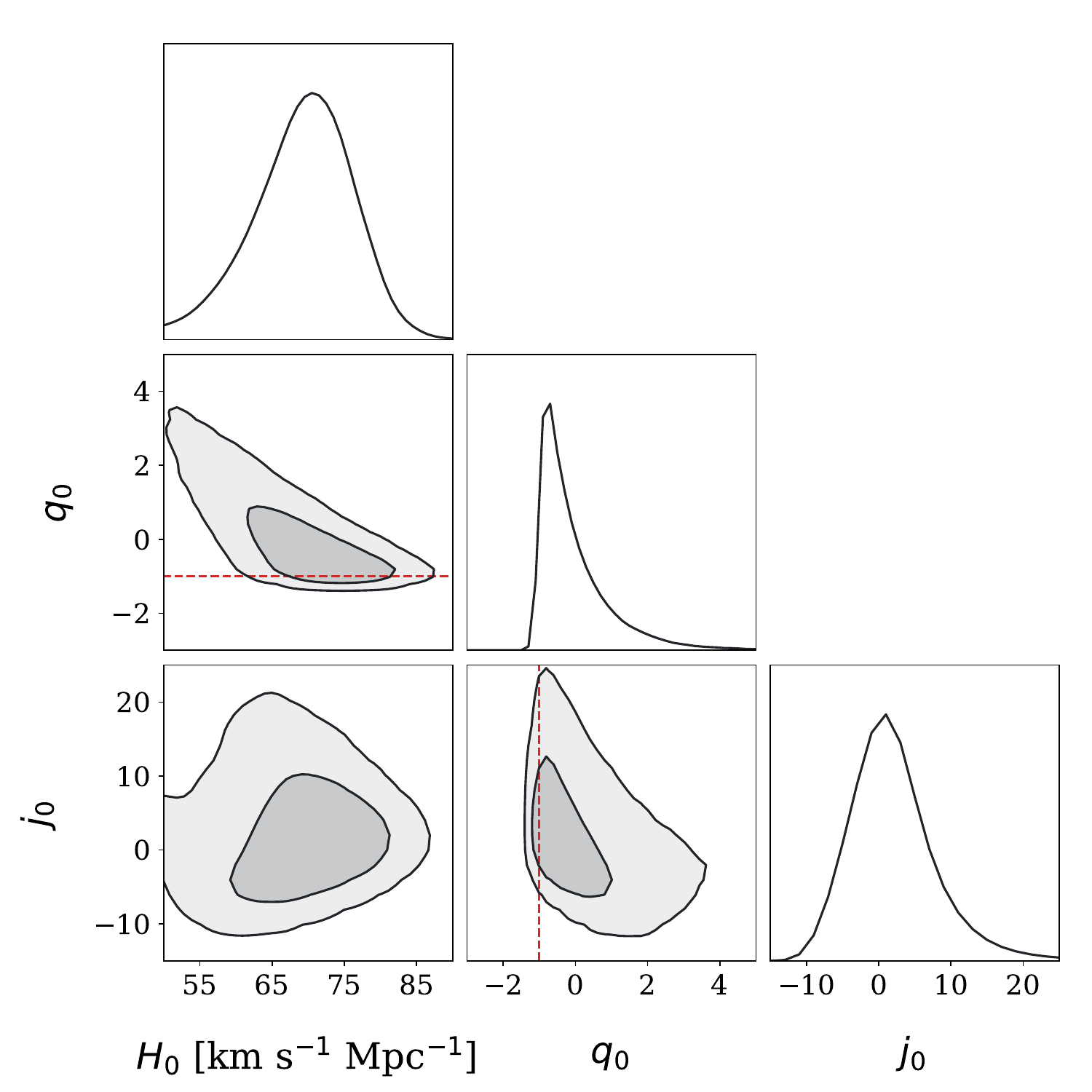}
    \caption{Sampling of the parameter space relative to the cosmographic parameters. On top, the general case in which the $dH/dz>0$ condition was removed, for clarity not only $q_0 = -1$ is drawn but also the whole $q_0 < -1$ region. On the bottom, the baseline case with the exclusion of the data in the range $0.15 < z < 0.20$.}
    \label{fig:cornersTMJ}
\end{figure}

In Sec. \ref{sec:RESULTS} we commented on the results of several cosmographic fits. Of those obtained from the TMJ-modelled ages, we showed the posterior distributions for the baseline (Fig. \ref{fig:cosmography_TMJ}) and the $w_0$CDM readout of our results (Fig. \ref{fig:corner_w0CDM}). Now we show some of the remaining corner plots that we did not present in the main text. In Fig. \ref{fig:cornersTMJ} we unveil the posterior distributions for the case without the $dH/dz > 0$ limitation (top) and omitting for the fit the data in the region of the great oscillation in $0.15 < z < 0.20$ (bottom). For the former, we shadowed the region with $q_0 < -1$, indicating all the parameter space that is limited by the condition on the derivative of $H$ in the baseline fit. For the later, we highlight how the exclusion of the data affected by the greater oscillation in the index-redshift trend does not particularly improve the overall fit.

Finally, we show an example of the full sampling of the parameter space for the baseline scenario, including the ages at zero redshift for each velocity dispersion group, $t_{0, v}$, in Fig. \ref{fig:fullcosmoTMJ}. As commented in the main text, the posterior distributions for the present-time ages are well defined within the prior we set, with typical $1\sigma$ of the order of $0.3$ Gyr.

\begin{figure*}
	\includegraphics[width=\textwidth]{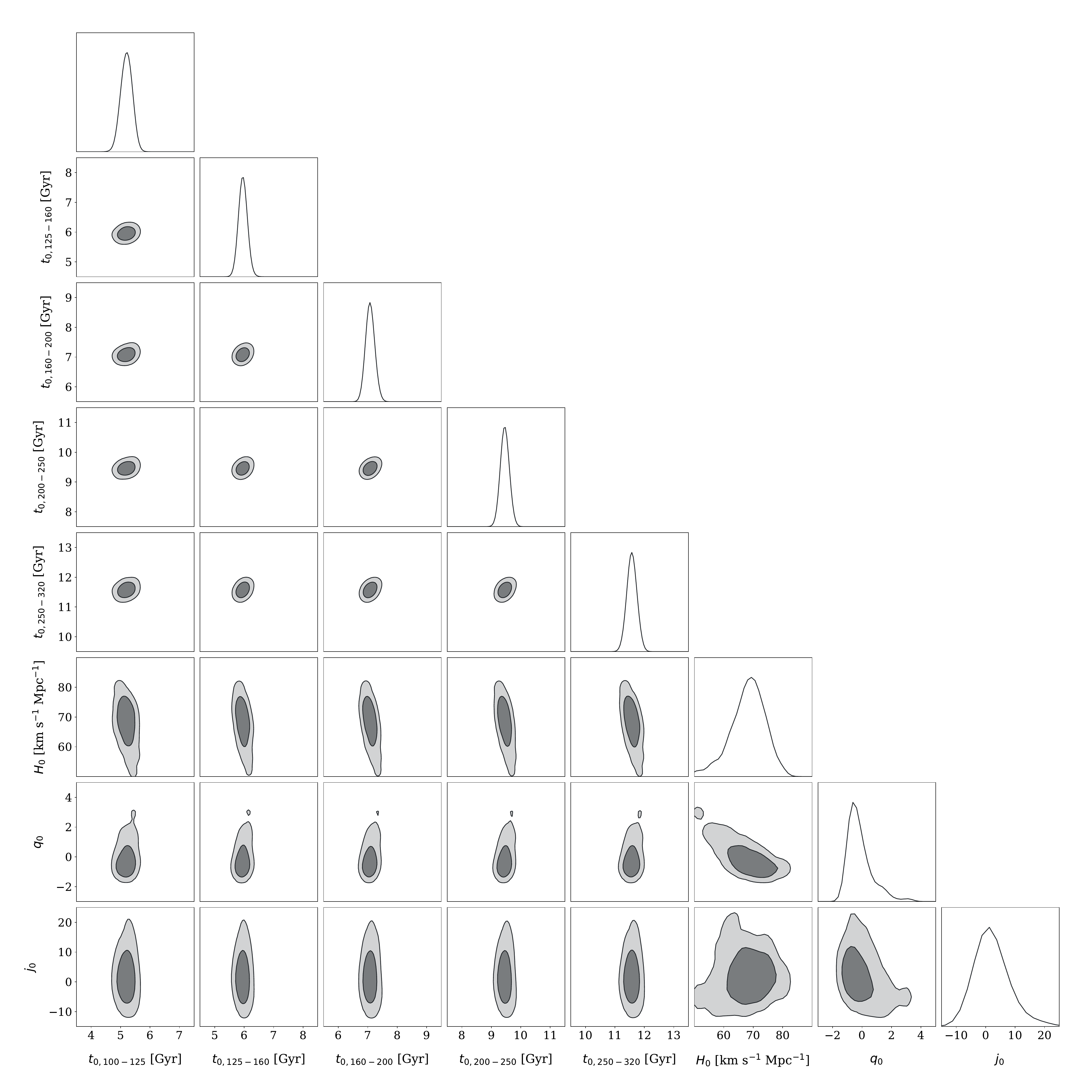}
    \caption{Full posterior probability distributions of the parameters included in the fit from Fig. \ref{fig:cosmography_TMJ}.}
    \label{fig:fullcosmoTMJ}
    \label{fig:fullcosmoTMJ}
\end{figure*}

\section{Scaling relations}
\label{sec:appendixA}

\begin{figure*}
	\includegraphics[width=0.49\textwidth]{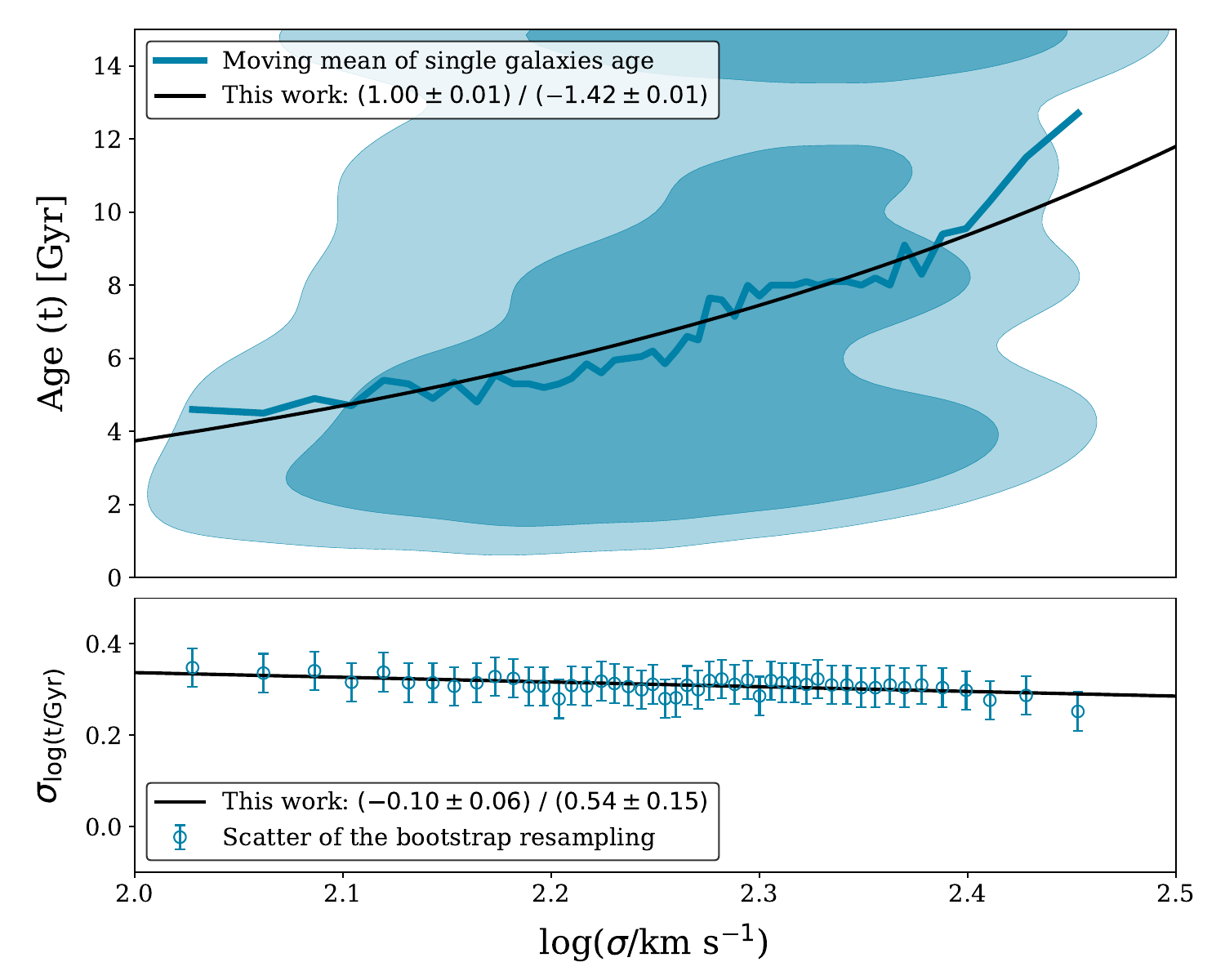}
    \includegraphics[width=0.49\textwidth]{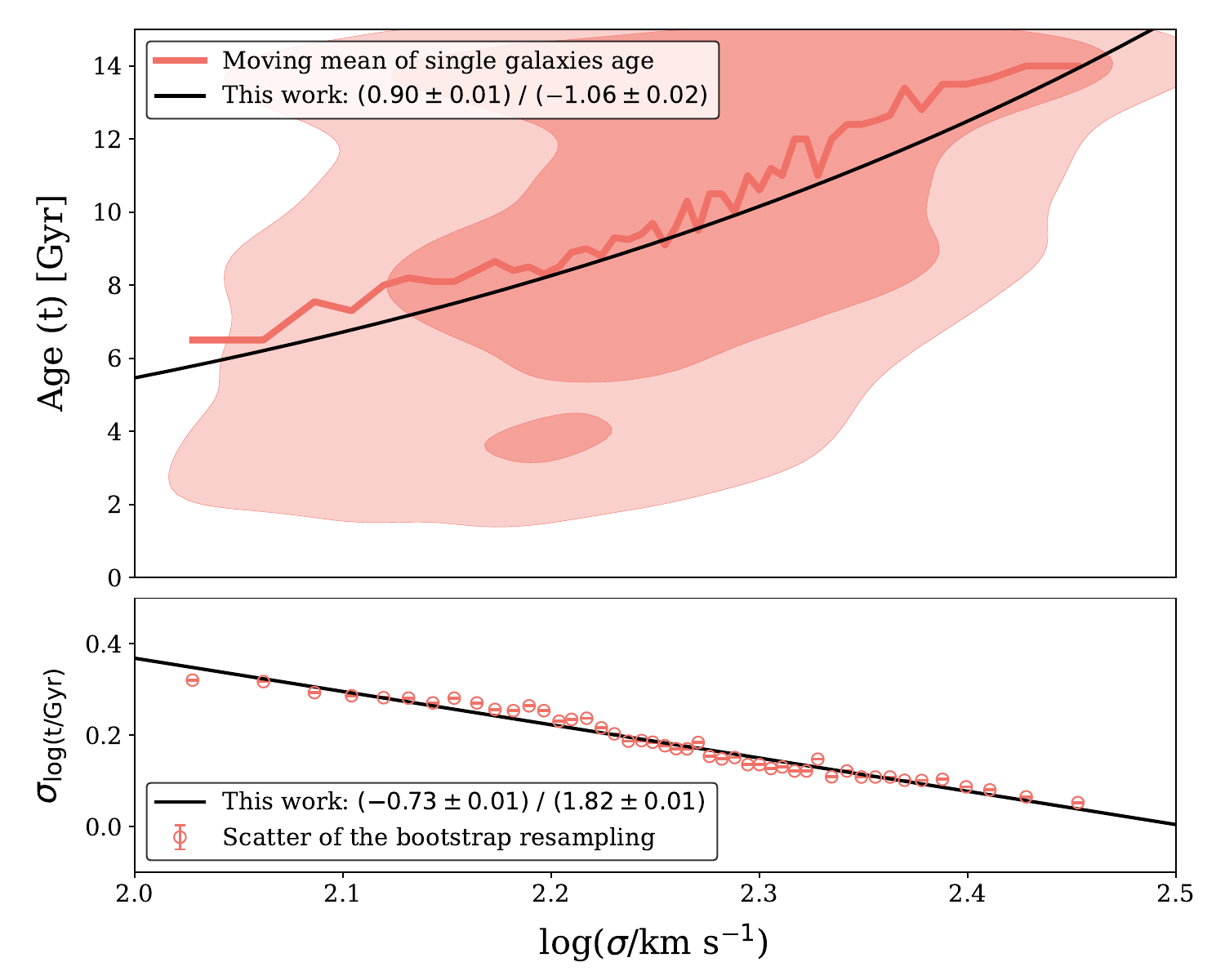}
    \includegraphics[width=0.49\textwidth]{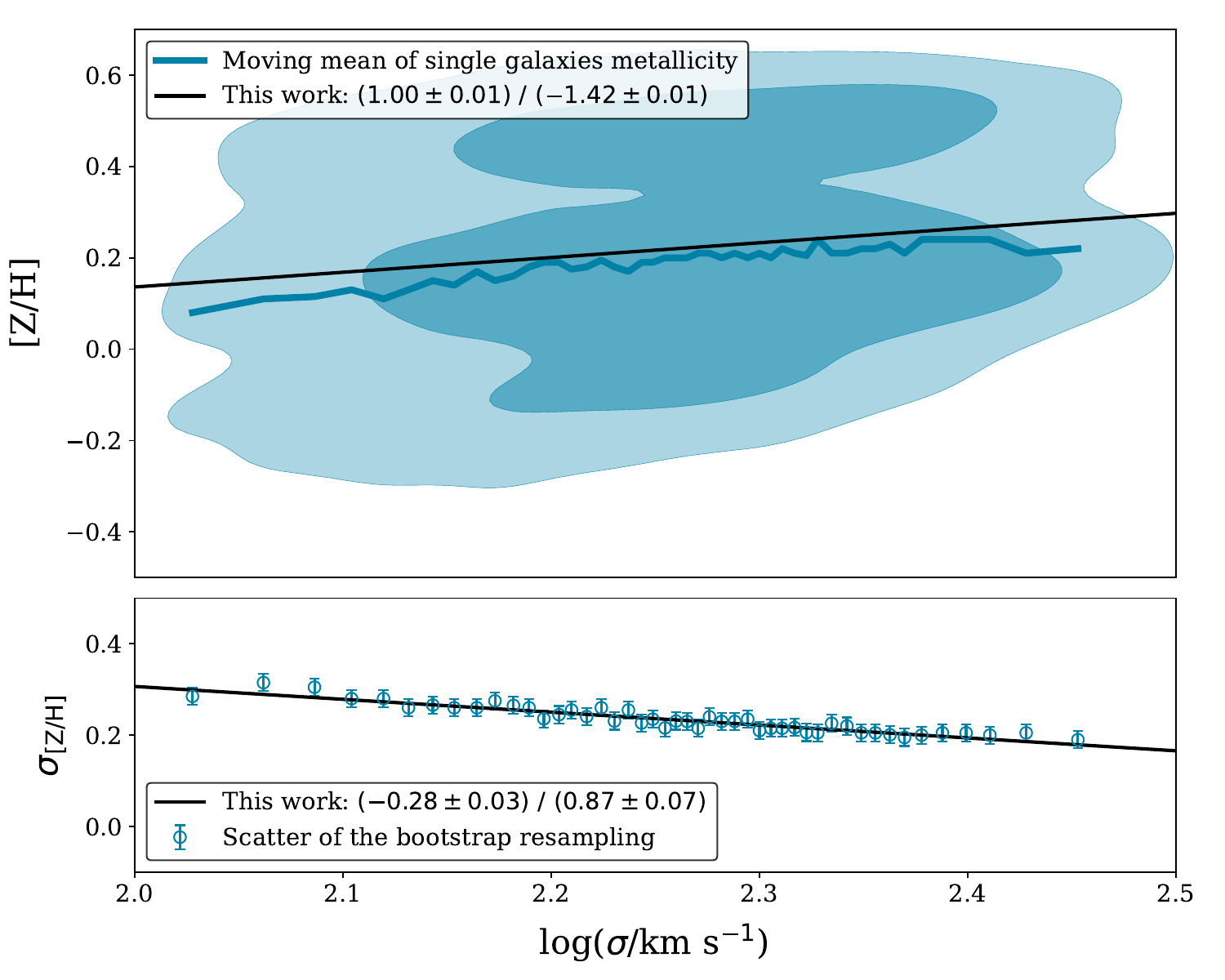}
    \includegraphics[width=0.49\textwidth]{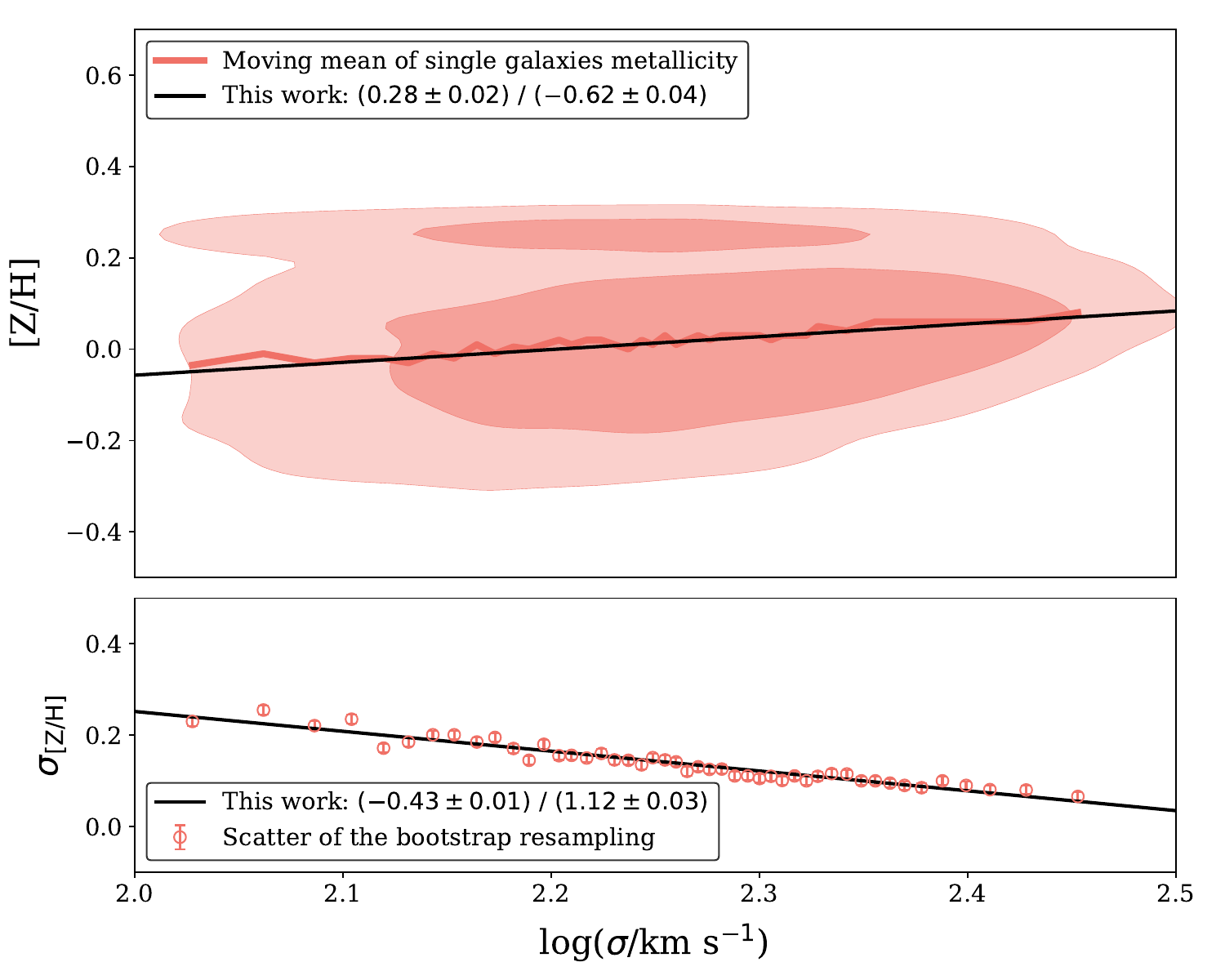}
    \includegraphics[width=0.49\textwidth]{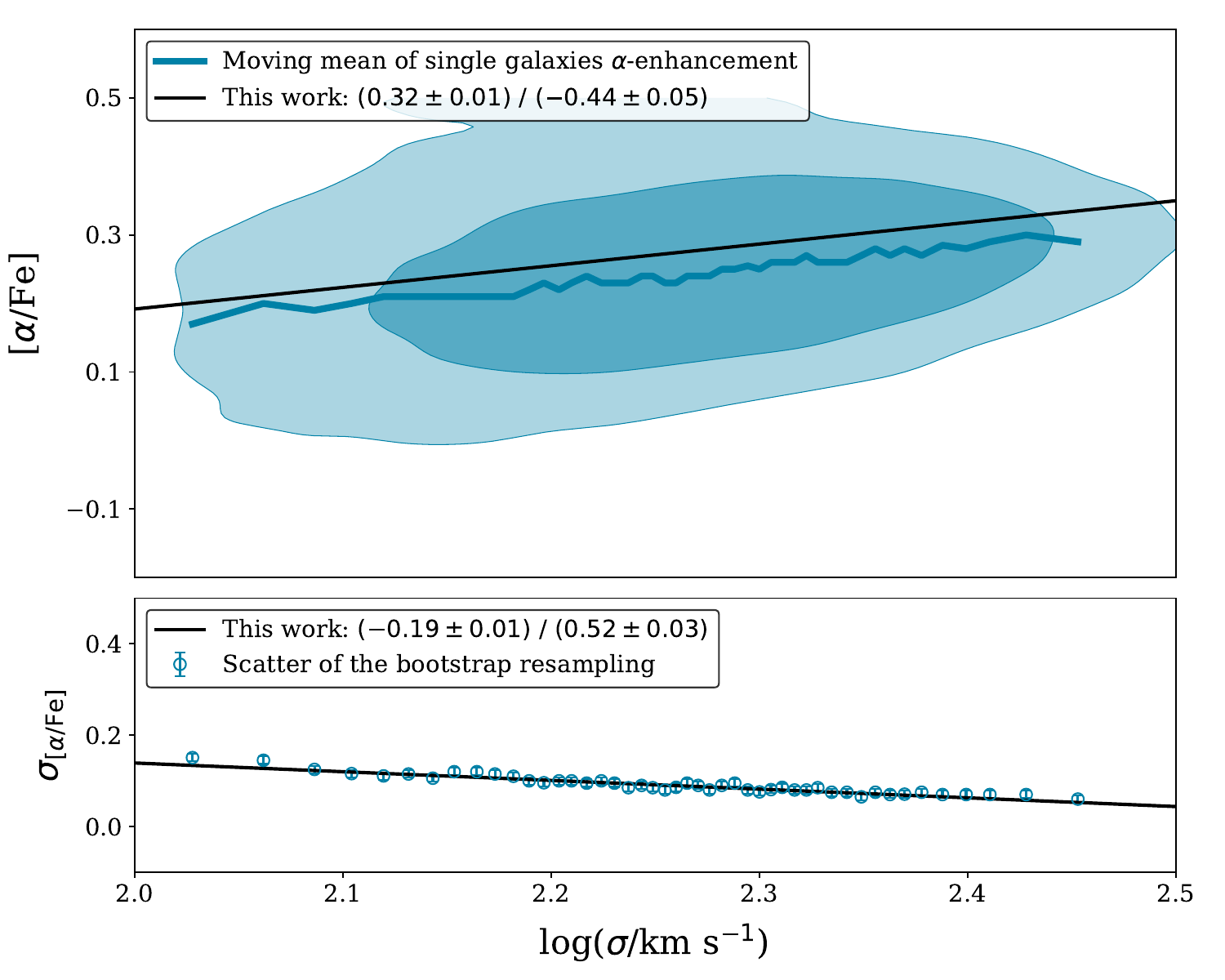} \hspace{0.2cm}
    \includegraphics[width=0.49\textwidth]{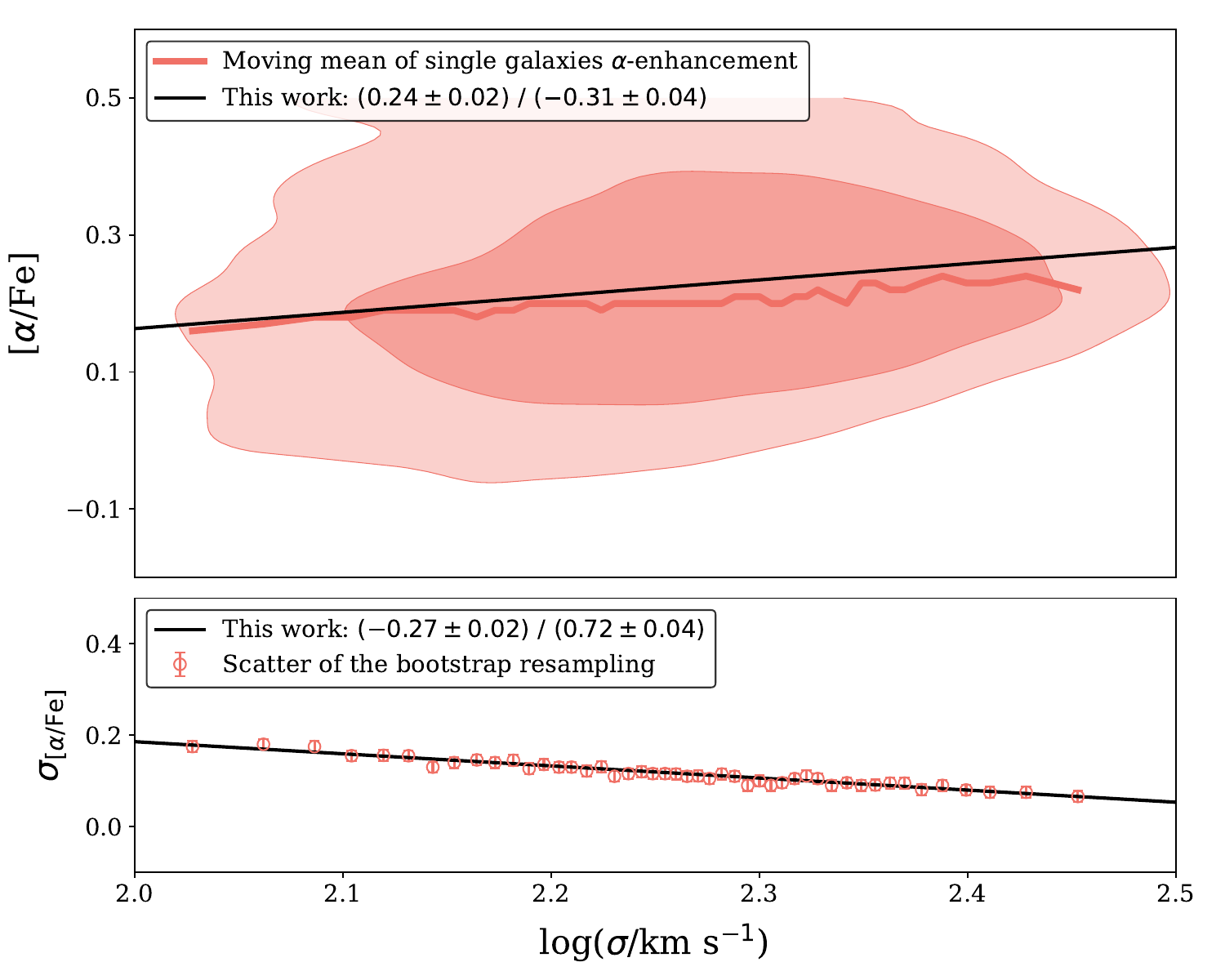}
    \caption{Density contours of the distribution age (upper panels), metallicity (middle panels) and $\alpha$-enhancement (lower panels) for single galaxies. The thick blue line represents the median, while the black line is the best log-log fit from the stacking procedure. The bottom subpanels of each panel show the scatter in the parameter distribution. Results for the TMJ and Knowles models are shown in blue (left panels) and pink (right panels).}
    \label{fig:bootst_TMJandK}
\end{figure*}

This appendix is included to show the distribution of single galaxies in the stellar parameter - velocity dispersion planes, as in Fig. \ref{fig:bootst_TMJandK}, where in blue we presented the results for the TMJ model and in light red for the Knowles model. In black we plot the tendency obtained from the custom stacks that are mapped in Fig. \ref{fig:archaeology_comparison}, showing the good agreement between the moving mean of the single galaxies (darker blue / red line in each case) with the fit coming from the stacks. In the lower panels, the bootstrap methodology explained in the main text give us a relation for the dispersion of the parameters and the velocity dispersion.

\section{Index strength oscillations}
\label{sec:indexoscilappendix}

\begin{figure*}
	\includegraphics[width=0.49\textwidth]{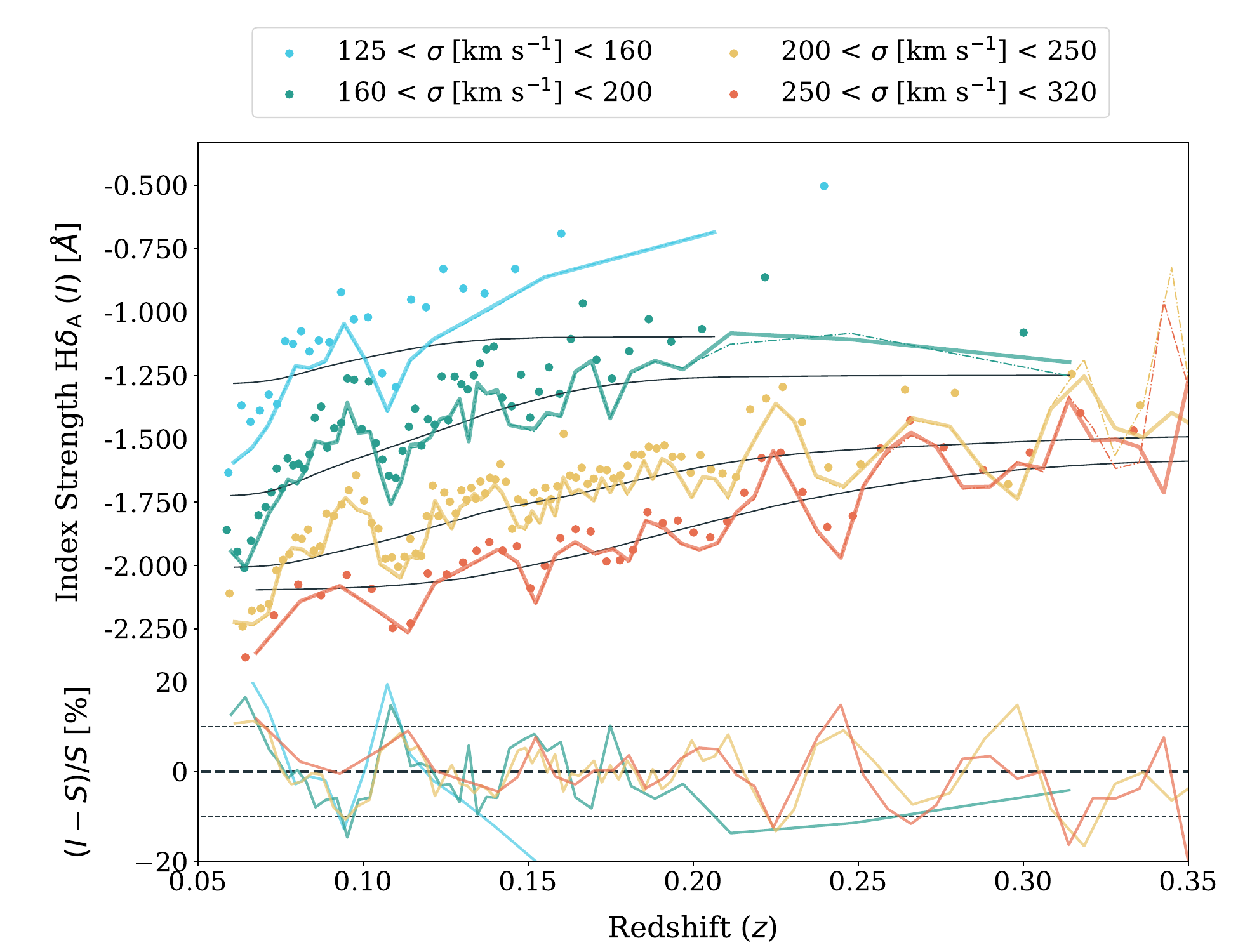}%
    \includegraphics[width=0.49\textwidth]{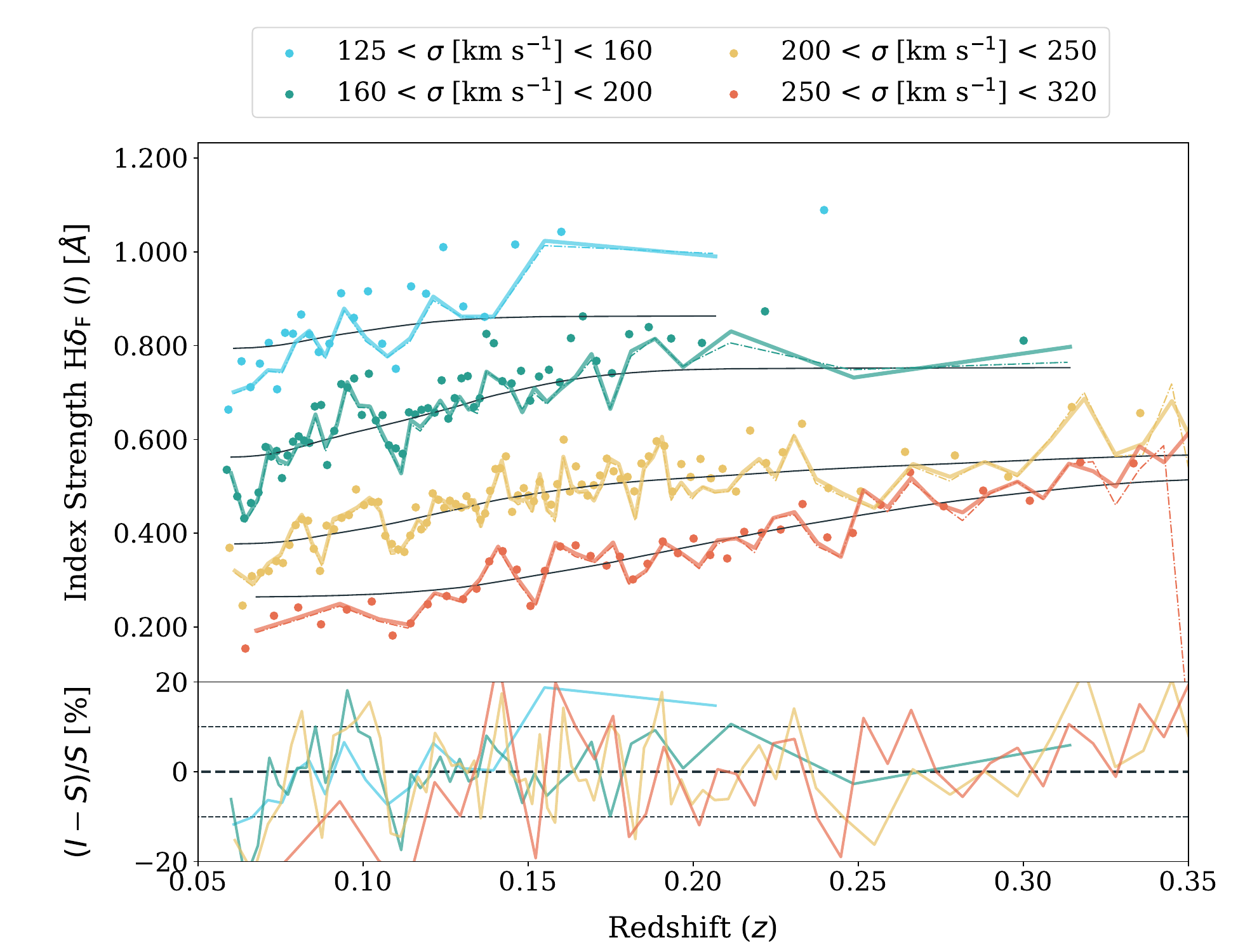}
    \includegraphics[width=0.49\textwidth]{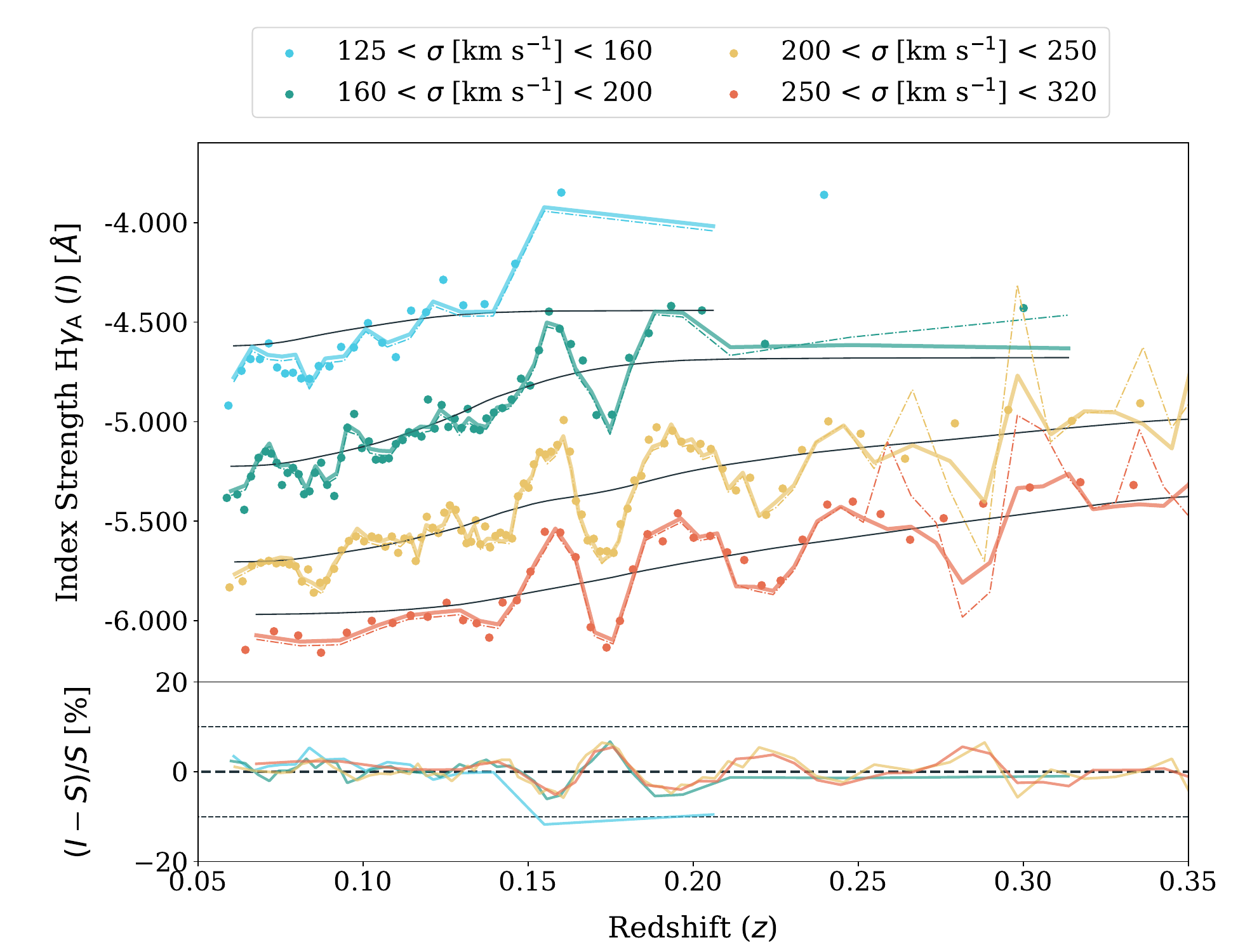}%
    \includegraphics[width=0.49\textwidth]{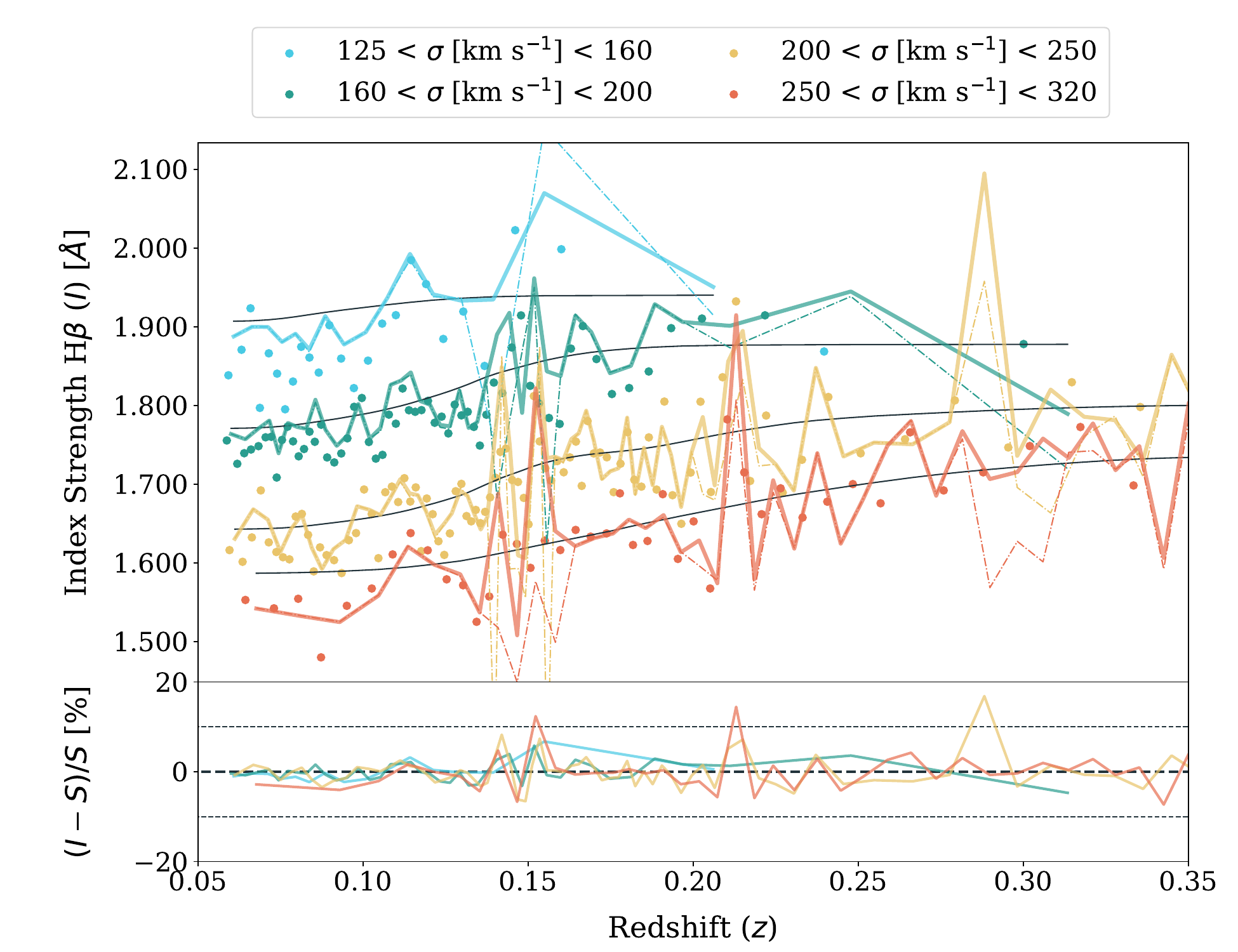}%
    
     \caption{Index-redshift relation for Balmer lines. On top H$\delta_\text{A}$ (left) and H$\delta_\text{F}$ (right). Below H$\gamma_\text{A}$ (left) and H$\beta$. The structure of each panel is equivalent to Fig. \ref{fig:HgF}.}
    \label{fig:otherindicesinfitBALMER}
\end{figure*}

\begin{figure*}
	\includegraphics[width=0.49\textwidth]{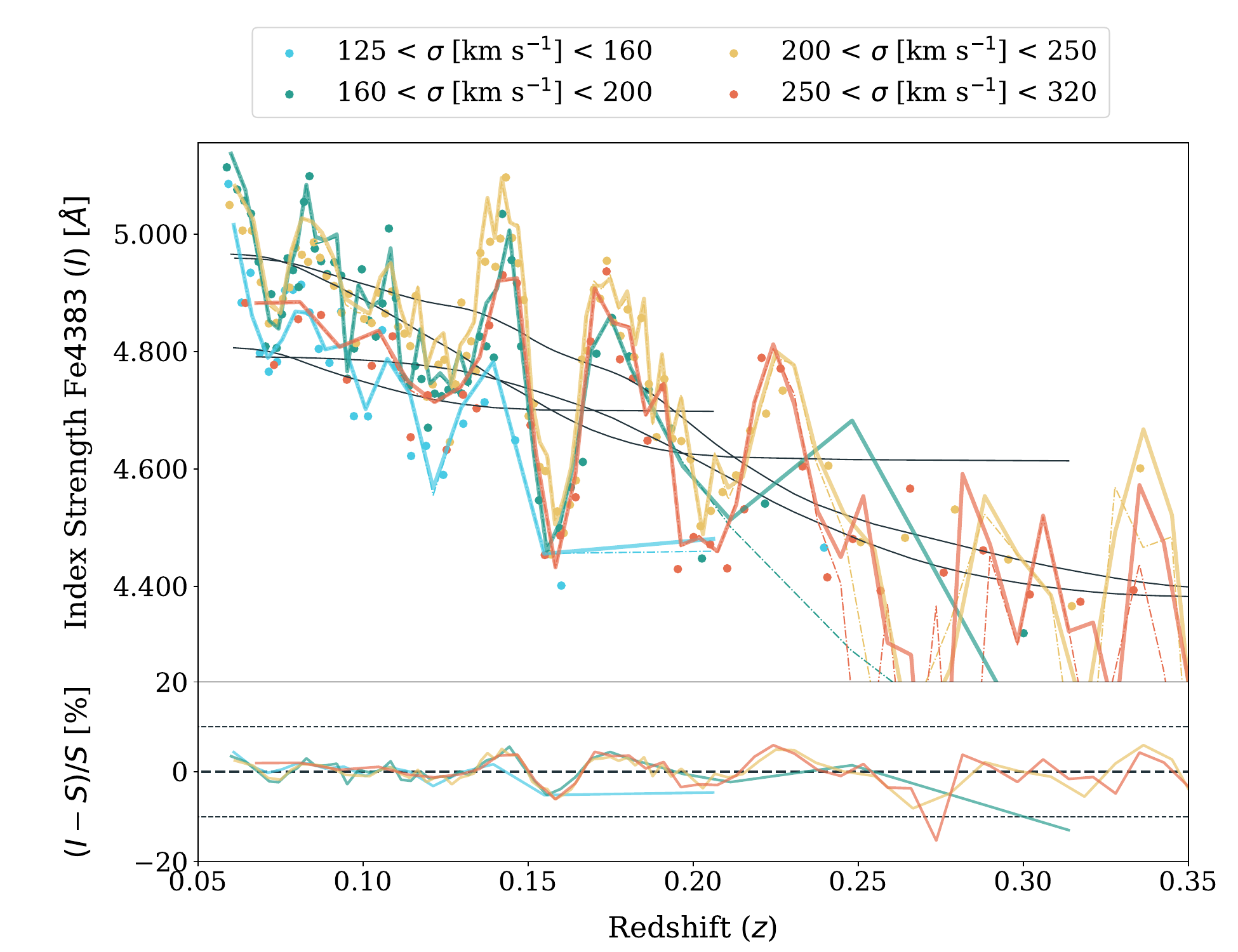}%
    \includegraphics[width=0.49\textwidth]{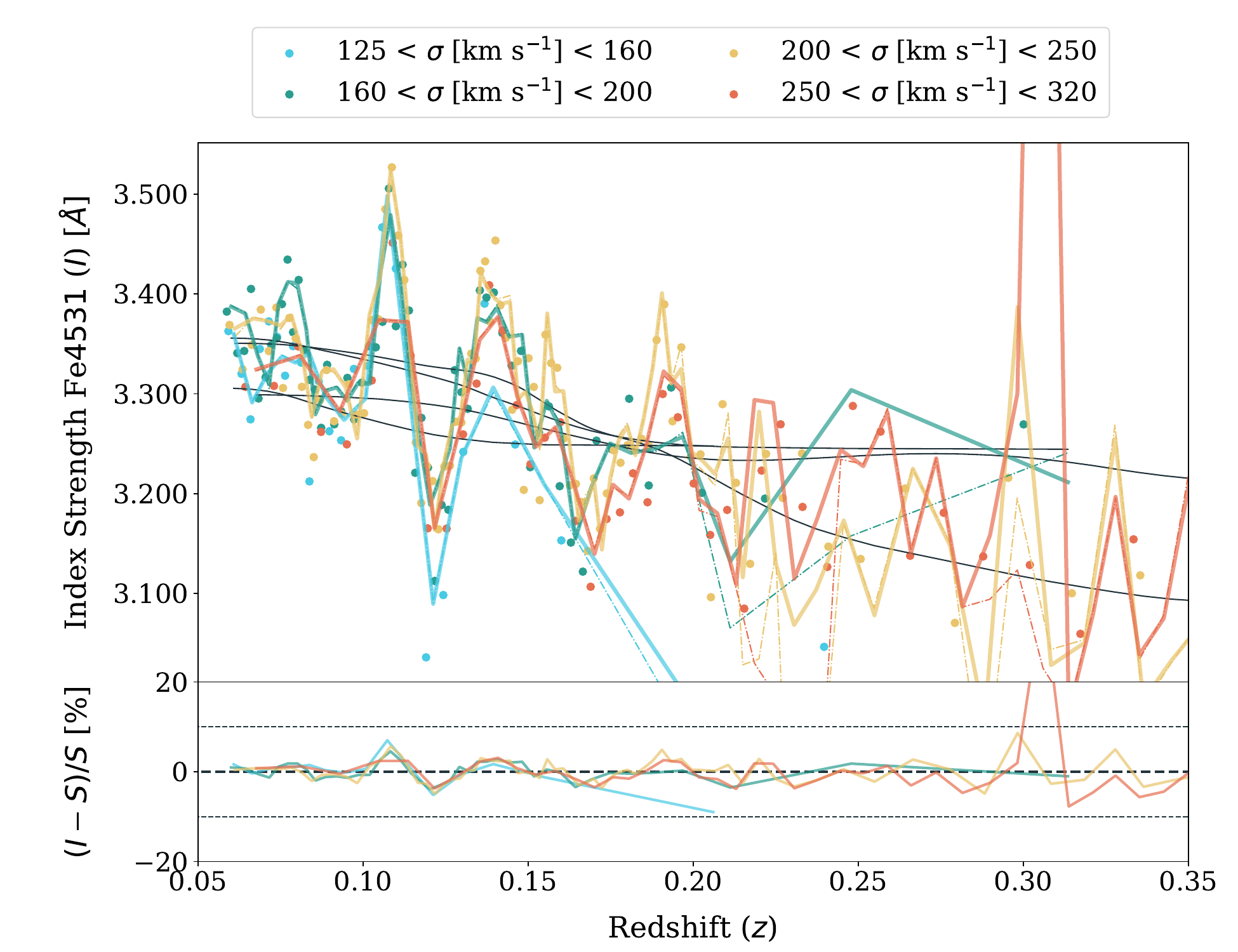}%
     \caption{Index-redshift relation for Iron lines Fe$4383$ (left) and Fe$4531$ (right). The structure of each panel is equivalent to Fig. \ref{fig:HgF}.}
    \label{fig:otherindicesinfitIRON}
\end{figure*}

This appendix is included to show the behaviour of several indices with redshift. We included the remaining Balmer indices (Fig. \ref{fig:otherindicesinfitBALMER}), as they present oscillations that represent deviations from the average value of the index of up to a $10\%$. However, the H$\delta_\text{F}$ index does not exhibit a clear coherence between the different velocity dispersion groups. Two iron indices (Fe$4383$, Fe$4531$) are presented (Fig. \ref{fig:otherindicesinfitIRON}) as the oscillations are clearly synchronous among the velocity dispersion groups. Even if the value of the fluctuation is relatively small relative to the index strength, the expected smooth behaviour is broken clearly.

\begin{figure*}
	\includegraphics[width=0.49\textwidth]{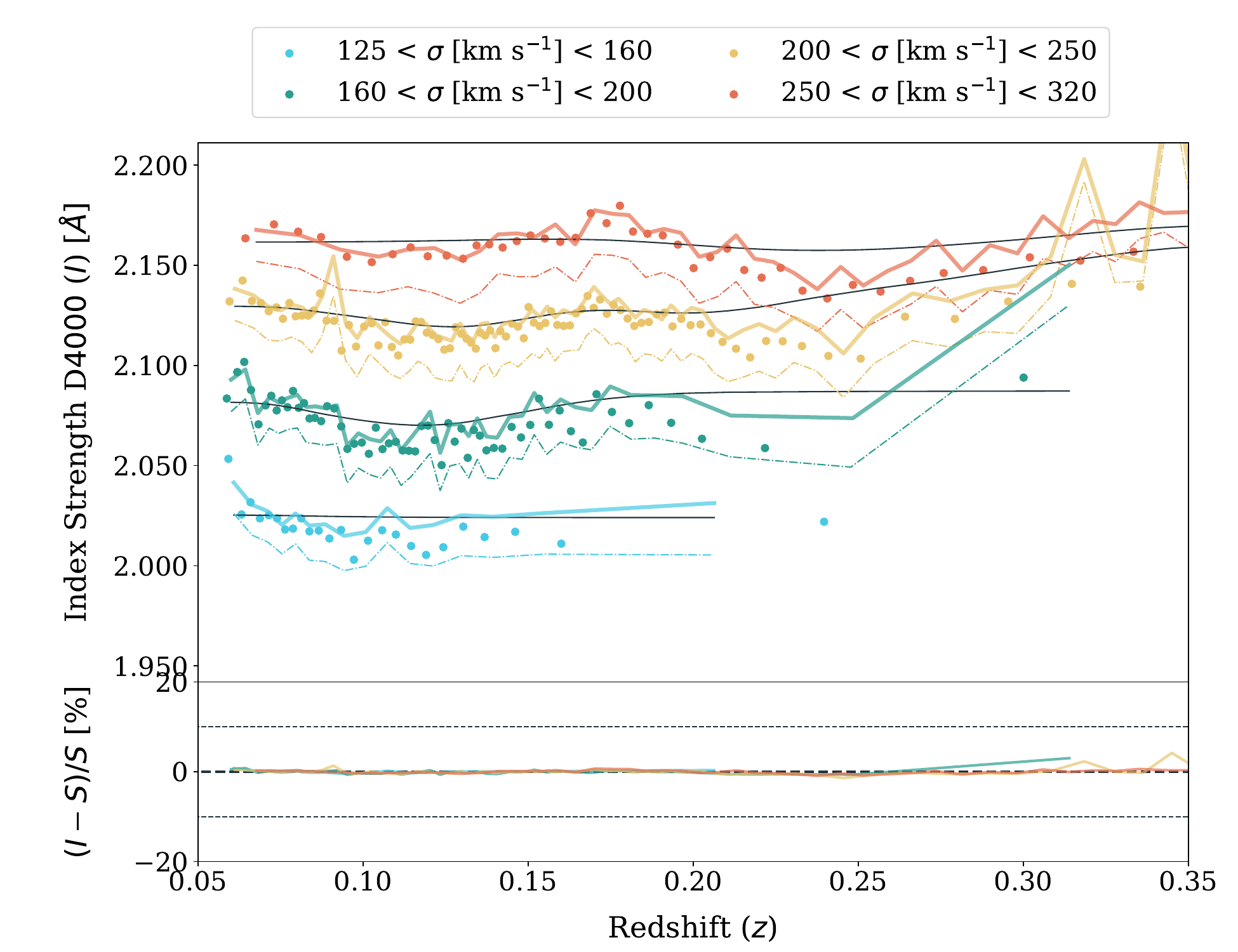}%
    \includegraphics[width=0.49\textwidth]{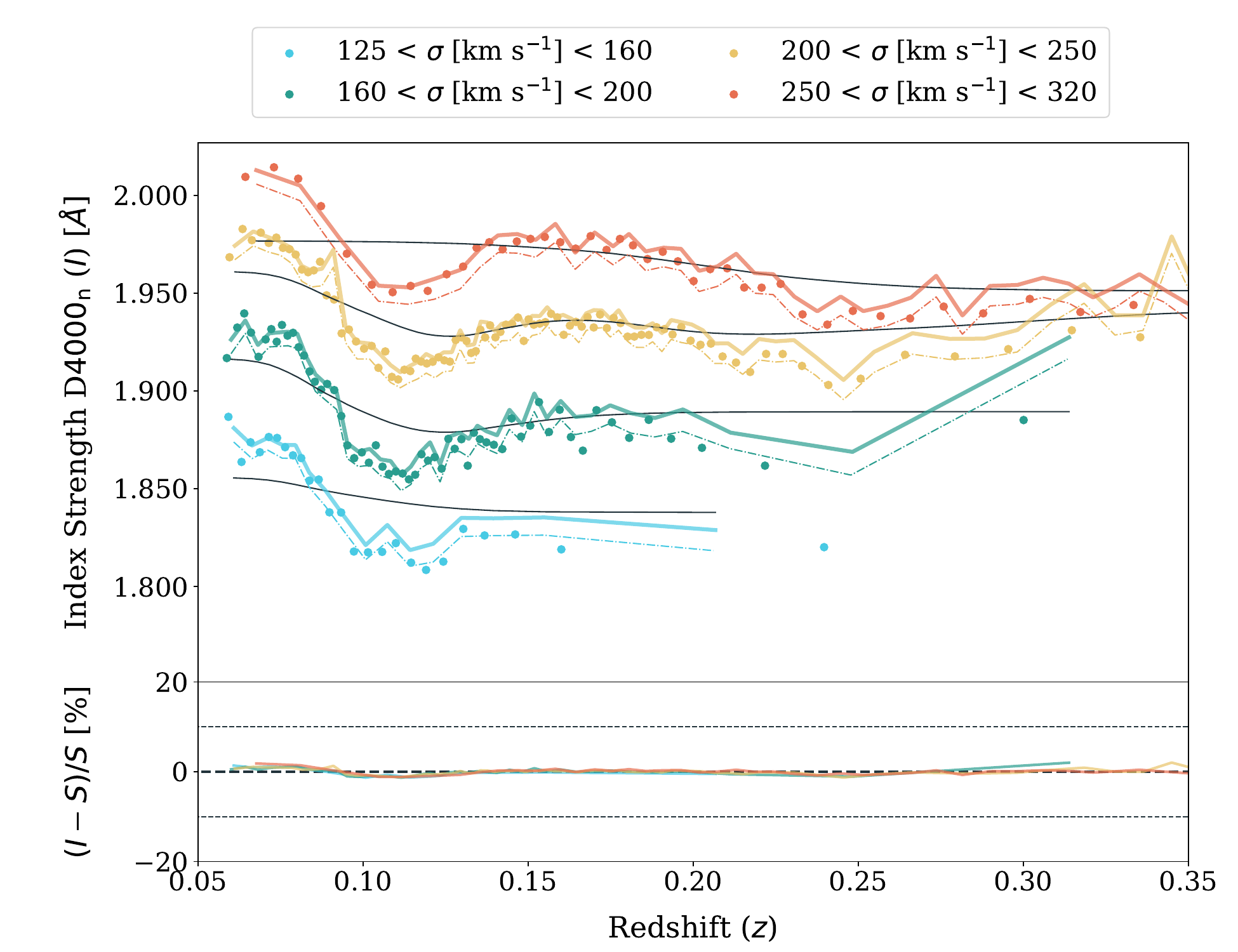}%
    \caption{Index-redshift relation for the discontinuity features, D$4000$ (left) and D$4000_\text{n}$ (right). The structure of each panel is equivalent to Fig. \ref{fig:HgF}.}
    \label{fig:otherindicesnofitD4000}
\end{figure*}

\begin{figure*}
    \includegraphics[width=0.49\textwidth]{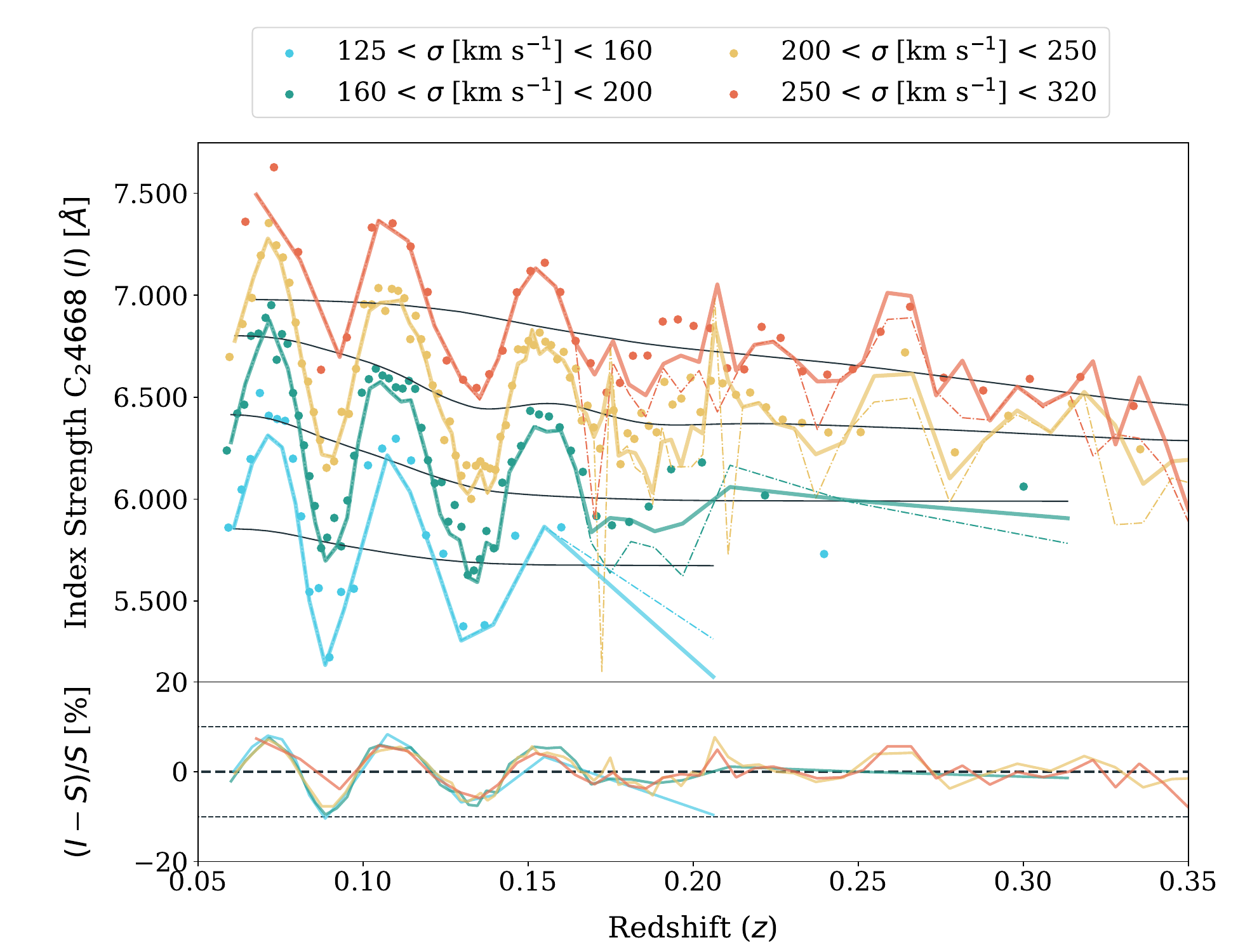}%
    \includegraphics[width=0.49\textwidth]{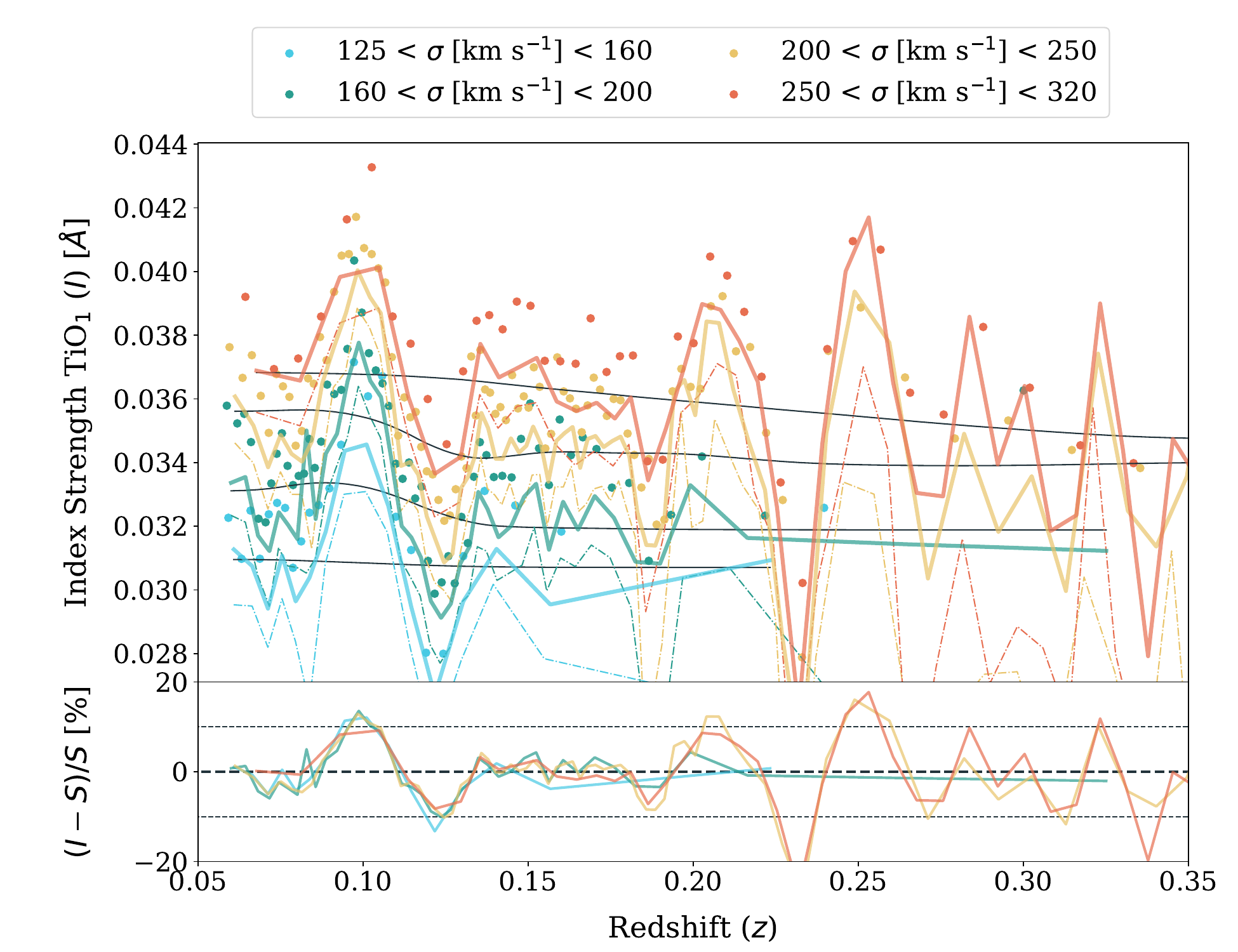}%
    \caption{Index-redshift relation for indices not included in the stellar parameter fits. C$_2 4668$ (left) and TiO$_1$ (right). The structure of each panel is equivalent to Fig. \ref{fig:HgF}.}
    \label{fig:otherindicesnofitC2TiO1}
\end{figure*}

In Fig. \ref{fig:otherindicesnofitD4000} we present the D$4000$ and D$4000_\text{n}$ spectral features, showing a clear non-monotonous behaviour. Particularly, from $z \sim 0.10$ to $z \sim 0.15$, the D$4000_\text{n}$ grows with redshift, opposing the expected soft decreasing trend. Then, in Fig. \ref{fig:otherindicesnofitC2TiO1}, the C$_2 4668$ and TiO$_1$ lines are shown to exhibit the case of an extreme oscillatory behaviour. In all cases the SDSS measurements (thin dashed lines), stacked data (dots) and single galaxies average (thick line) agree, backing our thesis that these are not spurious fluctuations arising from our treatment of the data.

\end{appendix}

\end{document}